%% file: LTC-BIP.tex
\documentclass[onefignum,onetabnum]{siamart190516}

\input{LTC-BIP_shared}

\ifpdf
\hypersetup{
  pdftitle={
 Bayesian inference of heterogeneous epidemic models: Application to COVID-19 spread accounting for long-term care facilities.
  },
  pdfauthor={P. Chen, K. Wu, and O. Ghattas}
}
\fi

\begin{document}

\maketitle

\begin{abstract}
We propose a high dimensional Bayesian inference framework for
learning heterogeneous dynamics of a COVID-19 model, with a specific
application to the dynamics and severity of COVID-19 inside
and outside long-term care (LTC) facilities. 
We develop a heterogeneous compartmental model that accounts for the
heterogeneity of the time-varying spread and severity of COVID-19
inside and outside LTC facilities, which is characterized by
time-dependent stochastic processes and time-independent parameters in $\sim$1500
dimensions after discretization. To infer these parameters, we use
reported data on the number of confirmed, hospitalized, and deceased
cases with suitable post-processing in both a deterministic inversion
approach with appropriate regularization as a first step, followed by
Bayesian inversion with proper prior distributions. To address the
curse of dimensionality and the ill-posedness of the high-dimensional
inference problem, we propose use of a dimension-independent projected
Stein variational gradient descent method, and demonstrate the intrinsic
low-dimensionality of the inverse problem. We present inference
results with quantified uncertainties for both New Jersey and Texas,
which experienced different epidemic phases and patterns. Moreover, we
also present forecasting and validation results based on the empirical
posterior samples of our inference for the future trajectory of
COVID-19.

\end{abstract}



\section{Introduction}
\label{sec:introduction}

The coronavirus disease 2019 (COVID-19) and resulting pandemic has had
a massive impact on worldwide public health (1.2 million deaths and 46
million reported infections as of 31 October
2020\footnote{\url{https://coronavirus.jhu.edu/map.html}}),
nations' economies
(a projected 4.5\% reduction in global GDP in 2020 despite over \$18
trillion in
stimulus\footnote{\url{https://www.oecd.org/economic-outlook/}}),
and the downstream consequences on human welfare.

Since SARS-CoV-2, the virus underling COVID-19, is a novel virus, much
remains uncertain about its transmission and spread. Computer modeling
is thus fundamental to the projections needed to inform decision
making and policy. A hallmark of the COVID-19 pandemic is that it is
characterized by severe heterogeneities across populations: in
infectiousness and susceptibility, in age- and co-morbidity-dependent
morbidity and mortality rates, in human behavior and government
policy, in medical care, and so on. Standard compartment-type
population models in the form of systems of ordinary differential
equations assume well-mixed homogeneous populations
\cite{KeelingRohani11} and thus are poorly suited to modeling strongly
heterogeneous disease spread. Fully accounting for heterogeneities in
infectious disease models, on the other hand, introduces numerous
parameters. Solution of the resulting large scale inverse problems
presents challenges due to the typically noisy, sparse, and
inconsistently reported epidemiological data.

Moreover, due to the uncertainties in both models and data,
quantification of uncertainties in parameter inference and subsequent
predictions is paramount. Adopting the framework of Bayesian inference
to account for these uncertainties leads to a high dimensional
Bayesian inverse problem governed by (for example) differential
equations. The twin curses of high dimensionality of parameter space
and complexity of the forward model have historically made Bayesian
inversion governed by differential equation models
prohibitive. However, a number of theoretical and algorithmic advances
over the past decade are making large-scale (partial) differential
equation-based Bayesian inversion tractable, based on exploitation of
the geometric structure and intrinsic low dimensionality of the
posterior. These include geometrically-informed MCMC methods
\cite{MartinWilcoxBursteddeEtAl12, PetraMartinStadlerEtAl14,
  CuiLawMarzouk16, CuiMartinMarzoukEtAl14, BeskosGirolamiLanEtAl17a,
  Bui-ThanhGirolami14, WangBui-ThanhGhattas18, CuiMarzoukWillcox16} as
well as more recent Stein variational methods
\cite{DetommasoCuiMarzoukEtAl18, ChenGhattas20, ChenGhattas20b,
  ChenWuChenEtAl19}.
These methods have made Bayesian inversion feasible for complex problems
in ice sheet dynamics and global seismology with as many as $O(10^6)$
parameters \cite{Bui-ThanhBursteddeGhattasEtAl12,
  Bui-ThanhGhattasMartinEtAl13, IsaacPetraStadlerEtAl15}.

Our goal in this paper is to specialize such methods---in particular
Stein variational methods, which are particularly appropriate for
highly nonlinear parameter-to-observable maps---to high-dimensional
heterogeneous infectious disease models, with specific application to
modeling COVID-19 accounting for long-term care (LTC) facilities.  A
large percentage (over
40\%\footnote{\url{https://www.nytimes.com/interactive/2020/us/coronavirus-nursing-homes.html}})
of COVID-19 deaths have occurred in nursing homes and other LTC
facilities in the United States (and many other countries), despite
the small percentage ($\sim$0.7\%) of the population in these
facilities. Modeling, inference, and prediction under uncertainty of
the spread and severity of COVID-19---while distinguishing between
populations both inside and outside LTC facilities as well as their
interactions---is crucial for implementing optimal strategies to
contain and mitigate the disease through both non-pharmaceutical and
pharmaceutical interventions.
We endow our COVID-19 epidemiological model (see below) with a large
number of uncertain parameters to account for variations in
transmission, testing, hospitalization ratio, and fatality ratio over
time, as well as other fixed parameters. This results in a challenging
Bayesian inverse problem governed by ODEs with $\sim$1500 parameters,
for which we employ the recently developed projected Stein variational
gradient descent method \cite{ChenGhattas20}. The success of this
method is predicated on the existence of an intrinsic low-dimensional
subspace on which the parameters influence the observables. We
investigate whether the inference problem for COVID-19 dynamics
satisfies this property for our model. 

There have been several different classes of epidemiological models
developed to infer the dynamics of COVID-19 and make
forecasts\footnote{See forecasts in
\url{https://www.cdc.gov/coronavirus/2019-ncov/covid-data/forecasting-us.html},
and \url{https://projects.fivethirtyeight.com/covid-forecasts/}},
including classical compartmental models in the form of ordinary
differential equations \cite{TangWangLiEtAl20,
  MoghadasShoukatFitzpatrickEtAl20, LinkaPeirlinckKuhl20} and partial
differential equations \cite{JhaCaoOden20, LeeLiuTembineEtAl20,
  ViguerieVenezianiLorenzoEtAl20}, network models
\cite{LiPeiChenEtAl20, ChinazziDavisAjelliEtAl20}, statistical
regression models \cite{COVIDMurrayothers20, WoodyTecDahanEtAl20,
  VerityOkellDorigattiEtAl20}, agent-based models
\cite{FergusonLaydonNedjatiGilaniEtAl20, RockettArnottLamEtAl20},
etc. See \cite{Hethcote00, KeelingRohani11, KissMillerSimonEtAl17} for
general references on mathematical modeling of infectious diseases.
Various
strategies to account for heterogeneities in models of the dynamics of
COVID-19 have been considered, including age, underlying health
condition, connectivity, susceptibility, infectiousness, and spatial
heterogeneity. However, given the significant expense of Bayesian
inversion in high dimensions with complex models, a common approach to
inference of parametrized models from data has been to invoke a low
dimensional ansatz on the form of the parametric variation (such as
assuming the variation in time of the basic reproduction number is
represented by a sigmoid function), resulting in just a handful of
parameters. One of our aims is to show that one can relax such
assumptions to better capture the complex dynamics by 
accounting for the time dependent parameter uncertainties via stochastic
processes, and attack the high dimensional inverse problem that
results upon discretization with specialized methods such as projected
Stein variational gradient descent.

\textbf{Contributions}.  In this work, we propose to address the
challenging ill posed inverse problem of Bayesian inference of
COVID-19 models with high dimensional parameter spaces from noisy
data, by exploiting the intrinsic low-dimensionality of the
parameter-to-observable map and through the use of the recently
developed projected Stein variational gradient descent method
\cite{ChenGhattas20}, which requires a number of model solutions that
is independent of the parameter dimension.  We show that it is
feasible to infer models of COVID-19 dynamics in very high parameter
dimensions by aggressively exploiting problem structure. As a testbed
application that is important in its own right,
we model heterogeneity and severity of COVID-19 accounting for disease
dynamics inside and outside LTC facilities and their interactions. We
are not aware of any model in the literature that specifically address
the heterogeneity associated with LTCs and the general population.
We employ a heterogeneous compartmental model with the compartments of
susceptible, exposed, infectious, recovered, hospitalized, deceased,
confirmed, and unconfirmed cases to describe the epidemics both inside
and outside LTC facilities. To better capture the dynamics, we
consider a subset of model parameters to be continuously varying in
time; these include (1) an effective virus transmission reduction
factor that represents a compound effect of, e.g., social distancing,
isolation, quarantining, and mask wearing; (2) an infection
confirmation ratio that reflects the testing capacity and strategy;
(3) an infection hospitalization ratio, and (4) a hospitalization
fatality ratio, the latter two of which relate to time dependent
changes in hospital capacity (ICU beds, ventilators), effectiveness of
treatment, infection variations with age and health condition of the
population, etc. This gives rise to an infinite dimensional Bayesian
inverse problem \cite{Stuart10}, which when discretized in time,
results in a high dimensional problem.

To infer the model parameters, including the $\sim$1500 time-dependent
parameters, we use reported data including the daily number of
deceased cases from inside and outside LTC facilities separately, as
well as the daily number of confirmed cases and the number of
currently hospitalized cases for both inside and outside LTC facilities combined. 
We present both a deterministic inversion 
approach with a suitable regularization, and a Bayesian inference
approach with proper prior distributions to learn these parameters
with quantified uncertainties.
We present the results of inference 
for both approaches, with data from both New Jersey and Texas that
represent different development stages of COVID-19. Our model can
capture well all reported data with 90\% credible intervals
covering the 7-day averaged data. Moreover, we present a future forecast
and validation of the number of confirmed, infected, hospitalized, and
deceased cases using our inference results. In a follow-on article, 
we use these posterior predictives as a basis for stochastic optimal
control to determine optimal mitigation strategies for COVID-19 under
uncertainty.

The paper is structured as follows. We present the heterogeneous
compartment model in Section \ref{sec:model}, followed by presentation
of the available data and their post-processing in Section
\ref{sec:data}. Section \ref{sec:deterministic} is devoted to
presentation of a deterministic inversion formulation with appropriate
regularization of the parameters and the inference results. This
serves as a first step for a Bayesian inference formulation with
proper prior distributions in Section \ref{sec:Bayesian}. We introduce
the projected Stein variational gradient descent method for the
Bayesian inference problem and the results in Section
\ref{sec:Bayesian}. Finally, we draw some conclusions and perspectives
in Section \ref{sec:conclusions}.


\section{Model}
\label{sec:model}

\begin{figure}[!htb]
	\centering
	\includegraphics[width=\linewidth]{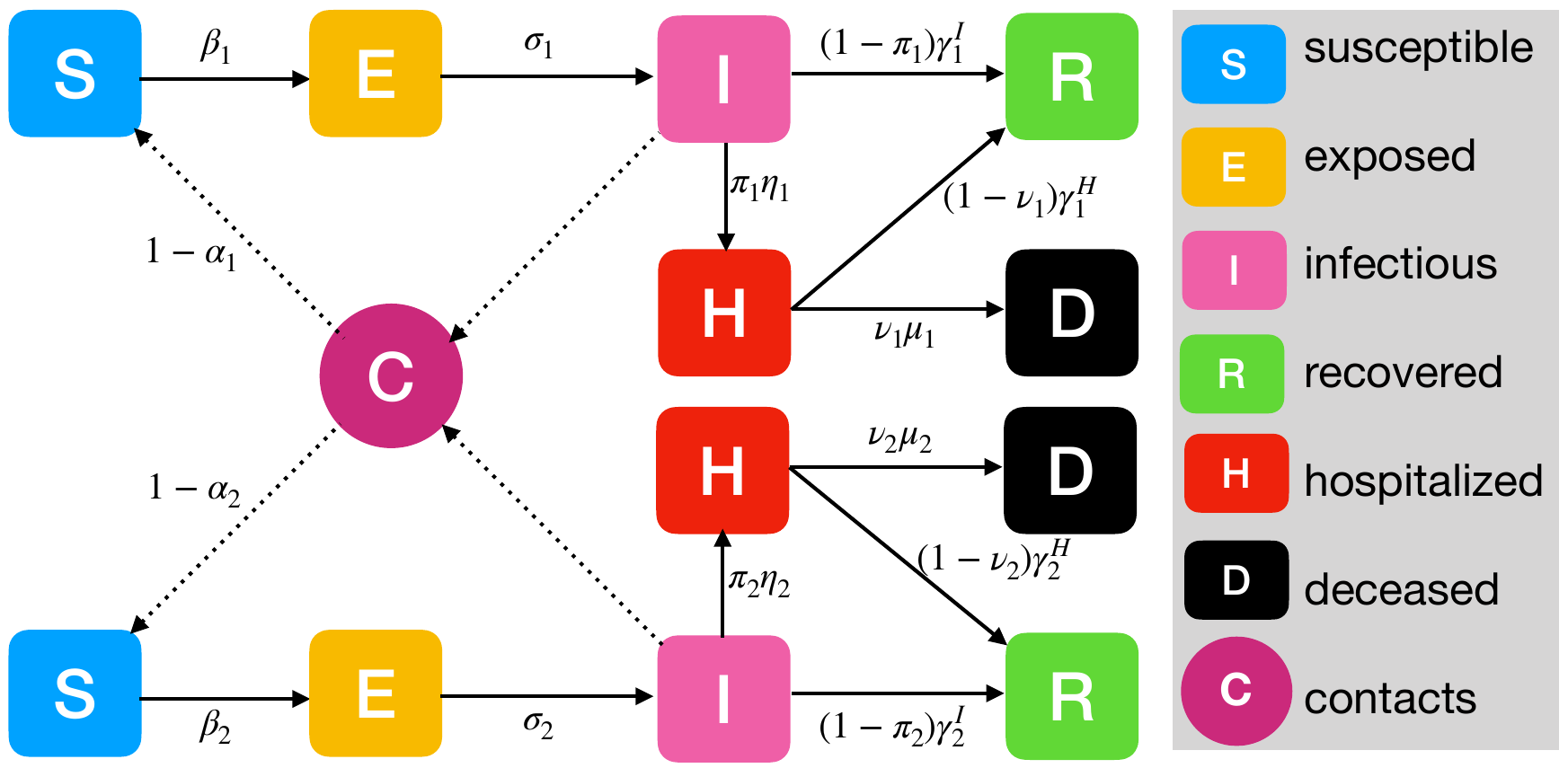}
	\caption{Diagram of a compartment model for the epidemics of COVID-19 inside and outside long-term care facilities.}\label{fig:SEIRHD}
\end{figure}

To learn the epidemics of COVID-19 both inside and outside long-term care facilities (LTC), we present a data-driven modeling of the epidemics, where the model is constructed by the principle of (1) being able to explain the available reported data,
specifically the number of daily deaths inside and outside LTC for a certain period of time, the number of currently hospitalized cases, and the number daily confirmed cases, and (2) as simple as possible to avoid over-parametrization and overfitting. Under such principle, we consider the following compartmental model, as shown in Figure \ref{fig:SEIRHD}, with two groups to describe the epidemics of COVID-19 inside and outside LTC within a time period $t \in (t_{\text{start}}, t_{\text{end}}]$

\begin{equation}\label{eq:LTC-model}
\begin{split}
\frac{dS_i(t)}{dt} & = - \beta_i C_i(t)S_i(t),
\\
\frac{dE_i(t)}{dt} & = \beta_i C_i(t)S_i(t) - \sigma_i E_i(t),
\\
\frac{dI_i(t)}{dt} & = \sigma_i E_i(t) - \pi_i(t) \eta_i I_i(t) - (1-\pi_i(t)) \gamma^I_i I_i(t),
\\
\frac{dH_i(t)}{dt} & = \pi_i(t) \eta_i I_i(t)  - \nu_i(t) \mu_i H_i(t) - (1-\nu_i(t)) \gamma^H_i H_i(t),
\\
\frac{dR_i(t)}{dt} & = (1-\pi_i(t)) \gamma_i I_i(t) + (1-\nu_i(t)) \gamma^H_i H_i(t),
\\
\frac{dD_i(t)}{dt} & = \nu_i(t) \mu_i H_i(t),
\\
\frac{dP^c_i(t)}{dt} & = \tau_i(t) \sigma_i E_i(t),
\\
\frac{dP^u_i(t)}{dt} & = (1-\tau_i(t))\sigma_i E_i(t),
\end{split}
\end{equation}
where the compartmental variables $S_i, E_i, I_i, H_i, R_i, D_i$ represent the number of susceptible, exposed, infectious, hospitalized, recovered, and deceased cases inside LTC ($i = 1$) and outside LTC ($i = 2$), and $P^c_i$ and $P^u_i$ represent the accumulated number of positive (infected) cases that are confirmed/reported by viral test or medical diagnosis, and unconfirmed/unreported inside LTC ($i = 1$) and outside LTC ($i = 2$). The effective transmission factor $C_i(t)$ is given by
\begin{equation}
C_i(t)  = (1-\alpha_i(t)) \, \sum_{i} \bC_{ij}\frac{I_j(t)}{N_j}.
\end{equation}
Here the numbers of population inside LTC and outside LTC are represented by $N_1$ and $N_2$. The time-dependent parameter $\alpha_i(t) \in (0, 1)$ accounts for an effective reduction of the transmission at time $t$ for group $i$ due to, e.g., social distancing, isolation, and quarantine. The contact matrix $\bC\in \bR^{2\times 2}$ represents the number of contacts inside ($\bC_{11}$) and outside ($\bC_{22}$) LTC, as well as the interactions between them ($\bC_{12}$ and $\bC_{21}$) through LTC and hospital staff, and family visitors to LTC. Specifically, $\bC_{12}$ represents the number of contacts between susceptible LTC residents and infectious others, while $\bC_{21}$ represents the number of contacts between infectious LTC residents and susceptible others. The parameters $\beta_i, \sigma_i, \eta_i, \mu_i, \gamma^I_i, \gamma^H_i$ are the transmission rate, the latency rate, the hospitalization rate, the deceased rate, and the recovery rate from infectious and hospitalized stages, which are the inverse of the transmission period, the latency period (exposed but not infectious yet), from infectious to hospitalized period, from hospitalized to deceased period, from infectious to recovery (not infectious) period, and from hospitalized to recovery (leaving hospital) period. The parameter $\pi_i(t)$ is related to the infection hospitalization ratio (IHR) denoted as $\zeta_i(t) \in (0, 1)$, i.e., the proportion of the infected cases that are hospitalized
\begin{equation}
\frac{\pi_i(t) \eta_i}{(1-\pi_i(t))\gamma^I_i} = \frac{\zeta_i(t)}{1-\zeta_i(t)} \Longrightarrow
\pi_i(t) = \frac{\gamma^I_i\zeta_i(t)}{\eta_i + (\gamma^I_i - \eta_i)\zeta_i(t)},
\end{equation}
where we assume that the IHR is time dependent possibly due to (1) the dynamic change of the severity of the infected population (e.g., age and health condition) and (2) hospital availability. 
Similarly, the parameter $\nu_i(t)$ is related to the hospitalization fatality ratio (HFR) denoted as $\xi_i(t) \in (0, 1)$, i.e., the proportion of the hospitalized cases that are deceased
\begin{equation}
\frac{\nu_i(t) \mu_i}{(1-\nu_i(t))\gamma^H_i} = \frac{\xi_i(t)}{1-\xi_i(t)} \Longrightarrow
\nu_i(t) = \frac{\gamma^H_i\xi_i(t)}{\mu_i + (\gamma^H_i - \mu_i)\xi_i(t)},
\end{equation}
where we assume that the HFR is also time dependent possibly due to (1) the dynamic change of the severity of the hospitalized population (e.g., age and health condition) and (2) medical resources (e.g., ICU bed, ventilator, and new drugs).
Finally, the parameter $\tau_i(t)$ represent the infection confirmation ratio (ICR), or the proportion of infected cases that are confirmed/reported positive by viral test or medical diagnosis, which depends on the test strategy and capacity. 

\section{Data}
\label{sec:data}

\begin{figure}[!htb]
	\centering	
	\includegraphics[width=0.51\linewidth]{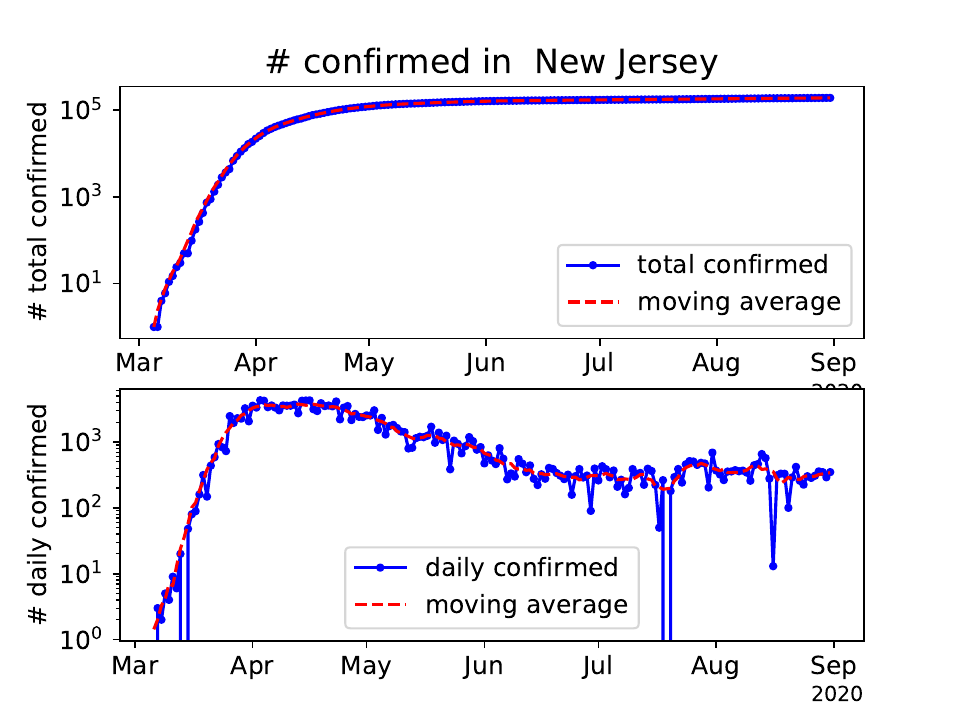}\hspace*{-0.5cm}
	\includegraphics[width=0.51\linewidth]{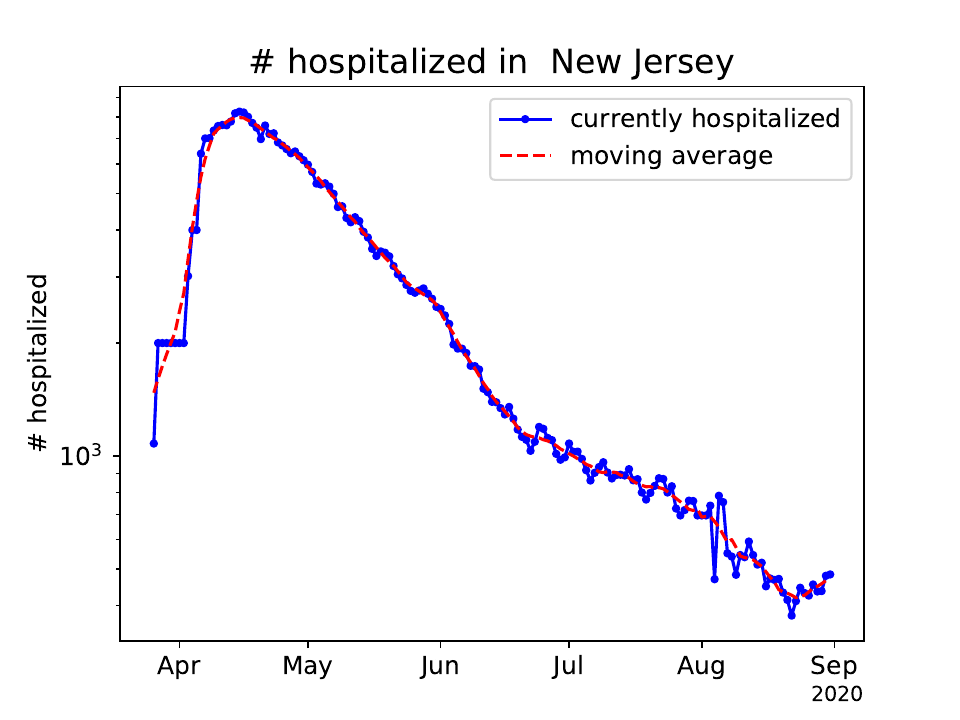}
	
	\vspace*{-0.2cm}
	
	\includegraphics[width=0.51\linewidth]{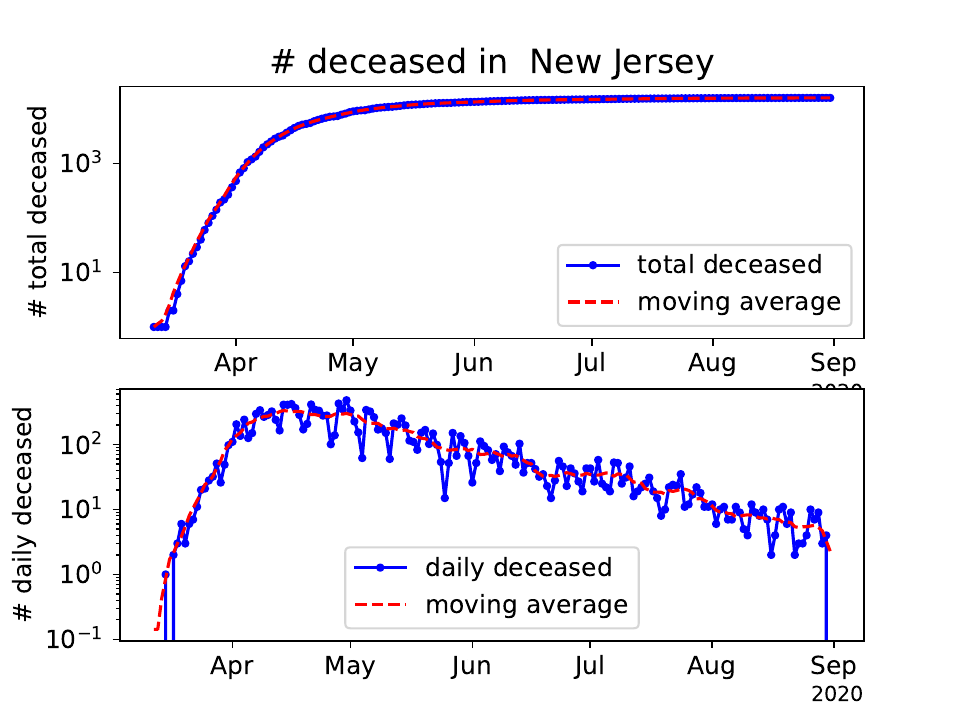}\hspace*{-0.5cm}
	\includegraphics[width=0.51\linewidth]{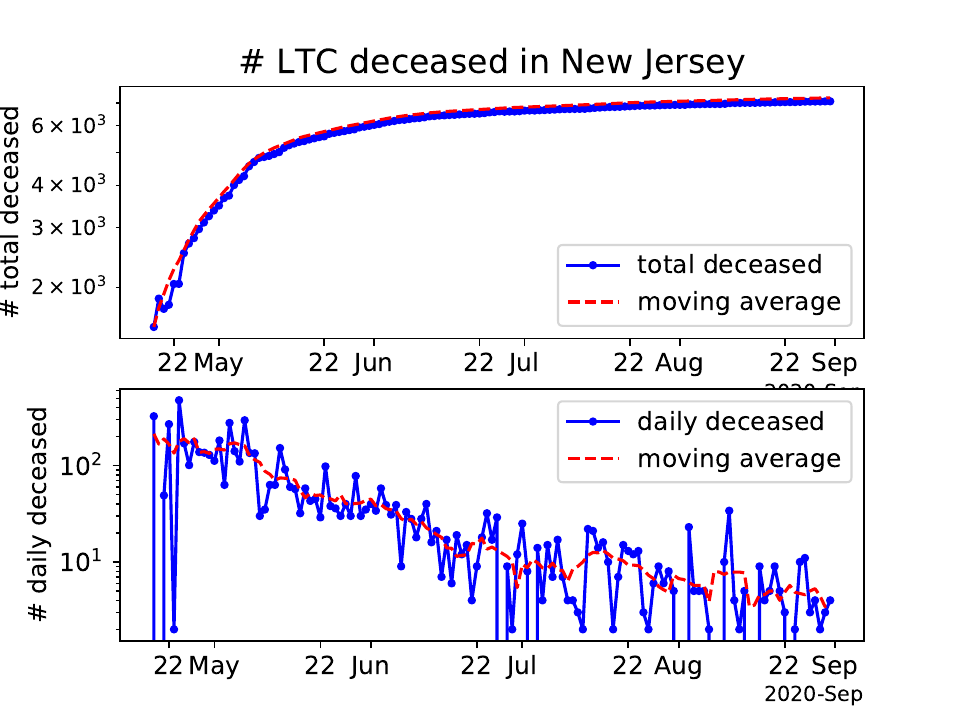}
	
	\caption{The number of total (accumulated) and daily confirmed cases (top-left), and deceased cases (bottom-left), and the number of currently hospitalized cases (top-right), as well as their seven day moving average. Bottom-right: the number of total and daily deceased cases, and their seven day moving average in long-term care (LTC) facilities. All the data are presented in logarithmic scale, which shows an early exponential growth.
	}\label{fig:data-NJ}
\end{figure}

\begin{figure}[!htb]
	\centering
	\includegraphics[width=0.51\linewidth]{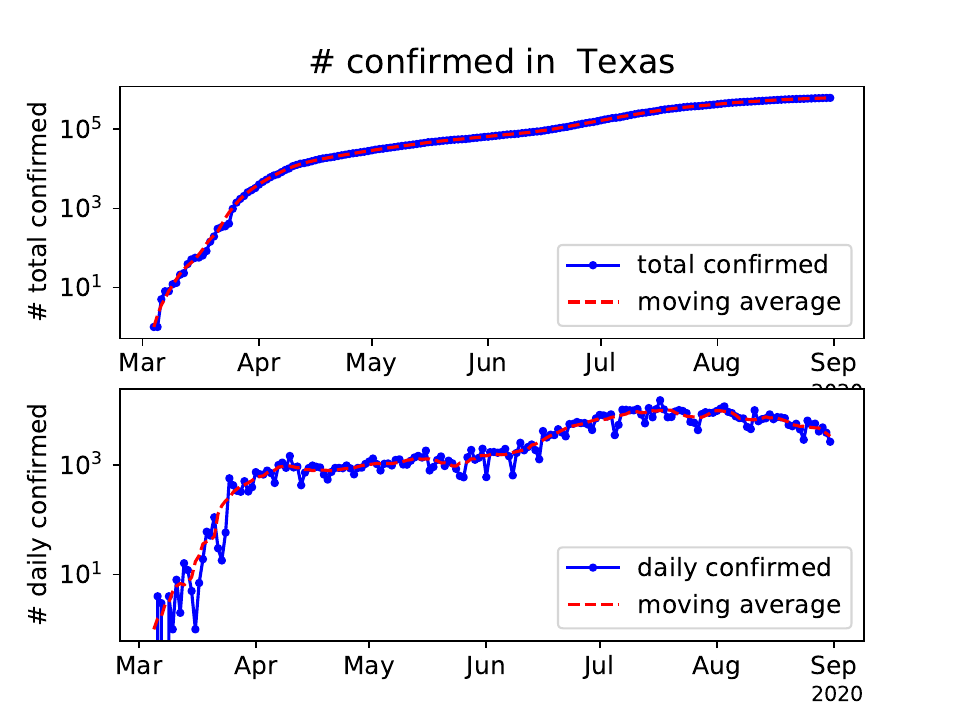}\hspace*{-0.5cm}
	\includegraphics[width=0.51\linewidth]{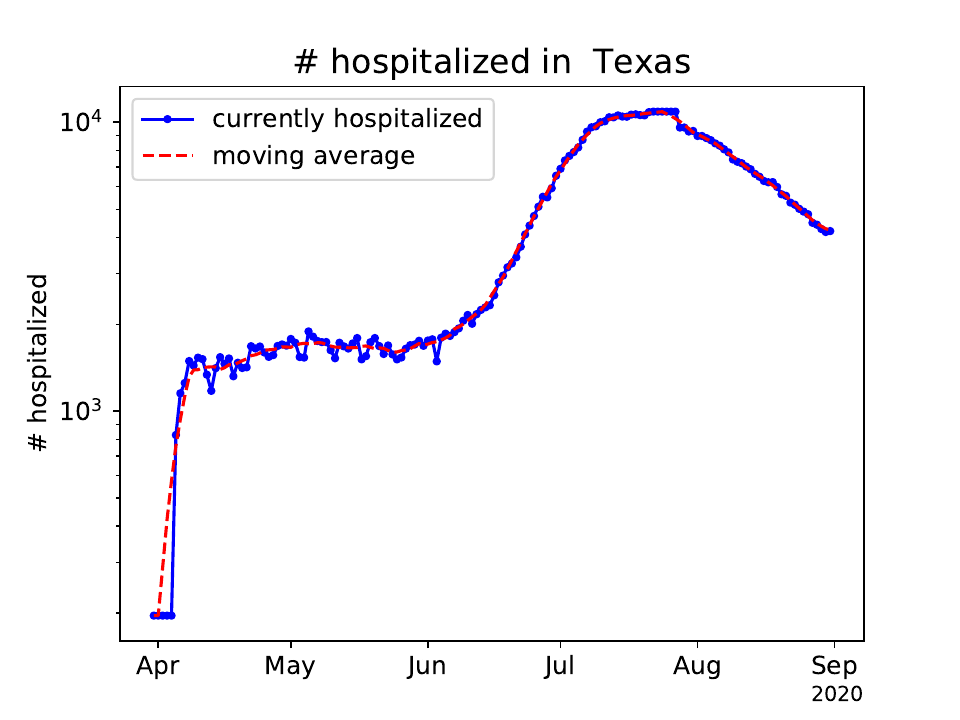}
	
	
	\includegraphics[width=0.51\linewidth]{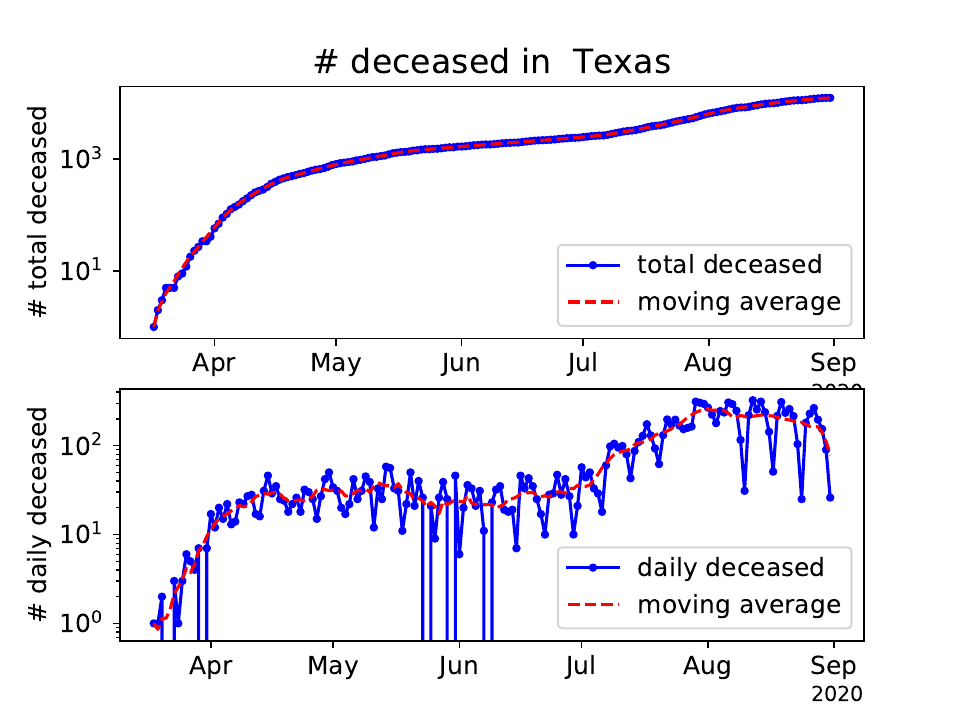}\hspace*{-0.5cm}
	\includegraphics[width=0.51\linewidth]{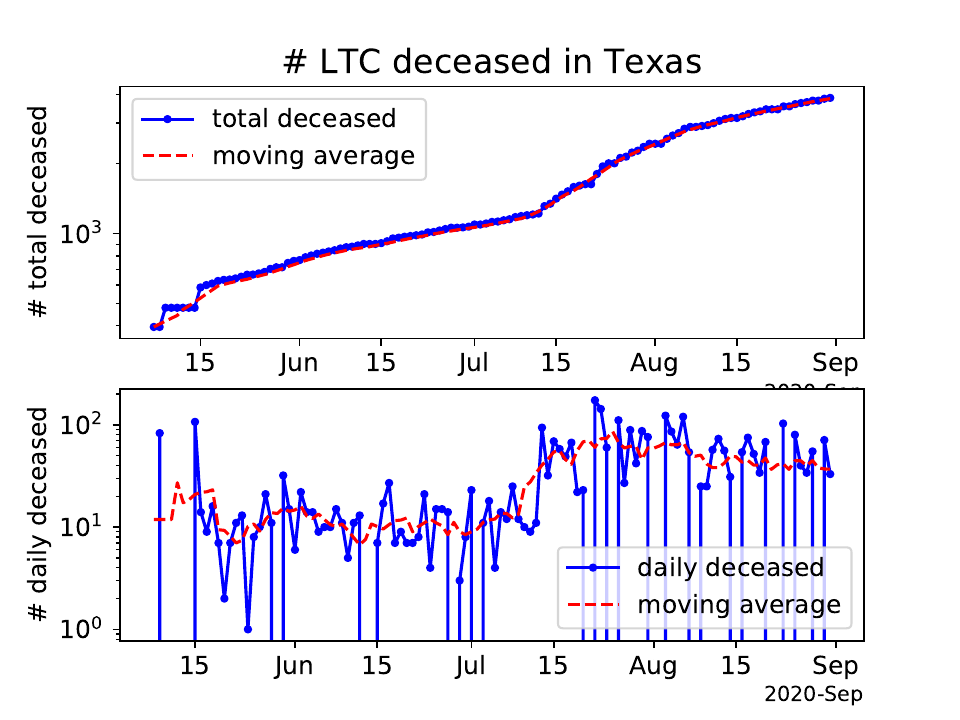}

	\caption{The same as in Figure \ref{fig:data-NJ} for Texas here. The increase of the number of confirmed, deceased, and hospitalized cases in Texas was slower than that of New Jersey in the early stage, and got accelerated before flattened lately.}\label{fig:data-TX}
\end{figure}

We choose the data of New Jersey (NJ) and Texas (TX) as two examples, whose LTC share of total COVID deaths are about 50\% and 31\% as of August 31. The LTC share of New Jersey is close to the US average and that in many European countries. The LTC share of Texas is close to 50\% early on and decreasing to 31\% when the LTC became strongly protected. Moreover, the two states have distinctive development of the epidemics: New Jersey experienced fast outbreak in an early stage (March and April) with a large number of deaths, and has since flatten the curve by implementing strict mitigation policy of lockdown, social distancing, etc., while Texas had slower spread of the virus than New Jersey early on (March and April) with a much smaller number of deaths, yet turned to an accelerated increase of cases after quickly relaxing the mitigation (from May to July) and flattened the curve since mid-July. These epidemic developments can be observed in Figure \ref{fig:data-NJ} for New Jersey and Figure \ref{fig:data-TX} for Texas.

The number of COVID deaths inside and outside LTC by state of US is reported in \emph{Nursing Home/LTC COVID Deaths By State}\footnote{\url{https://docs.google.com/spreadsheets/d/1ETm51GayRjlnoaRVtUOWfkolEeAQZ-zPhXkCbVe4_ik/edit\#gid=435667374}} for a certain period of time. The bottom-right part of Figure \ref{fig:data-NJ} displays the reported number of accumulated and daily deaths inside LTC in {New Jersey from April 17 to August 31}. The number of daily deceased cases has some negative values and is very oscillatory showing a 7 day periodic pattern with much smaller number of deaths reported during the weekend than during the week. We treat this oscillatory pattern as report error due to report delay, and possibly incompleteness and incorrectness. To correct the report error and make the daily numbers smoother in a given time interval $(t_1, t_2)$, we compute a seven day moving average $d^r_i$, $i = 1, 2$, from the reported daily data $\tilde{d}^r_i$, as 
\begin{equation}
d^r_i(t) = \frac{1}{7} \sum_{k = t-3}^{t+3} \tilde{d}^r_i(k), 
\end{equation}
where we set $\tilde{d}^r_i(k) = \tilde{d}^r_i(t_1)$ if $t-3 <= t_1$ and $\tilde{d}^r_i(k) = \tilde{d}^r_i(t_2)$ if $t+3 >=t_2$. 

The smoothed data by seven day moving average are also shown in Figure \ref{fig:data-NJ}. Note that the number of daily deceased cases is already relatively big on the first day of report, which can not reflect earlier stage of the epidemics. To learn the early dynamics, we use another set of data from the \emph{COVID19 Tracking} project\footnote{\url{https://github.com/COVID19Tracking}}, as shown in the other three figures in Figure \ref{fig:data-NJ}, which reports the number of daily deceased cases, infected cases that are confirmed by viral test or medical diagnosis from early March, as well as the number of currently hospitalized cases. Note that these data are aggregated for two groups of inside and outside LTC, which are not available separately.  To smooth these oscillatory data, we also compute a seven day moving average $p^r(t)$, $d^r(t)$, and $H^r(t)$ for the number of daily confirmed cases $\tilde{p}^r(t)$, daily deceased cases $\tilde{d}^r(t)$, and currently hospitalized cases $\tilde{H}^r(t)$. As the two sets of data are slightly inconsistent for the reported number of deceased cases, we use $d^r_1(t)$ for the number of deaths inside LTC, and $d^r_2(t) = d^r(t) - d^r_1(t)$ as that outside LTC to keep a consistent $d^r(t)$ in the entire time period.
We denote the number of accumulated confirmed and deceased cases for the smoothed data as
\begin{equation}
P^r(t) = \sum_{s = t_{\text{start}}}^t p^r(s), \quad  D^r(t) = \sum_{s = t_{\text{start}}}^t d^r(s), \quad \text{ and } D_i^r(t) = \sum_{s = t_{\text{start}}}^t d_i^r(s).
\end{equation}



\section{Deterministic inversion}
\label{sec:deterministic}

As the compartmental variables of the dynamic model \eqref{eq:LTC-model} is very sensitive to the parameters due to their exponential growth in time, the uncertainty on the (pre-inferred) parameters would lead to large uncertainty or even blowup in the model output. To avoid such blowup, we first deterministically infer the parameters at which the model can optimally fit the data in this section, and then quantify the uncertainty for the inferred parameters in next section. We formulate the deterministic inversion problem, and present the results for New Jersey and Texas.

\subsection{Formulation}

To infer the parameters in \eqref{eq:LTC-model} given data presented above, we minimize an objective function with respect to the parameters, which consists of a misfit function between the model output and the data, and a penalty function that imposes certain constraints on the parameters.

By $D(t)$ at $t \in [t_{\text{start}}, t_{\text{end}}]$ we denote the number of total deaths from the model \eqref{fig:data-NJ}, given as 
\begin{equation}
D(t) = D_1(t) + D_2(t).
\end{equation}
Then we compute the number of daily deaths from the model \eqref{fig:data-NJ} at $t \in  [t_{\text{start}}+1, t_{\text{end}}]$ as
\begin{equation}
d_1(t) = D_1(t) - D_1(t-1), \; d_2(t) = D_2(t) - D_2(t-1), \; \text{ and } \; d(t) = D(t) - D(t-1).
\end{equation}
Similarly, we denote $P^c(t)$ as the number of total infected cases that are confirmed, given by 
\begin{equation}
P^c(t) = P^c_1(t) + P^c_2(t),
\end{equation}
and the number of daily confirmed cases as 
\begin{equation}
p^c(t) = P^c(t) - P^c(t-1).
\end{equation}

As the early data (deaths and confirmed cases) are subject to randomness of the epidemics in the early stage, we use the data from the time when the numbers are relatively big. Let $t^p_1$ denote the first day that the number of total confirmed cases surpassed 100, $t_1^h$ denote the first day that the number of currently hospitalized cases surpassed 10, $t_1^d$ denote the first day that the number of total deaths surpassed 10, $t_2^d$ denote the day before the number of LTC deaths is reported, and $t_{\text{end}}$ the last day the data is available. To this end, we define the misfit function as 
\begin{equation}\label{eq:misift}
\begin{split}
F(\theta) &= \sum_{t = t_1^h}^{t_{\text{end}}} \left(\log(H(t,\theta)) - \log(H^r(t))\right)^2\\
& + \sum_{t = t_1^p}^{t_{\text{end}}} \left(\log(P^c(t,\theta)) - \log(P^r(t))\right)^2 + 0.1\sum_{t = t_1^p+1}^{t_{\text{end}}} \left(\log(p^c(t,\theta)) - \log(p^r(t))\right)^2\\
& + \sum_{t = t_1^d}^{t_2^d} \left(\log(D(t,\theta)) - \log(D^r(t))\right)^2 + 0.1\sum_{t = t_1^d+1}^{t_2^d} \left(\log(d(t,\theta)) - \log(d^r(t))\right)^2\\
& + \sum_{i=1}^2\sum_{t = t_2^d+1}^{t_{\text{end}}} \left(\log(D_i(t,\theta)) - \log(D^r_i(t))\right)^2 + 0.1\sum_{i=1}^2\sum_{t = t_2^d+2}^{t_{\text{end}}} \left(\log(d_i(t,\theta)) - \log(d^r_i(t))\right)^2,
\end{split}
\end{equation}
where we 
use a small weight
$0.1$ for the daily data as they are more oscillatory thus hard to match and probably less reliable than the accumulated data. Moreover, these weights are roughly proportional to the inverse of the standard deviation of the data (the difference between the reported number and seven day averaged number) that reflects the uncertainty or reliability of the reported data.

The aggregated parameter set $\theta$ in \eqref{eq:misift} collects all the parameters appearing in model \eqref{eq:LTC-model} as
\begin{equation}\label{eq:theta}
\theta = \bigcup_{i=1}^{2} \left\{\left(\alpha_i(t)\right)_{t\in t_\text{set}}, \left(\tau_i(t)\right)_{t\in t_\text{set}}, \left(\zeta_i(t)\right)_{t\in t_\text{set}}, \left(\xi_i(t)\right)_{t\in t_\text{set}}, \; \beta_i, \sigma_i, \eta_i, \mu_i,  \gamma^I_i, \gamma^H_i, \bC_{i1}, \bC_{i2} \right\},
\end{equation}
where $t_{\text{set}} = (t_0, t_1, \dots, t_k)$ is a sequence of equispaced time points with step size $\delta t$. 
The parameters $\alpha_i(t), \tau_i(t), \zeta_i(t), \xi_i(t)$ at other time $t \notin t_{\text{set}} $ are obtained by piecewise linear interpolation using values of $\left(\alpha_i(t)\right)_{t\in t_\text{set}}, \left(\tau_i(t)\right)_{t\in t_\text{set}}, \left(\zeta_i(t)\right)_{t\in t_\text{set}}, \left(\xi_i(t)\right)_{t\in t_\text{set}}$. Such an interpolatory definition with a relatively big step size, e.g., $\delta t = 7$, is employed to prevent over-parametrization that may lead to daily oscillatory and large variation in the transmission reduction effect, which plays a role of regularization in the deterministic inversion by solving a relatively low-dimensional optimization problem. The optimizer can be interpolated and used as an initial guess to refine the optimization with smaller step size, e.g., $\delta t = 1$, which  is often easier than directly solving a high-dimensional optimization problem. In the misfit function defined in \eqref{eq:misift}, we use data of the number of both accumulated and daily confirmed and deceased cases, as we find that the accumulated data can prevent over-fitting for the oscillatory (even after smoothing by moving average) daily data. Since these numbers are different by several orders of magnitude, we use their logarithmic scale to make them comparable. Moreover, we explicitly write $\theta$ as one argument for the quantities in the misfit function \eqref{eq:misift} that depends on $\theta$ through the model \eqref{eq:LTC-model}.

We consider the following penalty for the parameters
\begin{equation}\label{eq:penalty}
\begin{split}
G(\theta) & = \sum_{i=1}^2 \sum_{j=1}^k \left(\frac{\alpha_i(t_j) - \alpha_i(t_{j-1})}{\delta t} \right)^2 \delta t + \sum_{j=1}^k  \max\left(0, \frac{\alpha_1(t_{j-1}) - \alpha_1(t_{j})}{\delta t}\right)^2 \delta t
\\
& + \sum_{i=1}^2 \sum_{j=1}^k \left(\frac{\tau_i(t_j) - \tau_i(t_{j-1})}{\delta t} \right)^2 \delta t + \sum_{i=1}^2 \sum_{j=0}^k \left(\frac{\tau_i(t_j) - \bar{\tau}_i}{s_t}\right)^2 \delta t
\\
& + \sum_{i=1}^2 \sum_{j=1}^k \left(\frac{\zeta_i(t_j) - \zeta_i(t_{j-1})}{\delta t} \right)^2 \delta t + \sum_{i=1}^2 \sum_{j=0}^k \left(\frac{\zeta_i(t_j) - \bar{\zeta}_i}{s_t}\right)^2 \delta t
\\
& + \sum_{i=1}^2 \sum_{j=1}^k \left(\frac{\xi_i(t_j) - \xi_i(t_{j-1})}{\delta t} \right)^2 \delta t + \sum_{i=1}^2 \sum_{j=0}^k \left(\frac{\xi_i(t_j) - \bar{\xi}_i}{s_t}\right)^2 \delta t
\\
&  +\sum_{i=1}^2 \left(\frac{\beta_i-\bar{\beta}_i}{s \bar{\beta}_i}\right)^2 + \sum_{i=1}^2 \left(\frac{\sigma_i-\bar{\sigma}_i}{s \bar{\sigma}_i}\right)^2 + \sum_{i=1}^2 \left(\frac{\eta_i-\bar{\eta}_i}{s \bar{\eta}_i}\right)^2  +\sum_{i=1}^2 \left(\frac{\mu_i-\bar{\mu}_i}{s\bar{\mu}_i}\right)^2  \\
&  +\sum_{i=1}^2 \left(\frac{\gamma^I_i-\bar{\gamma}^I_i}{s\bar{\gamma}^I_i}\right)^2+ \sum_{i=1}^2 \left(\frac{\gamma^H_i-\bar{\gamma}^H_i}{s\bar{\gamma}^H_i}\right)^2 + \sum_{i=1}^2 \sum_{j=1}^2 \left(\frac{\bC_{ij} - \bar{\bC}_{ij}}{s\bar{\bC}_{ij}}\right)^2,
\end{split}
\end{equation}
where the first term in the first line represents the change of the transmission reduction factor $\alpha_i(t)$ in time $t$, an approximation to the variational integral $\int_{t_0}^{t_k} (d\alpha_i(t))^2 dt$ by a trapezoidal rule, which is penalized to prevent its large variation; the second term in the first line is used to penalize decreasing $\alpha_1(t)$, which reflects an increasing transmission reduction inside LTC as strict mitigation policy has been implemented and likely maintained in preventing excessive death that has accounted for a significant proportion of total deaths despite a small population, about 0.7\% of total population in average in US. Similar penalizations on the change of the ICR, IHR, HFR in time are considered. In addition, we consider a penalization on the deviation of ICR, IHR, HFR from their reference values weighted by a scaling parameter $s_t > 0$, which is set to be large that reflects less confidence and allows relatively large deviation from their reference values.
In the last two lines of of the penalty function \eqref{eq:penalty}, the deviation of $\beta_i, \sigma_i, \eta_i, \mu_i, \gamma^I_i, \gamma^H_i, \bC_{ij}$ from their reference values $\bar{\beta}_i, \bar{\sigma}_i, \bar{\eta}_i, \bar{\mu}_i, \bar{\gamma}^I_i, \bar{\gamma}^H_i,\bar{\bC}_{ij}$ is penalized, which is weighted with a relative scaling parameter $s > 0$. 

Given the misfit function \eqref{eq:misift} and the penalty function \eqref{eq:penalty}, we define an objective function 
\begin{equation}\label{eq:objective}
J(\theta) = F(\theta) + \lambda \, G(\theta),
\end{equation}
where $\lambda > 0$ is a scaling parameter that balances the two terms. The deterministic inversion problem can then be formulated as the optimization problem: find $\theta^* \in \Theta$, such that 
\begin{equation}\label{eq:inference-deterministic}
\theta^* = \argmin_{\theta \in \Theta} J(\theta),
\end{equation}
where $\Theta$ is an admissible parameter range. We start the optimization with a relatively big step size $\delta t = 7$, which leads to a 224-dimensional parameter $\theta$ for both New Jersey and Texas, including dimension 26 for each of the time dependent variables, and dimension 16 for the rest of the time independent parameters. To refine the optimization with smaller step size $\delta t = 1$, we use the interpolated optimizer as the initial guess, which leads to a 1520-dimensional parameter $\theta$ for New Jersey (1488 for Texas), including dimension 188 for New Jersey (184 for Texas) for each of time dependent variables. To solve the high-dimensional optimization problem \eqref{eq:inference-deterministic} with bounded constraint on the parameter $\theta \in \Theta$, we use a Python implementation of an optimizer SLSQP, a Sequential Least SQuares Programming algorithm, which is a gradient-based quasi Newton method with a BFGS update for the approximation of the Hessian \cite{Kraftothers88}. The gradient of the objective function $J(\theta)$ with respect to the high-dimensional parameter $\theta$ is computed by automatic differentiation using an \emph{Autograd}\footnote{\url{https://github.com/HIPS/autograd}} library.  We also develop an accelerated computation of the gradient by an adjoint approach, for which the expressions of involved derivatives are first symbolically computed by a symbolic library \emph{SymPy}\footnote{\url{https://www.sympy.org/en/index.html}} and then accelerated by a high performance Python compiler \emph{Numba}\footnote{\url{https://numba.pydata.org/}}. See an abstract formulation of the adjoint-based computation of the gradient in Appendix \ref{sec:adjoint}, which is used in next section for Bayesian inference, about 100X faster than the automatic differentiation by \emph{Autograd}.

\subsection{Results}

We present the setup and the results of the deterministic inversion for New Jersey and Texas. We initialize the model \eqref{eq:LTC-model} with the number of susceptible cases $S_1$ and $S_2$ the same as the population inside and outside LTC; we set  $E_1 = 1$ exposed case inside LTC and $E_2 = 100$ exposed cases outside LTC, and all other compartments as zero. 
To initialize the parameters, we set $\alpha_1(t) = \alpha_2(t) = 0,, t \in t_{\text{set}}$, and set a bound as $\alpha_i(t_0) \in [0, 0.1]$ to have a small transmission reduction at the beginning, and $\alpha_i(t) \in [0, 0.9], t = t_1, \dots, t_k$, where the upper bound $0.9$ assumes that the transmission can not be completely prevented by the mitigation policy. For the reference values of the parameters presented in the following, besides a few assumptions, we consider those available from \cite{VerityOkellDorigattiEtAl20, LiPeiChenEtAl20, FergusonLaydonNedjatiGilaniEtAl20}, \emph{CDC COVID-19 Pandemic Planning Scenarios}\footnote{\url{https://www.cdc.gov/coronavirus/2019-ncov/hcp/planning-scenarios.html}} and references therein. More specifically, we initialize ICR $\tau_1(t) = 40\%$ and $\tau_2(t) = 10\%$, IHR $\zeta_1(t) = 25\%$ and $\zeta_2(t) = 5\%$, HFR $\xi_1(t) = 40\%$ and $\xi_2(t) = 10\%$, which lead to an initial infection fatality ratios (IFR = IHR$\times$HFR) $10\%$ inside LTC and $0.5\%$ outside LTC. We also use these values as the reference values in the penalty term \eqref{eq:penalty}. We consider the bound $[\bar{\rho}/4, 2\bar{\rho}] \subset [0, 1]$, with $\bar{\rho}$ representing the initial values of the ratios $\tau_i(t), \zeta_i(t), \xi_i(t)$.
For the rest of parameters in $\theta$, we initialize them at their reference values $\bar{\rho}$ as reported in Table \ref{tab:parameters}, and consider the bound $ [\bar{\rho}/2, 2\bar{\rho}]$ to reflect relatively large uncertainty. Specifically, for the latency rate we use $\sigma_1 = \sigma_2 = 1/4$ (inverse of the latency period, exposed but not infectious yet, 4 days), for hospitalization rates we use $\eta_1 = \eta_2 = 1/8$ (inverse of the period from hospitalization to death, 8 days), for the recovery rates we use $\gamma^I_1 = \gamma^I_2 = 1/5$ (inverse of the infectious period, 5 days) and $\gamma^H_1 = \gamma^H_2 = 1/7$ (inverse of the hospital stay period, 7 days). For the contact matrix we use $\bC_{11} = 4, \bC_{12} = 0.5, \bC_{21} = 0.1, \bC_{22} = 2$, which reflect relatively more frequent contacts inside LTC than outside LTC, and larger infection influence of people from outside to inside LTC than  those from inside to outside LTC. The contact matrix is rescaled by the transmission parameters $\beta_1 = \beta_2 = 0.3$ such that the initial doubling time of infection and death is about $2 \sim 3$ days.


\begin{table}[!htb]
	\caption{Reference values and optimal values of the time independent parameters in $\theta$ collected in \eqref{eq:theta}. The optimal values are obtained by solving the deterministic inversion problem \eqref{eq:inference-deterministic} }\label{tab:parameters}
	\centering
	\begin{tabular}{|c|c|c|c|c|c|c|c|c|}
		\hline
		\multirow{2}{*}{$\theta$}			& \multicolumn{2}{c|}{reference values} & \multicolumn{2}{|c|}{optimal values (NJ)} &  \multicolumn{2}{|c|}{optimal values (TX)}	\\
		\cline{2-7}
		&$i=1$ & $i=2$ & $i=1$ & $i=2$ & $i=1$ & $i=2$ \\
		\hline
		$\beta_i$ &0.30  & 0.30 & 0.39 & 0.28 & 0.30 & 0.33  \\
		\hline
		$\sigma_i$ & 0.25 & 0.25 & 0.30 & 0.27 & 0.28 & 0.28   \\
		\hline
		$\eta_i$ & 0.13 & 0.13 & 0.12 & 0.13 & 0.13  & 0.12   \\
		\hline
		$\mu_i$ & 0.13 & 0.13 & 0.13 & 0.13 & 0.13 & 0.13  \\
		\hline
		$\gamma^I_i$ & 0.20 & 0.20 & 0.19  & 0.22 & 0.21  & 0.21  \\
		\hline
		$\gamma^H_i$ & 0.14 & 0.14 & 0.15 & 0.17 & 0.15 & 0.16  \\
		\hline
		$\bC_{i1}/\bC_{i2}$ & 4.00/0.50 & 0.10/2.00 & 5.15/0.50 & 0.11/1.62 & 4.09/0.51 & 0.10/2.10  \\
		\hline
	\end{tabular}
\end{table}

\begin{figure}[!htb]
	\centering
	\includegraphics[width=0.51\linewidth]{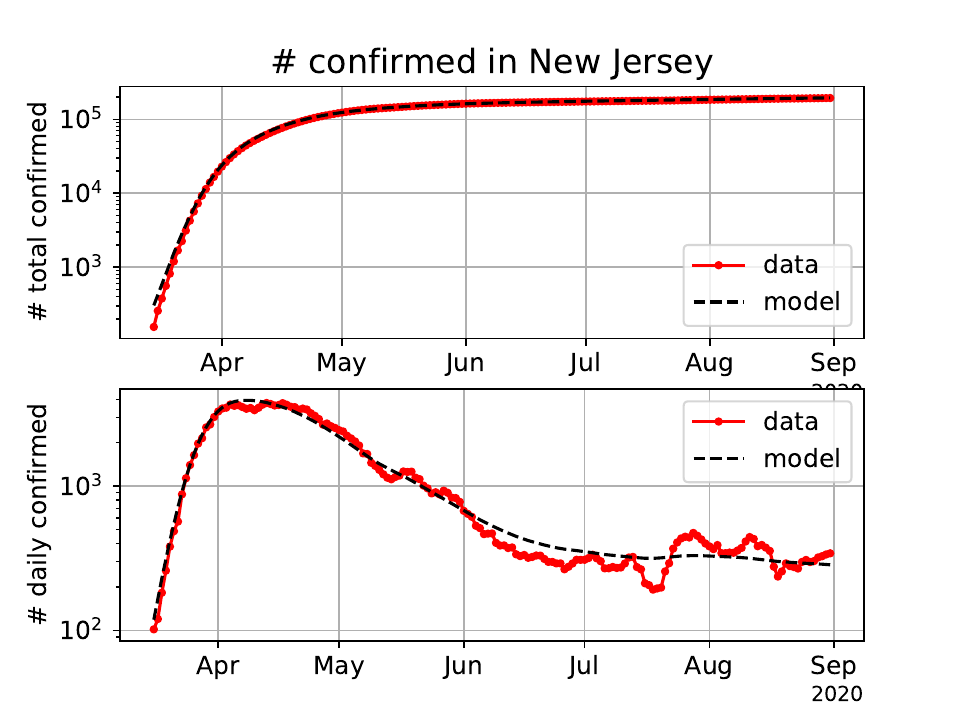}\hspace*{-0.5cm}
	\includegraphics[width=0.51\linewidth]{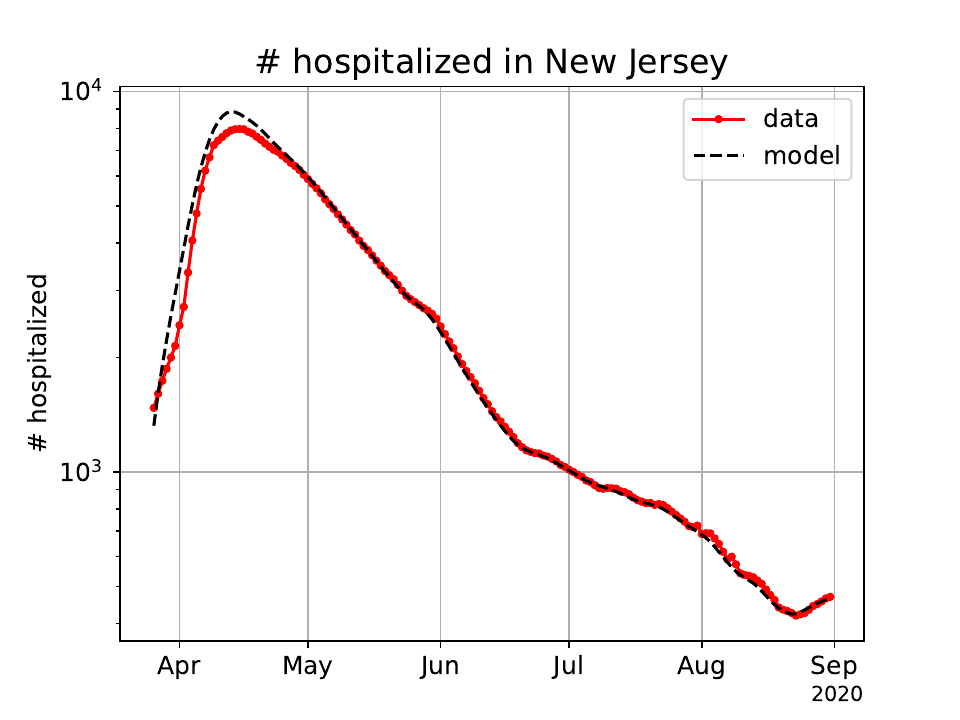}
	
	\includegraphics[width=0.51\linewidth]{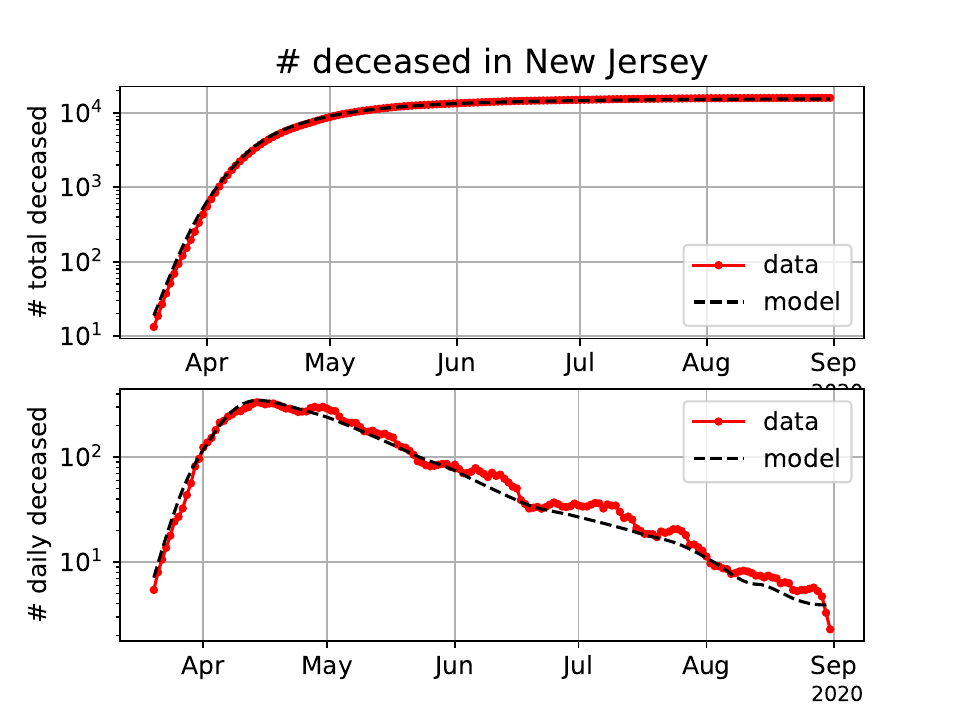}\hspace*{-0.5cm}
	\includegraphics[width=0.51\linewidth]{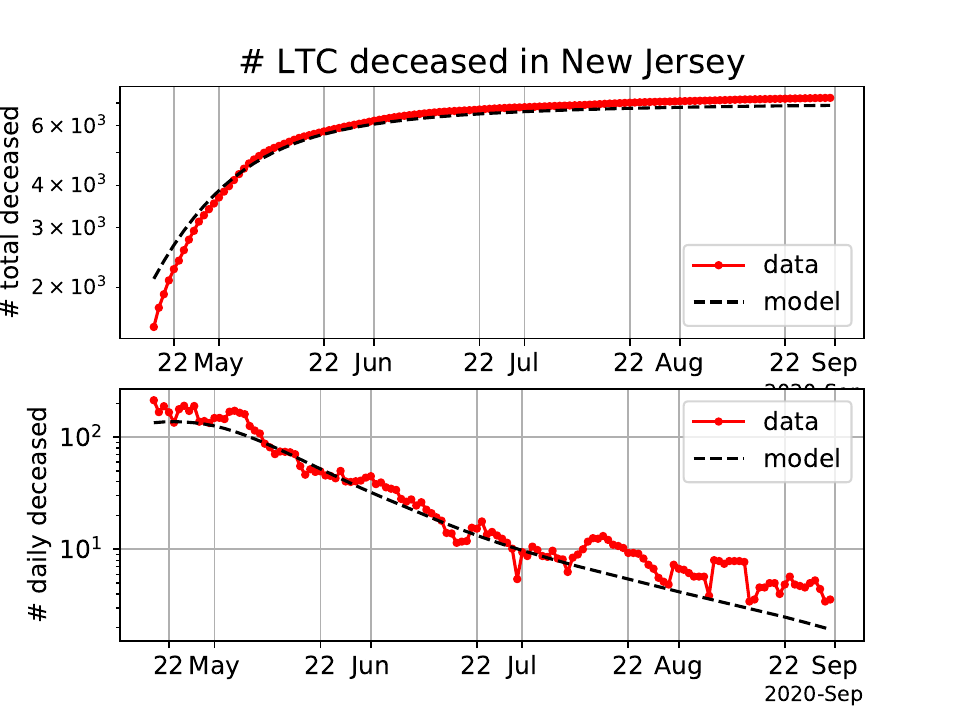}
	
	\caption{The same as in Figure \ref{fig:data-NJ}. The data here is the same as the seven day moving average data in Figure \ref{fig:data-NJ}. The model output is the result of model \eqref{eq:LTC-model} at the optimal parameter value obtained by deterministic inversion.}\label{fig:deterministic-NJ}
\end{figure}

\begin{figure}[!htb]
	\centering
	\includegraphics[width=0.51\linewidth]{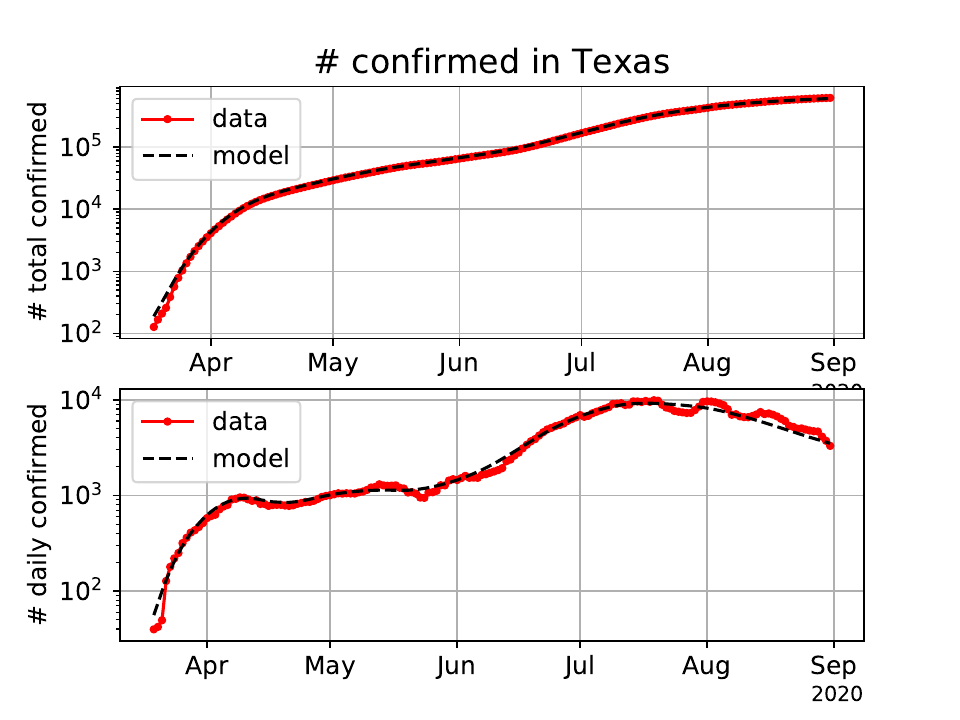}\hspace*{-0.5cm}
	\includegraphics[width=0.51\linewidth]{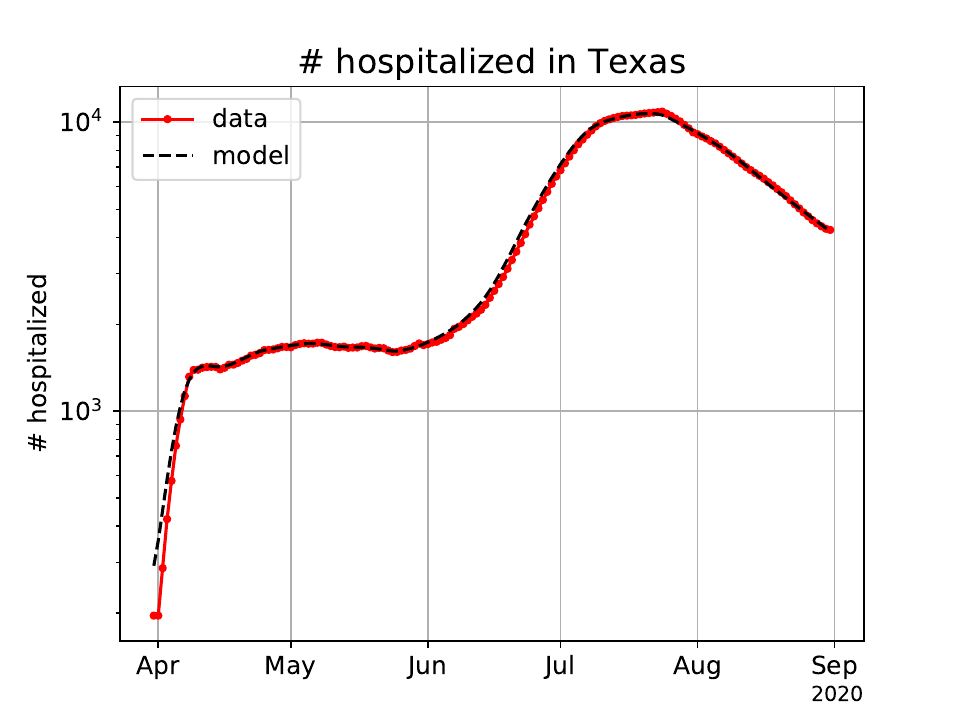}
	
	\includegraphics[width=0.51\linewidth]{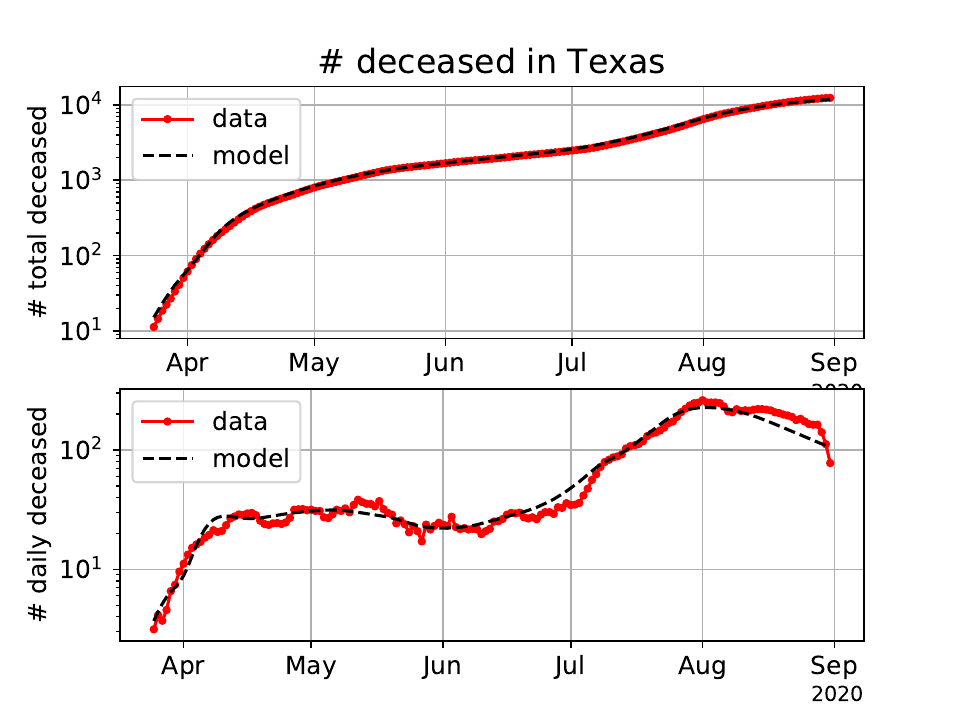}\hspace*{-0.5cm}
	\includegraphics[width=0.51\linewidth]{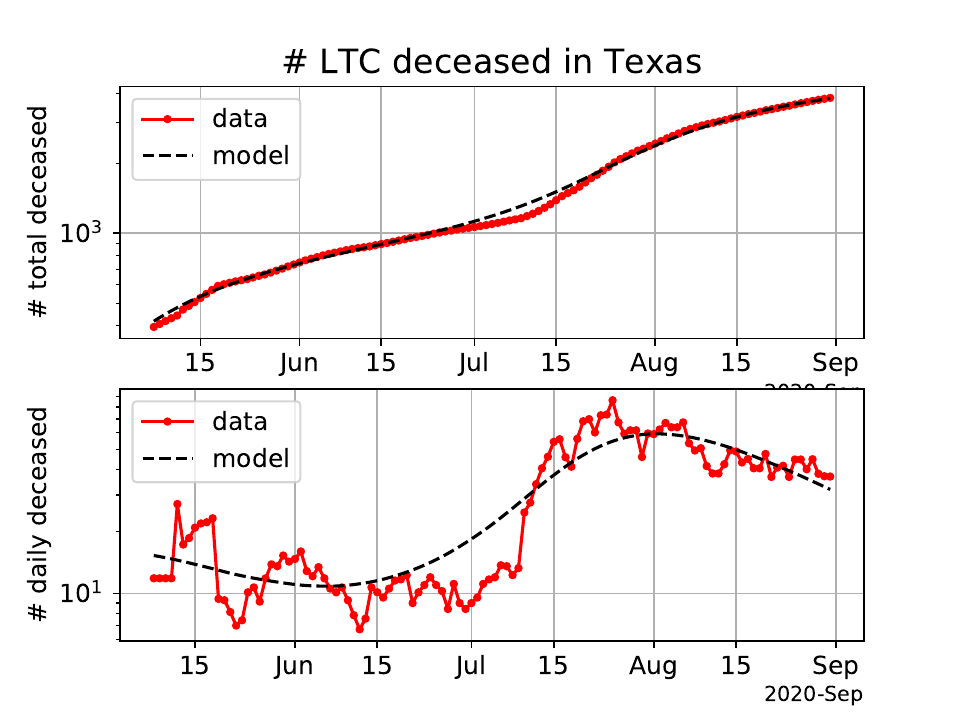}
	
	\caption{The same as in Figure \ref{fig:data-TX}. The data here is the same as the seven day moving average data in Figure \ref{fig:data-TX}. The model output is the result of model \eqref{eq:LTC-model} at the optimal parameter value obtained by deterministic inversion.}\label{fig:deterministic-TX}
\end{figure}

\begin{figure}[!htb]
	\centering
	\includegraphics[width=0.51\linewidth]{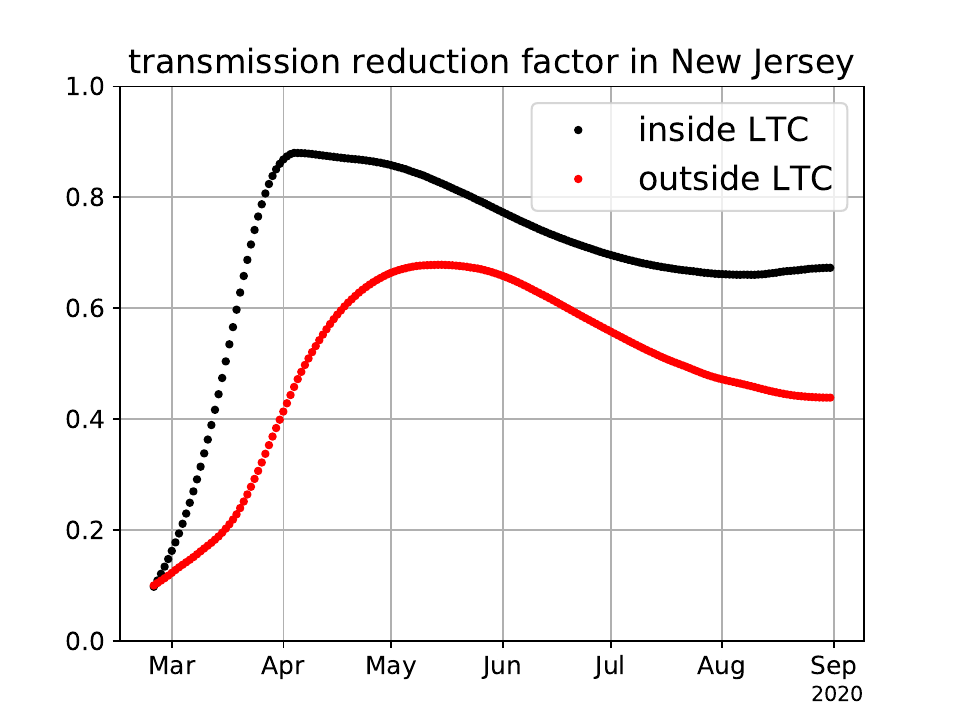}\hspace*{-0.5cm}
	\includegraphics[width=0.51\linewidth]{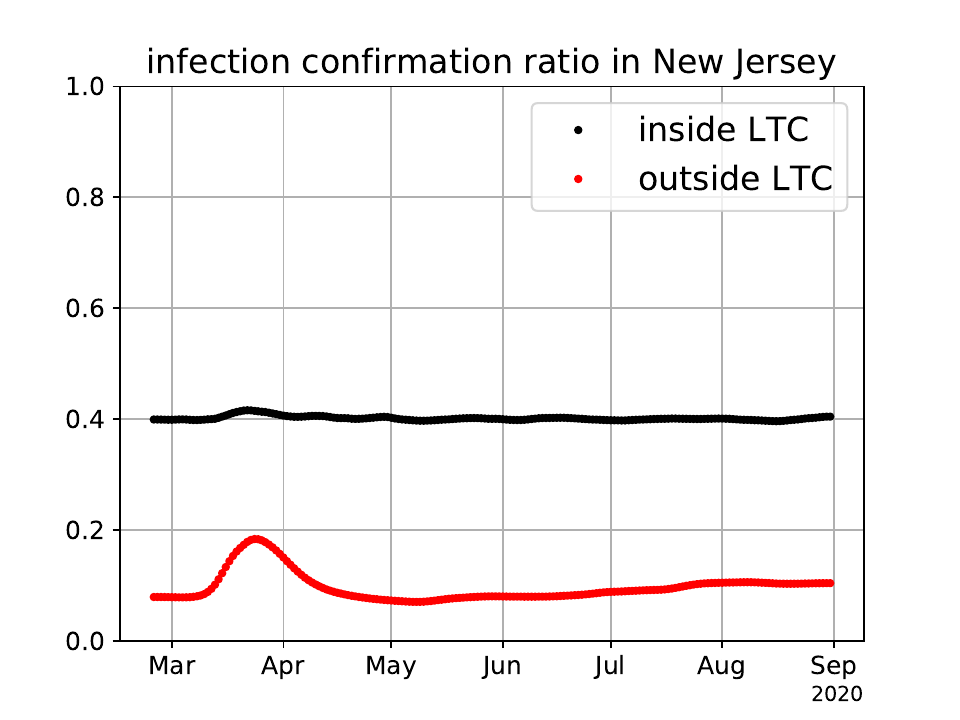}	
	\includegraphics[width=0.51\linewidth]{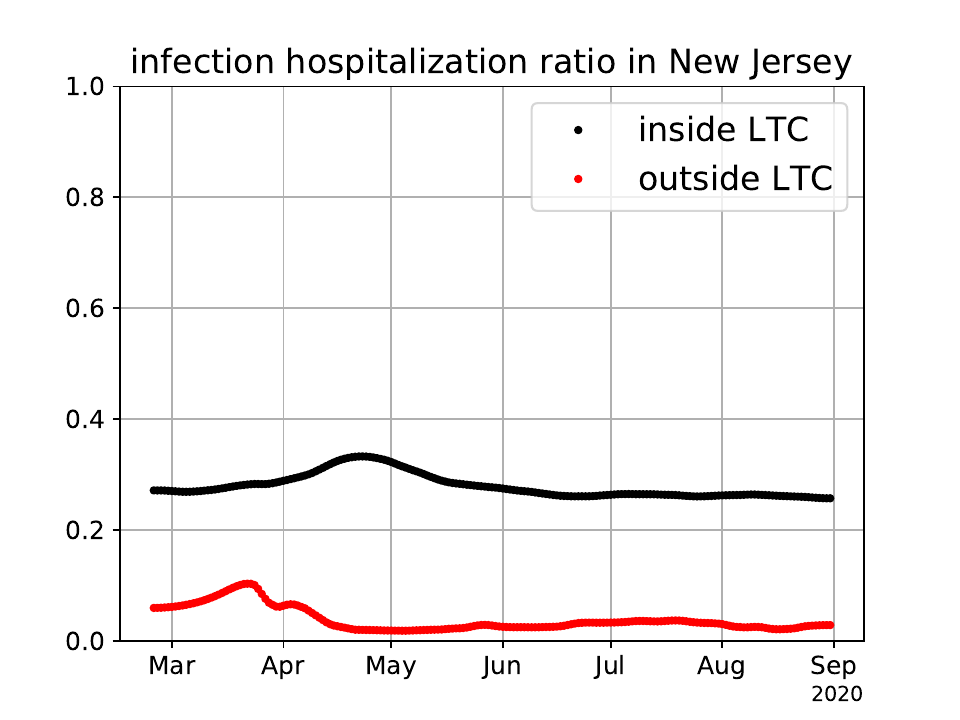}\hspace*{-0.5cm}
	\includegraphics[width=0.51\linewidth]{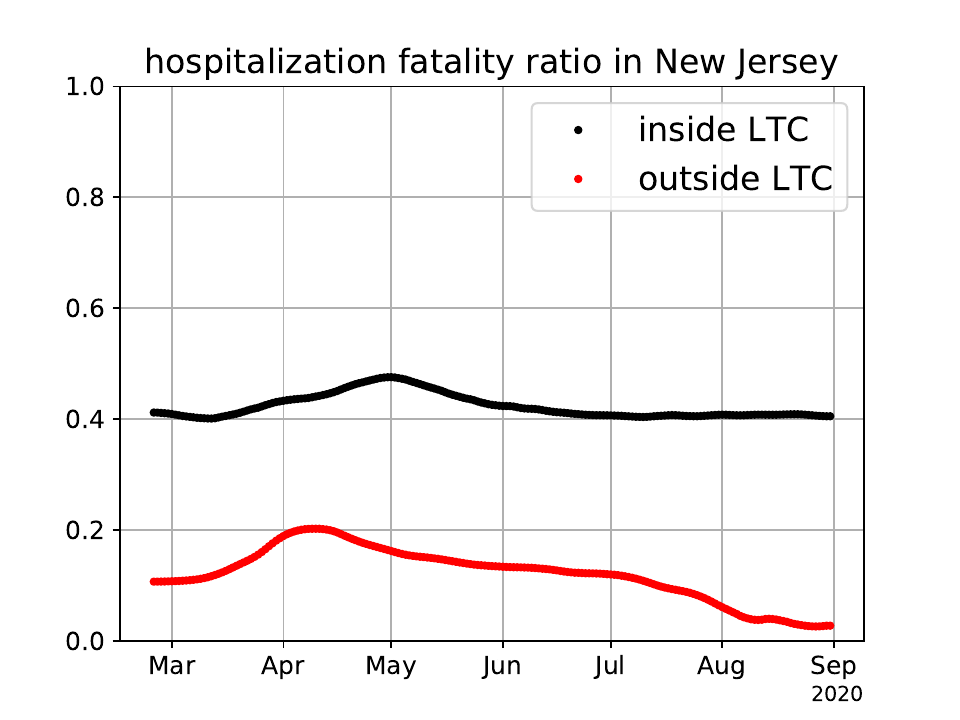}
	\caption{Optimal time dependent parameters inside and outside LTC in New Jersey by deterministic inversion \eqref{eq:inference-deterministic}.}\label{fig:deterministic-NJ-time}
\end{figure}

\begin{figure}[!htb]
	\centering
	\includegraphics[width=0.51\linewidth]{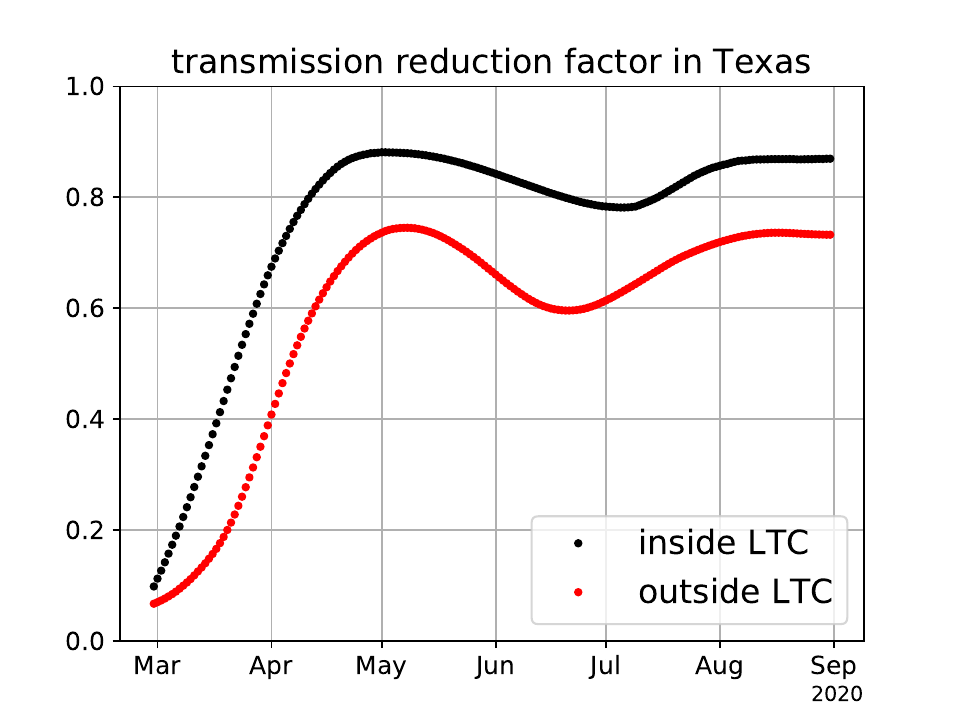}\hspace*{-0.5cm}
	\includegraphics[width=0.51\linewidth]{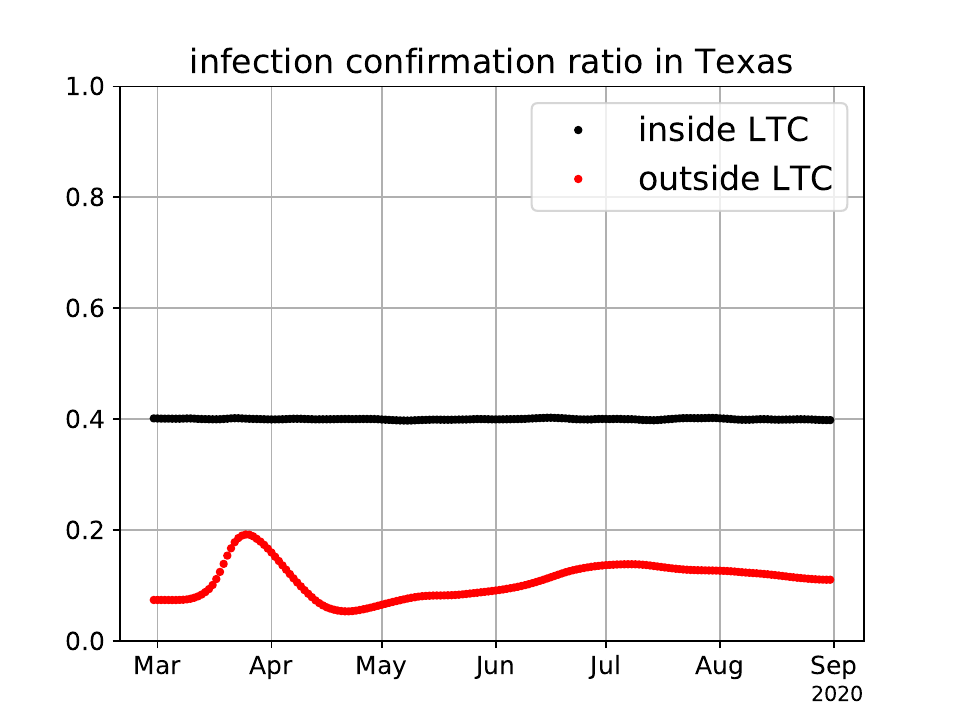}	
	\includegraphics[width=0.51\linewidth]{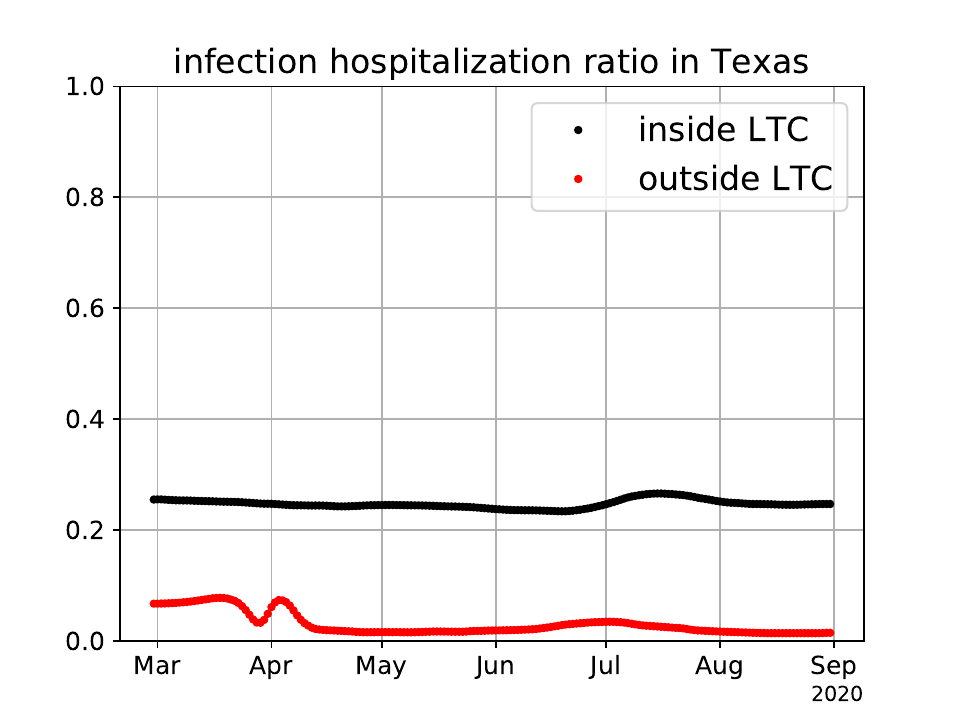}\hspace*{-0.5cm}
	\includegraphics[width=0.51\linewidth]{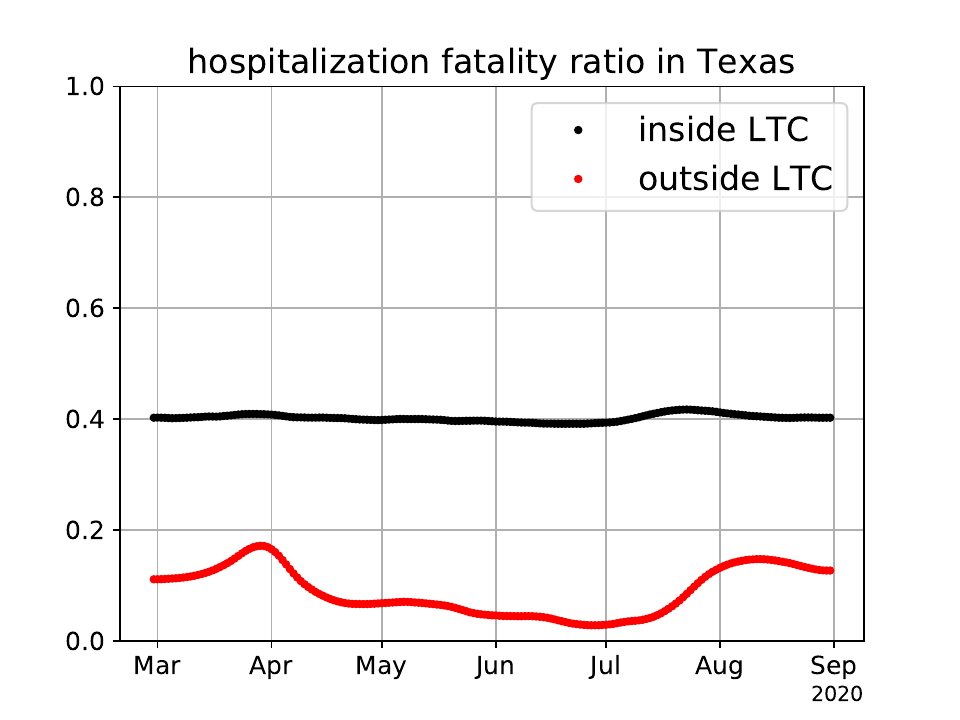}
	\caption{Optimal time dependent parameters inside and outside LTC in Texas by deterministic inversion \eqref{eq:inference-deterministic}.}\label{fig:deterministic-TX-time}
\end{figure}

We solve the optimization problem \eqref{eq:inference-deterministic} for deterministic inversion of the parameters with the scaling parameter $\lambda = 100$ to balance the misfit and penalty terms. We use relatively large scaling parameters $s_t = 10, s = 5$ in the penalty term \eqref{eq:penalty} such that the optimal parameters can be easily deviated from the reference values. We run the SLSQP optimization algorithm and stop it at 100 iterations, when the objective function \eqref{eq:objective} does not decrease much (less than $10^{-4}$ compared to the decrease at the first step). We plot the comparison between the model output and the smoothed data (with seven day moving average) in Figure \ref{fig:deterministic-NJ} for New Jersey and Figure \ref{fig:deterministic-TX} for Texas. An overall very high match between the model and all the data can be observed for both states, thanks to the flexibility of the parametrization especially in the time dependent transmission reduction factors, infection confirmation ratios, infection hospitalization ratios, and hospitalization fatality ratios for both inside and outside of LTC, whose optimal values are shown in Figure \ref{fig:deterministic-NJ-time} for New Jersey and Figure \ref{fig:deterministic-TX-time} for Texas. This demonstrates the effectivity of both the parametrization and the optimization algorithm in the framework of deterministic inversion. We remark that some data and model mismatch can be observed, e.g., the LTC deceased cases in both states can be probably accounted by the poor quality of the data due to delay or misreport, which can be supported by the very oscillatory (already smoothed by 7 day moving average) and the abruptly changed (in early July) number of daily deaths especially in Texas. This slight mismatch can also be explained by the use of the relatively small weight for the daily data in the misfit function \eqref{eq:misift}, which is to prevent catching the artifact of oscillation and overfitting that may lead to too much and unrealistic oscillation in the time dependent parameters. Nevertheless, the data and model output are matched rather well for the number of accumulated confirmed and deceased cases.
We also remark that the data and model match for the number of hospitalized cases is very high, even for the sharp change, e.g., in mid-April and late August in New Jersey, and in early April, mid-June and late July in Texas.

We make a few observations and comments for the deterministic inversion results of parameters. The optimal transmission reduction factor inside and outside LTC are shown in the top-left part of Figure \ref{fig:deterministic-NJ-time} for New Jersey and Figure \ref{fig:deterministic-TX-time} for Texas. We can see a sharp increase in late March and early April for both inside and outside LTC in both states, when increasingly stricter mitigation policy was implemented, e.g., school closure, prohibitive large group gathering, lockdown, limited or prohibitive family visiting to LTC, increasing use of personal protective equipment (PPE) inside LTC. The transmission reduction is relaxed by the gradual lift of the strict mitigation measures with several phases starting from May. However, it is observed that the lift was faster in Texas than in New Jersey in May and June outside LTC, which led to the sharp increase of the number of confirmed and hospitalized cases in Texas, and continuing decrease in New Jersey. The transmission reduction factor is increased in July in Texas probably because the lift of effective mitigation measures were reversed, while wearing masks was suggested and later mandated. We remark that the relatively more effective transmission reduction inside LTC than outside has been maintained probably by large-scale testing and widely use of PPE. We observe that the infection hospitalization ratios and hospitalization fatality ratios outside LTC in both states has decreased from their initial values, which may be explained by the shift of the new infections from the general population to the younger and healthier population, which is built into the penalization \eqref{eq:penalty}. We remark that the unexpected increase of the hospitalization fatality ratio in Texas from mid-July is probably due to the change of the reported fatality from July 27 and on.\footnote{\url{https://dshs.texas.gov/news/releases/2020/20200727.aspx}}
The optimal values for the other time independent parameters are reported in Table \ref{tab:parameters}, some of which are evidently different from the reference values. The slightly different values between New Jersey and Texas imply slightly different transmission/confirmation/hospitalization/death/recovery pattern. However, we caution that these optimal values may differ from the reality even though they provide the best fit for the data because (1) the inference problem is ill-posedness, i.e., the data are noisy, limited, and can only effectively inform the most data sensitive parameters, (2) the model is inadequate that may not capture the heterogeneity in age and health condition, different susceptibility, infectivity, and symptoms of different people, and (3) the partial information embedded in the penalization on the model parameters may not represent the reality.

\section{Bayesian inference}
\label{sec:Bayesian}

In learning the epidemics through assimilating the available data to the model \eqref{eq:LTC-model} by solving the optimization problem \eqref{eq:inference-deterministic}, various inevitable uncertainties are neglected, including the parameter uncertainty and data uncertainty, besides the uncertainty in the model itself. Without accounting for these uncertainties, the parameters inferred from the smoothed data may not capture the reality, which may lead to incorrect forecast results for the future transmission of the virus and ineffective optimal mitigation design based on the deterministic inversion.
In this section, we deal with the parameter and data uncertainties in a Bayesian framework for statistical inference of the parameters equipped with probability distributions and provide confidence for the model output. In particular, to solve the Bayesian inference problem with high-dimensional and heterogeneous parameters (time-dependent and time-independent), we employ our recently developed pSVGD method \cite{ChenGhattas20}, which exploits the intrinsic low dimensionality of the parameter-to-observable map and efficiently draw samples from the posterior distribution by solving an optimal transport problem with parameter projection into low-dimensional subspaces.

\subsection{Formulation}
The parameter $\theta$ defined in \eqref{eq:theta} is a set of heterogeneous parameters, including the discrete sequences of the four types of time dependent parameters $\left(\alpha_i(t)\right)_{t \in t_{\text{set}}}, \left(\tau_i(t)\right)_{t \in t_{\text{set}}}, \left(\zeta_i(t)\right)_{t \in t_{\text{set}}}, \left(\xi_i(t)\right)_{t \in t_{\text{set}}}$, $i = 1, 2$, the time independent parameters $\beta_i, \sigma_i, \eta_i, \mu_i,  \gamma^I_i, \gamma^H_i$, $ i = 1, 2$, and a contact matrix $\bC$. We present prior probability distributions for these parameters. For simplicity, we assume that these parameters are independently distributed. 

We construct the prior distributions for $\left(\tau_i(t)\right)_{t \in t_{\text{set}}}, \left(\zeta_i(t)\right)_{t \in t_{\text{set}}}, \left(\xi_i(t)\right)_{t \in t_{\text{set}}}$ in the same way as for $\left(\alpha_i(t)\right)_{t \in t_{\text{set}}}$, and only present it for $\left(\alpha_i(t)\right)_{t \in t_{\text{set}}}$ for simplicity. The sequences $\left(\alpha_i(t)\right)_{t \in t_{\text{set}}}$ are obtained as a temporal discretization of  
the transmission reduction factor $\alpha_i(t)$ for $t\in (t_0, t_k)$. To equip a prior probability distribution for $\alpha_i(t)\in (0, 1)$,  we assume that it is given by the form 
\begin{equation}\label{eq:tanh}
\alpha_i(t) = \frac{1}{2}\left(\tanh( g_i(t) ) + 1\right),
\end{equation}
where $\tanh$ is the hyperbolic tangent function so that the constraint $\alpha_i(t)\in (0, 1)$ is always satisfied. Moreover, the time-dependent function $g_i(t) \in \cN(g_i^*, \mathcal{C}_g)$ is assumed to be a Gaussian process with mean $g_i^*$ and covariance $\mathcal{C}_g$. We set the mean as
\begin{equation}
g_i^*(t) = \arctanh(2\alpha_i^*(t) - 1),
\end{equation}
where $\alpha_i^*(t)$ for $t\in (t_0, t_k)$ is defined as piecewise linear interpolation of the optimal sequences $(\alpha_i^*(t))_{t \in t_{\text{set}}}$, which is obtained as the solution of the optimization problem \eqref{eq:inference-deterministic} for deterministic inversion. We assume the covariance $\mathcal{C}_g$ is given by 
\begin{equation}\label{eq:C-g}
\mathcal{C}_g = \cB^{-1},
\end{equation}
where $\cB $ is an elliptic differential operator in time domain $(t_0, t_k)$ given by
\begin{equation}
\cB = -s_g \Delta_t + s_I I,
\end{equation}
equipped with homogeneous boundary condition, where $s_g > 0, s_I > 0$ are scaling parameters,  $\Delta_t$ is the Laplacian in time, and $I$ is identity. Such a covariance \eqref{eq:C-g} is often used in defining Gaussian process/random field with $s_g$ and $s_I$ controlling its correlation length and variance.
By a temporal discretization at equispaced points $t_0, \dots, t_k$ using finite difference, we have
\begin{equation}
\Delta_t g(t) \approx  \frac{g(t - \delta t) - 2g(t) + g(t+\delta t)}{(\delta t)^2},
\end{equation}
for any twicely differentiable function $g: [t_0, t_k] \to \bR$, we obtain a discrete Laplacian $\mathbb{L}$ with homogeneous Neumann boundary condition, i.e., $g'(t_0) = g'(t_k) = 0$, such that
\begin{equation}
\bsg^T \bC_g^{-1} \bsg  = \bsg^T (-s_g\mathbb{L} + s_I \bI) \bsg = s_g \sum_{j = 1}^{k} \left(\frac{g(t_j) - g(t_{j-1})}{\delta t}\right)^2 + s_I ||\bsg||_2^2,
\end{equation}
for $\bsg = (g(t_0), \dots, g(t_k))^T$, where the first term has the same form as that in the penalty term \eqref{eq:penalty} up to a constant. To this end, we obtain the prior distribution for the vector $\bsg_i = (g_i(t_0), \dots, g_i(t_k))^T$ from the discretization of the Gaussian process $g_i \sim \cN(g_i^*, \mathcal{C}_g)$, as $\bsg_i \sim \cN(\bsg_i^*, \bC_g)$ with the mean vector $\bsg_i^* = (g^*_i(t_0), \dots, g^*_i(t_k))$ and covariance matrix $\bC_g = (-s_g \mathbb{L} + s_I \bI)^{-1}$. 

By slight abuse of notation, we denote $\bsg$ as a collection of all the $4 \times 2$ (4 for different ratios, 2 for inside and outside LTC) time dependent parameters, with a total dimension of $8(k+1)$. Meanwhile, by slight abuse of notation we denote $\bC_g$ as a block diagonal covariance matrix, where each of the eight blocks is the covariance matrix $(-s_g \mathbb{L} + s_I \bI)^{-1}$. 

For the time independent parameters, including the scalar parameters of various rates $\beta_i, \sigma_i, \eta_i, \mu_i,  \gamma^I_i, \gamma^H_i$, $ i = 1, 2$, and the contact matrix $\bC$, we collect them as a vector of parameters 
\begin{equation}
\bsf = (\beta_1, \beta_2, \sigma_1, \sigma_2, \eta_1, \eta_2, \mu_1, \mu_2, \gamma^I_1, \gamma^I_2, \gamma^H_1, \gamma^H_2, \bC_{11}, \bC_{12}, \bC_{21}, \bC_{22})^T,
\end{equation}
and assume that it has a prior distribution as 
\begin{equation}\label{eq:logh}
\log(\bsf) = \bsh \sim \cN(\bsh^*, \bC_h),
\end{equation}
where $\bsh^* = \log(\bsf^*)$ for $\bsf^*$ defined as the optimal values of $\bsf$ obtained from the deterministic inversion problem \eqref{eq:inference-deterministic}. For the covariance $\bC_h$, we assume that it is diagonal with the diagonal entries given as the entries of the vector $(s_h \bsh^*)^2$ with a scaling parameter $s_h > 0$. Under this assumption, we have 
\begin{equation}\label{eq:C-h}
(\bsh-\bsh^*)^T \bC_h^{-1} (\bsh-\bsh^*) = \sum_{l = 1}^{22}\left(\frac{h_l - h_l^*}{s_hh_l^*}\right)^2, 
\end{equation}
which has the same form as that in the penalty term \eqref{eq:penalty} up to a constant. Finally, we collect all the parameters as $x = (\bsg, \bsh) \in \bR^{d_x}$ of dimension $d_x$. Then the prior density of $x$, denoted as $p_0(x)$, satisfies
\begin{equation}
p_0(x) \propto \exp\left(- \frac{1}{2} (\bsg-\bsg^*)^T \bC^{-1}_g (\bsg-\bsg^*) - \frac{1}{2} (\bsh-\bsh^*)^T \bC_h^{-1} (\bsh-\bsh^*) \right).
\end{equation}
Note that $-\log(p_0(x))$ has similar form as part of that in the penalty term \eqref{eq:penalty}. However, the parameter $x$ here, after the nonlinear transformation by the function in \eqref{eq:tanh} and the logarithmic function in \eqref{eq:logh}, 
is different from $\theta$ in \eqref{eq:penalty}. We remark that the deterministic inversion problem \eqref{eq:inference-deterministic} can also be formulated as an optimization problem with respect to the transformed parameter $x$. However, we observe from tests that it is much more difficult to solve than the one with respect to the parameter $\theta$, mainly due to the stiffness of the nonlinear transformation, especially \eqref{eq:tanh}, in a major part of the whole parameter space. Nevertheless, by defining the prior distribution of $x$ centered at the transformed optimal values of the parameter $\theta$, and use a proper covariance for $x$, the stiffness can be effectively alleviated in the Bayesian inference problem.

Based on the above definition of the transformed parameter $x$ and its prior distribution, we assume that the data and model output can be formulated as 
\begin{equation}
y = u(x) + \omega,
\end{equation}
where the data $y$ represents a collection of all the data as given in the misfit function \eqref{eq:misift}, i.e., 
\begin{equation}
\begin{split}
y = & \Big(\{\log(H^r(t))\}, \{\log(P^r(t))\}, \{0.1\log(p^r(t))\}, \\
& \{\log(D^r(t))\}, \{0.1\log(d^r(t))\}, \{\log(D_i^r(t))\}, \{0.1\log(d_i^r(t))\}\Big),
\end{split}
\end{equation}
and the parameter-to-observable map $u(x)$ is given by 
\begin{equation}
\begin{split}
u(x)  = & \Big(\{\log(H(t, x))\}, \{\log(P^c(t, x))\}, \{0.1\log(p^c(t, x))\},  \\
& \{\log(D(t,x))\}, \{0.1\log(d(t,x))\}, \{\log(D_i(t,x))\}, \{0.1\log(d_i(t,x))\} \Big),
\end{split}
\end{equation}
where each set of data in $y$ and model output in $u(x)$ has a time span corresponding to that in \eqref{eq:misift}. Note that in $u(x)$ the model output depends on the parameter $x$ through $\theta$ by the transformation \eqref{eq:tanh} and \eqref{eq:logh}, as well as the model \eqref{eq:LTC-model} which depends on the parameter $\theta$. For simplicity, we assume that the additive noise $\omega \sim \cN(0, \bC_\omega)$ with zero mean and a symmetric positive definite covariance matrix $\bC_\omega$. Then the likelihood function, denoted as $f(x)$, satisfies
\begin{equation}
f(x) \propto \exp\left(
-\frac{1}{2} (y - u(x))^T \bC_\omega^{-1} (y - u(x))
\right).
\end{equation}
By Bayes' rule, the posterior density of the parameter $x$ is given by 
\beq\label{eq:Bayes}
p(x) = \frac{1}{Z} f(x) p_0(x), 
\eeq
where $Z$ is a normalization constant given by
\beq
Z = \int_{\bR^{d_x}} f(x) p_0(x) dx,
\eeq
which is typically intractable to compute, especially for a large dimension $n$. The tasks of Bayesian inference is to draw samples of the parameter $x$ from its posterior distribution, and compute some statistical quantity of interest, e.g., mean, variance, and confidence interval of the parameter and the model output. Several challenges are faced for these tasks, including (1) curse of dimensionality --- the computational complexity grows rapidly (typically exponentially) for conventional methods with respect to the parameter dimension, and (2) the geometry of the posterior density is complex, e.g., non Gaussian, multimodal, concentrating in local regions, etc. We have developed several approaches to address the curse of dimensionality, including \cite{ChenSchwab15, ChenSchwab16, ChenSchwab16b, ChenVillaGhattas17, ChenWuChenEtAl19, ChenGhattas20, ChenGhattas20b} that exploit the sparsity, low dimensionality, and geometry of the parameter-to-observable map. In this work, we employ our recently developed projected Stein variational gradient descent (pSVGD) method \cite{ChenGhattas20}, which can break or considerably alleviate the curse of dimensionality by exploiting the intrinsic low-dimensionality of the inference problem using gradient information matrix, and effectively deal with the complex geometry by transporting samples in parallel from the prior distribution to the posterior through optimization.

%

\subsection{pSVGD}
We briefly present SVGD from \cite{LiuWang16} and then introduce the projected SVGD from \cite{ChenGhattas20} to tackle high-dimensional inference problems. As a variational inference method, SVGD seeks an approximate density $q \in \cQ$ in a density function space $\cQ$ for the posterior density $p$ by minimizing a Kullback--Leibler (KL) divergence that measures the difference between two densities, i.e., 
\beq\label{eq:KL}
q^* = \argmin_{q \in \cQ} D_{\text{KL}} (q | p),
\eeq
where $D_{\text{KL}}(q|p) = \bE_{x \sim q}[\log(q/p)]$, which vanishes when $q = p$. A specific function class is $\cQ = \{T_\sharp p_0: T \in \cT\}$, where $T_\sharp$ is a pushforward map that pushes the prior density to a new density $q := T_\sharp p_0$ through a transport map $T(\cdot):\bR^{d_x} \to \bR^{d_x}$ defined in a space $\cT$.
For 
\beq\label{eq:transport}
T(x) = x + \epsilon \phi(x),
\eeq
where $\phi : \bR^{d_x} \to \bR^{d_x}$ is a differentiable perturbation map, and $\epsilon > 0$ is a learning rate, there holds
\beq\label{eq:directional-gradient}
\nabla_\epsilon D_{\text{KL}}(T_\sharp p_0 | p) \big|_{\epsilon = 0} = - \bE_{x\sim p_0} [\text{trace}(\cA_p \phi(x))],
\eeq
where $\cA_p$ is the Stein operator given by 
\beq
\cA_p \phi(x) = \nabla_x \log p(x) \phi(x)^T + \nabla_x \phi(x). 
\eeq
A practical SVGD algorithm was obtained by choosing $\cT = (\cH_{d_x})^{d_x} = \cH_{d_x} \times \cdots \times \cH_{d_x}$, a tensor product of a reproducing kernel Hilbert space (RKHS) $\cH_{d_x}$ with kernel $k(\cdot, \cdot): \bR^{d_x} \times \bR^{d_x} \to \bR$, which push the samples $x^0_1, \dots, x^0_N$ drawn from the prior $p_0$ step by step as
\beq
x_m^{\ell+1} = x_m^\ell + \epsilon_l \hat{\phi}_{\ell}^*(x_m^\ell), \quad m = 1, \dots, N, \ell = 0, 1, \dots  
\eeq
where $\hat{\phi}_{\ell}^*(x_m^\ell) $ is an approximate steepest direction given by
\beq\label{eq:gradient-saa}
\hat{\phi}_{\ell}^*(x_m^\ell) = \frac{1}{N} \sum_{n=1}^N \nabla_{x_n^\ell} \log p(x_n^\ell) k(x_n^\ell, x_m^\ell) + \nabla_{x_n^\ell} k(x_n^\ell, x_m^\ell).
\eeq
The kernel $k$ plays a critical role in pushing the samples to the posterior. One choice is Gaussian
\beq\label{eq:x-kernel}
k(x,x') = \exp\left(- \frac{||x-x'||^2_2}{h_x}\right),
\eeq
where $h_x$ is a bandwidth, e.g., $h = \text{med}^2/\log(N)$ with $\text{med}$ representing the median of sample distances \cite{LiuWang16}. 
However, it is known that the kernel suffers from the \emph{curse of dimensionality} for large dimension $d_x$, which leads to the samples that are not representative of the posterior, as observed in \cite{ZhuoLiuShiEtAl18, WangZengLiu18, ChenWuChenEtAl19, ChenGhattas20}. 

To address the curse of dimensionality, we exploit the property that the subspace in which the posterior differs from the prior is low dimensional due to the ill-posedness of the inference problem, as proven for some model problems and numerically demonstrated for many others \cite{BashirWillcoxGhattasEtAl08, FlathWilcoxAkcelikEtAl11,
	Bui-ThanhGhattas12a, 
	AlexanderianPetraStadlerEtAl14, IsaacPetraStadlerEtAl15, 
	MartinWilcoxBursteddeEtAl12, ChenVillaGhattas17, ChenVillaGhattas19, ChenGhattas19a, ChenWuChenEtAl19, ChenGhattas20}. 
We project the parameter $x$ into this subspace spanned by $r$ basis functions $X_r = \text{span}(\psi_1, \dots, \psi_r)$ as 
\beq\label{eq:projector}
P_r x := \sum_{i=1}^r \psi_i \psi_i^T x = \Psi_r w, \quad \forall x \in \bR^{d_x}, 
\eeq
where the matrix $\Psi_r := (\psi_1, \dots, \psi_r) \in \bR^{d_x\times r}$ represents the projection matrix and the vector $w := (w_1, \dots, w_r)^T \in \bR^r$ is the coefficient vector with element $w_i := \psi_i^T x$ for $i = 1, \dots, r$. The basis functions $\psi_1, \dots, \psi_r$ are obtained as the eigenvectors for the generalized eigenvalue problem 
\beq\label{eq:generalized-eigenproblem}
H \psi_i = \lambda_i \bC_x \psi_i,
\eeq
where $(\lambda_i, \psi_i)_{i=1}^r$ represent the dominant eigenpairs of $(H, \bC_x)$ correspond to the $r$ largest eigenvalues $\lambda_1 \geq \cdots \geq \lambda_{r}$. The matrix $\bC_x$ is the prior covariance of the parameter $x = (\bsg, \bsh)$, i.e., $\bC_x = \text{diag}(\bC_g,\bC_h)$, a block diagonal matrix with each diagonal block representing the covariance of $\bsg$ and $\bsh$. The matrix $H$ is chosen as a gradient information matrix, which is defined as the average of the outer product of the gradient of the log-likelihood w.r.t.\ the posterior, i.e.,  
\beq\label{eq:grad-grad-T}
H = \int_{\bR^{d_x}} (\nabla_x \log f(x)) (\nabla_x \log f(x))^T p(x) dx, 
\eeq
which is evaluated by sample average approximation to be specified later.  By the projection \eqref{eq:projector}, we consider a decomposition of the prior for the parameter $x= x^r + x^\perp$, where $x^r = P_r x$, as 
\beq
p_0(x) = p_0^r(x^r ) p_0^\perp(x^\perp|x^r),
\eeq
where $p_0^r$ and $p_0^\perp$ are marginal and conditional densities for $x^r$ and $x^\perp$. Moreover, we define a marginal likelihood function (an optimal approximation of the full likelihood function \cite{ChenGhattas20}) as
\beq\label{eq:optimal-profile}
f_r(x^r) = \int_{X_\perp} f(x^r + \tilde{x}) p^\perp_0(\tilde{x} | x^r) d \tilde{x},
\eeq
where $X_\perp$ is the complement of the subspace $X_r$. Since the uncertainty in $x^r = P_r x = \Psi_r w$ can be fully represented by the coefficient $w$, we consider the prior and posterior density for $w$, given by
\begin{equation}\label{eq:w-posterior}
\pi_0(w) = p_0^r(\Psi_r w), \quad \text{ and } \pi(w) = \frac{1}{Z_w} f_r(\Psi_r w) \pi_0(w), 
\end{equation}
where the normalization constant $Z_w = E_{w \sim \pi_0}[g(\Psi_r w)]$. 

To sample from the posterior $\pi$ in \eqref{eq:w-posterior}, we employ the SVGD method presented above for the projection coefficient. Specifically,
pSVGD pushes the samples $w_1^0, \dots, w_N^0$ from $\pi_0(w)$ as
\beq\label{eq:w-update-l}
w_m^{\ell+1} = w_m^\ell + \epsilon_l \hat{\phi}_{\ell}^{r, *}(w_m^\ell), \quad m = 1, \dots, N, \ell = 0, 1, \dots,
\eeq
with a step size $\epsilon_l$ and an approximate steepest direction 
\beq\label{eq:w-gradient-saa}
\hat{\phi}_{\ell}^{r, *}(w_m^\ell) = \frac{1}{N} \sum_{n=1}^N \nabla_{w_n^\ell} \log \pi(w_n^\ell) k^r(w_n^\ell, w_m^\ell) + \nabla_{w_n^\ell} k^r(w_n^\ell, w_m^\ell),
\eeq
where the kernel $k^r$ is specified as
\beq\label{eq:w-kernel}
k^r(w,w') = \exp\left(- \frac{1}{h_w} (w-w')^T (\Lambda + I) (w - w') \right).
\eeq
where the metric matrix $\Lambda + I$ with eigenvalues $\Lambda = \text{diag}(\lambda_1, \dots, \lambda_r)$ of \eqref{eq:generalized-eigenproblem} are used to account for data impact in different directions $\psi_1, \dots, \psi_r$. The pSVGD algorithm is presented in Algorithm \ref{alg:pSVGD}, which is implemented in parallel with samples distributed in multiple cores.

\begin{algorithm}[!htb]
	\caption{pSVGD in parallel}
	\label{alg:pSVGD}
	\begin{algorithmic}[1]
		\STATE{\bfseries Input:} samples $\{x_n^0\}_{n=1}^N$ in each of $K$ cores, basis $\Psi_r$, maximum iteration $L_{\text{max}}$, tolerance $w_{\text{tol}}$.
		\STATE{\bfseries Output:} posterior samples $\{x_n^*\}_{n=1}^N$ in each core.
		\STATE Set $\ell = 0$, project $w_n^0 = \Psi_r^T x_n^0$,  $x_n^\perp = x_n^0 - \Psi_r w_n^0$, and communicate for $\{w_n^0\}_{n=1}^N$.
		\REPEAT
		\STATE Compute gradients $\nabla_{w_n^\ell} \log \pi(w_n^\ell)$ for $n = 1, \dots, N$, and communicate.
		\STATE Compute the kernel values $k^r(w_n^\ell, w_m^\ell)$ and their gradients $\nabla_{w_n^\ell} k^r(w_n^\ell, w_m^\ell)$ for $n = 1, \dots, NK$, $m = 1, \dots, N$, and communicate for them.
		\STATE Update samples $w^{\ell+1}_m$ from $w^{\ell}_m$ by \eqref{eq:w-update-l} and \eqref{eq:w-gradient-saa} for $m = 1, \dots, N$, with $NK$ samples used for SAA in \eqref{eq:w-gradient-saa}, and communicate for $\{w_m^0\}_{m=1}^N$.
		\STATE Set $\ell \leftarrow \ell + 1$.
		\UNTIL{$\ell \geq L_{\text{max}}$ or $\text{mean} (||w_m^\ell - w_m^{\ell-1}||_2) \leq w_{\text{tol}}$.}
		\STATE Reconstruct samples $x_n^* = \Psi_r w_n^\ell + x_n^\perp$.
	\end{algorithmic}
\end{algorithm}
We remark that the projected likelihood in \eqref{eq:w-posterior} can be approximated by its definition \eqref{eq:optimal-profile} as 
\beq
f_r(\Psi_r w_n^\ell) \approx f(\Psi_r w_n^\ell + x_n^\perp),
\eeq
since its variation in $X_\perp$ is negligible.
We adaptively construct the projector $P_r$ with basis $\Psi_r$ for Algorithm \ref{alg:pSVGD}, for which we evaluate $H$ in \eqref{eq:grad-grad-T} by the sample average approximation
\beq\label{eq:a-H}
\hat{H}: = \frac{1}{M}\sum_{m=1}^M \nabla_x \log f(x^\ell_m) (\nabla_x \log f(x^\ell_m))^T,
\eeq
where the samples $x_1^\ell, \dots, x_M^\ell$ are adaptively transported from the prior samples $x_1^0, \dots, x_M^0$ by pSVGD. This procedure is summarized in Algorithm \ref{alg:apSVGD}. We remark that by the adaptive construction, we push the samples to their posterior in each subspace $X_r^{\ell_x}$ spanned by (possibly) different basis $\Psi_r^{\ell_x}$ with different $r$ for different ${\ell_x}$, during which the frozen samples $x^\perp_n$ in Algorithm \ref{alg:pSVGD} are also updated at each step $\ell_x$ of Algorithm \ref{alg:apSVGD}. 

\begin{algorithm}[!htb]
	\caption{Adaptive pSVGD in parallel}
	\label{alg:apSVGD}
	\begin{algorithmic}[1]
		\STATE{\bfseries Input:} samples $\{x_n^0\}_{n=1}^N$ in each of $K$ cores, $L_{\text{max}}^x, L^w_{\text{max}}$, $x_{\text{tol}}, w_{\text{tol}}$.
		\STATE{\bfseries Output:} posterior samples $\{x_n^*\}_{n=1}^N$ in each core.
		\STATE Set $\ell_x = 0$.
		\REPEAT
		\STATE Compute $\nabla_x \log f(x_n^{\ell_x})$ in \eqref{eq:a-H}, $n = 1, \dots, N$ in each core, and communicate. 
		\STATE Solve \eqref{eq:generalized-eigenproblem} with $H$ approximated as in \eqref{eq:a-H}, to get bases $\Psi_r^{\ell_x}$.
		\STATE Apply the pSVGD Algorithm \ref{alg:pSVGD}, i.e., 
		$$
		\{x_n^*\}_{n=1}^N = \text{pSVGD}(\{x_n^{\ell_x}\}_{n=1}^N, \Psi_r^{\ell_x}, L_{\text{max}}^w, w_{\text{tol}}).
		$$
		\STATE Set $\ell_x \leftarrow \ell_x + 1$ and $x_n^{\ell_x} = x_n^*$, $n = 1, \dots, N$.
		\UNTIL{$\ell_x \geq L^x_{\text{max}}$ or $\text{mean} (||x_m^{\ell_x} - x_m^{\ell_x-1}||_X) \leq x_{\text{tol}}$.}
	\end{algorithmic}
\end{algorithm}

\subsection{Results}

\begin{figure}[!htb]
	\centering
	\includegraphics[width=0.51\linewidth]{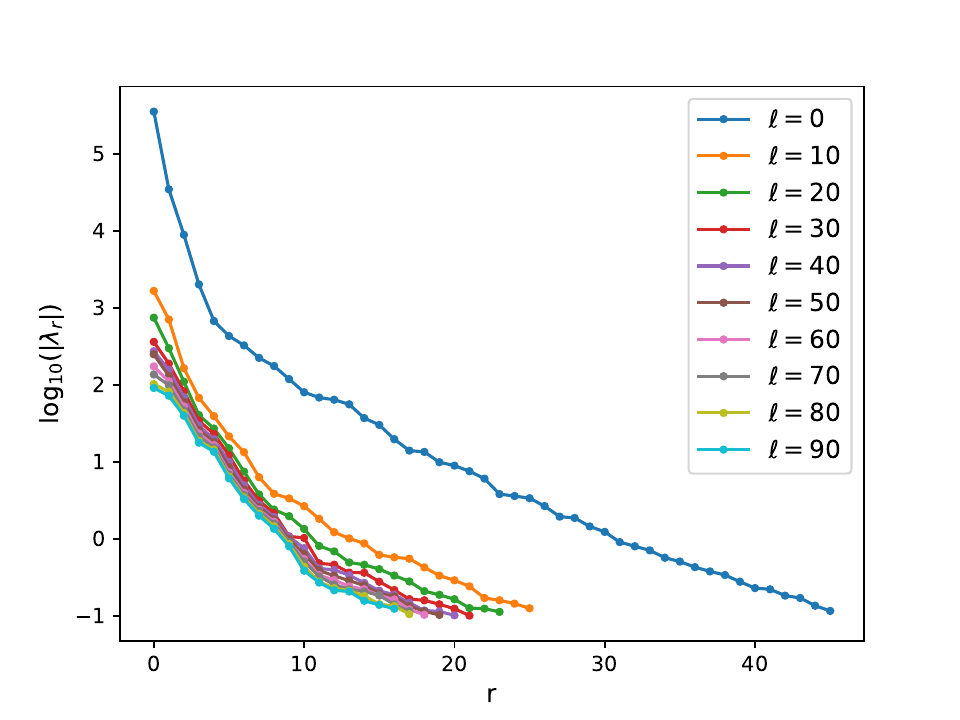}\hspace*{-0.5cm}	
	\includegraphics[width=0.51\linewidth]{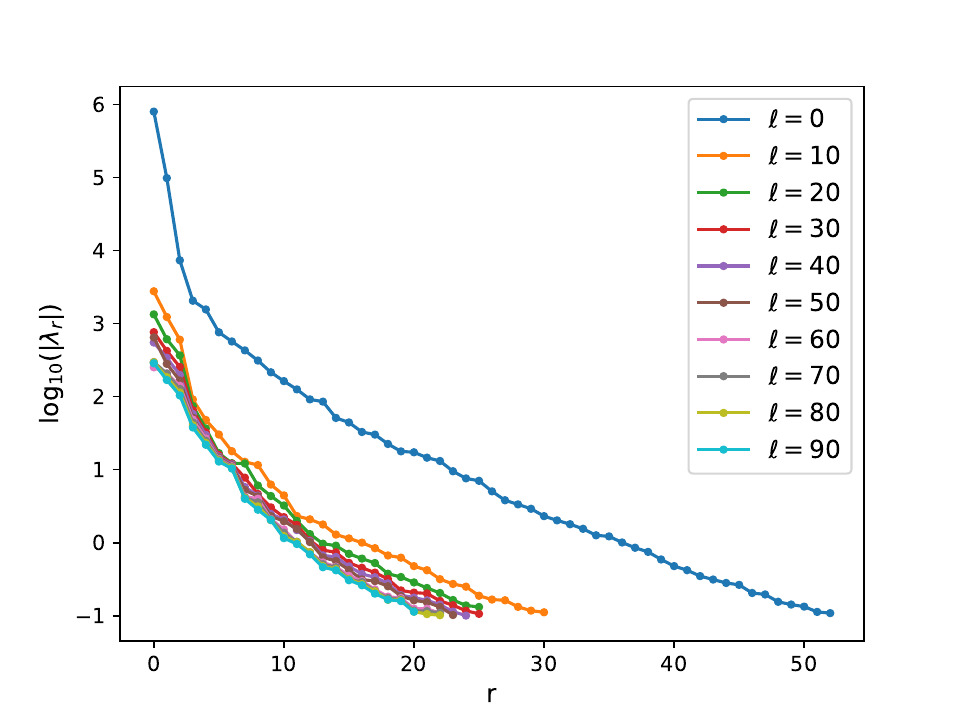}
	
	\caption{Decay of the eigenvalues (in logarithmic scale) of the generalized eigenvalue problem \eqref{eq:generalized-eigenproblem} at different iterations $\ell$ of pSVGD sample update, where $H$ is approximated by $\hat{H}$ in \eqref{eq:a-H}. Left: New Jersey. Right: Texas.}\label{fig:bayesian-eigenvalues}
\end{figure}

\begin{figure}[!htb]
	\centering
	\includegraphics[width=0.51\linewidth]{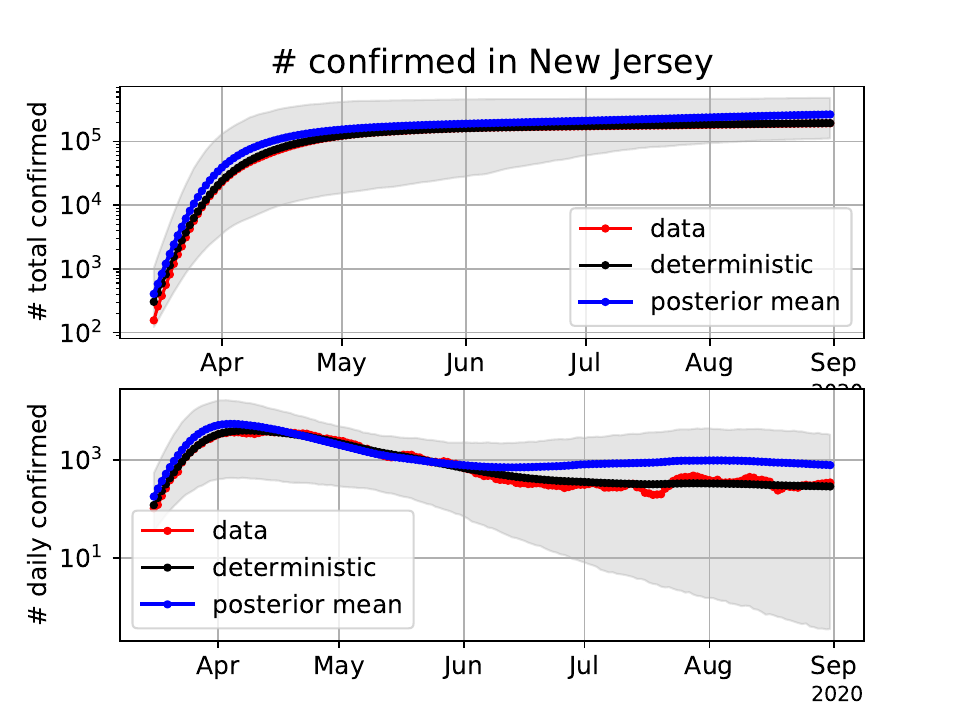}\hspace*{-0.5cm}
	\includegraphics[width=0.51\linewidth]{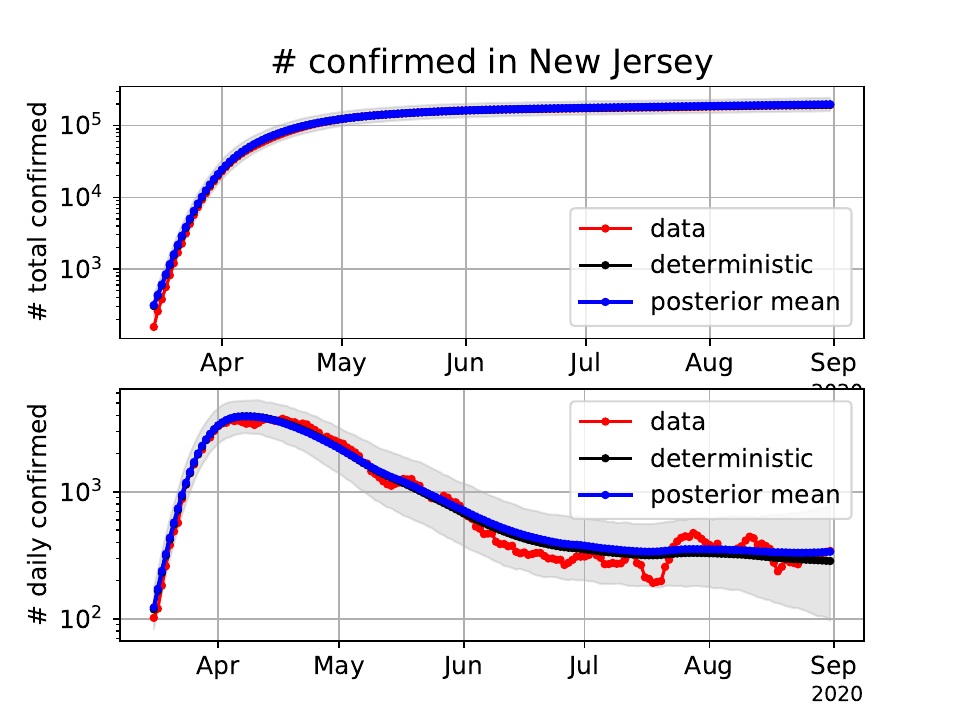}
	
	\vspace*{-0.3cm}
	
	\includegraphics[width=0.51\linewidth]{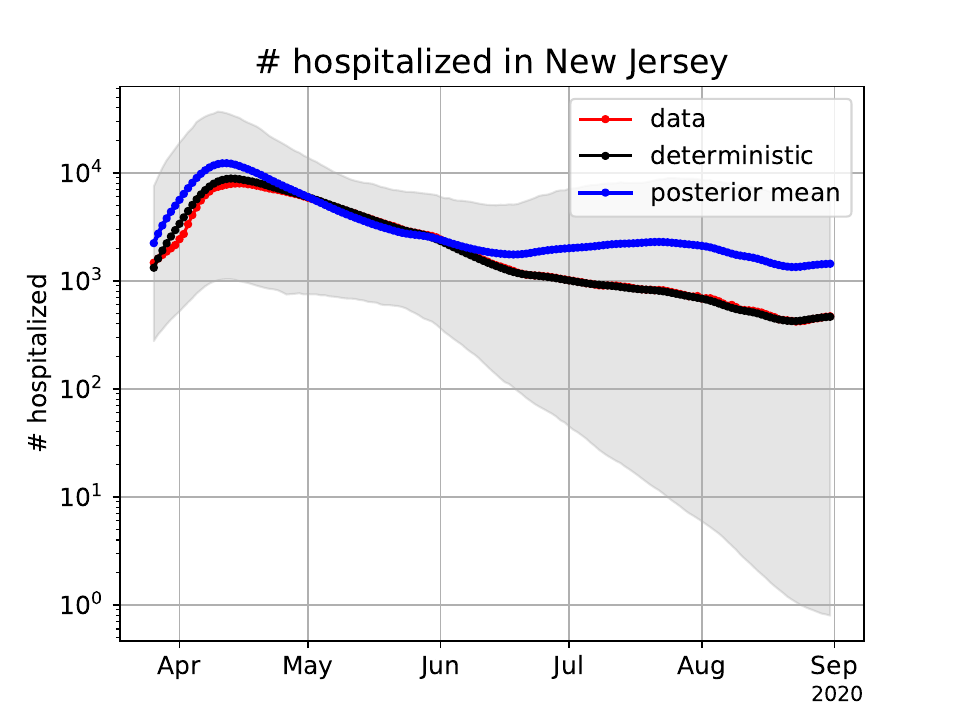}\hspace*{-0.5cm}
	\includegraphics[width=0.51\linewidth]{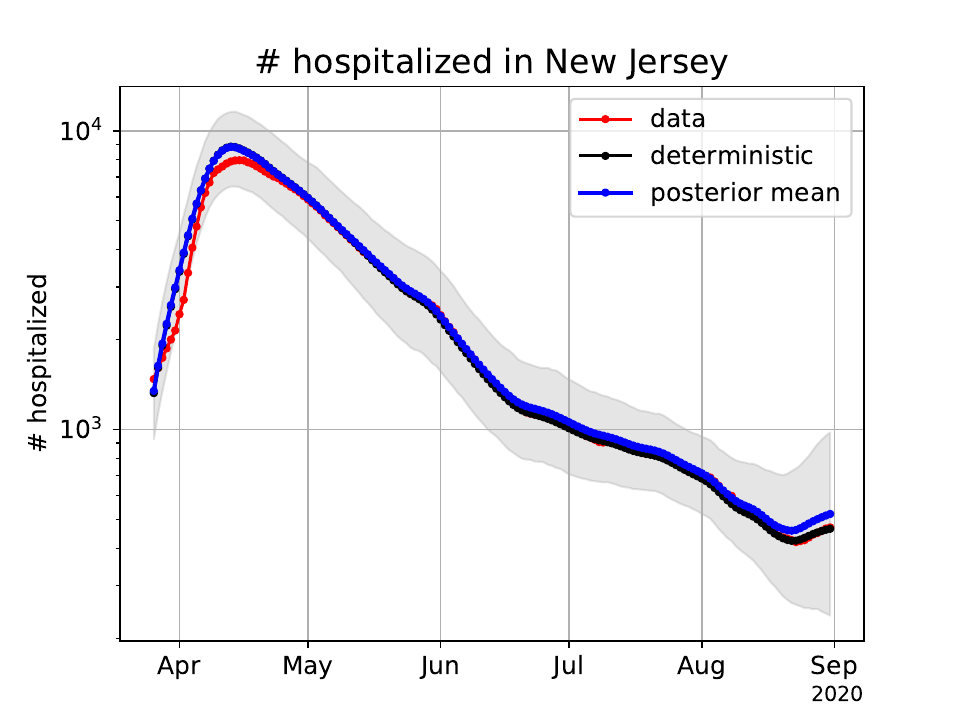}
	
	\caption{Comparison of the data and model outputs for confirmed and hospitalized cases by both the deterministic inversion and the Bayesian inference. The data here are the same as the seven day moving average data in Figure \ref{fig:data-NJ}. The deterministic model outputs are the results of model \eqref{eq:LTC-model} at the optimal parameter value obtained by the deterministic inversion \eqref{eq:inference-deterministic}. The sample mean and 90\% credible interval (90\% samples fall in this interval) of the model outputs are obtained as the results of model \eqref{eq:LTC-model} at the prior samples (left) and Bayesian posterior samples (right).}\label{fig:bayesian-NJ}
\end{figure}

\begin{figure}[!htb]
	\centering
	\includegraphics[width=0.51\linewidth]{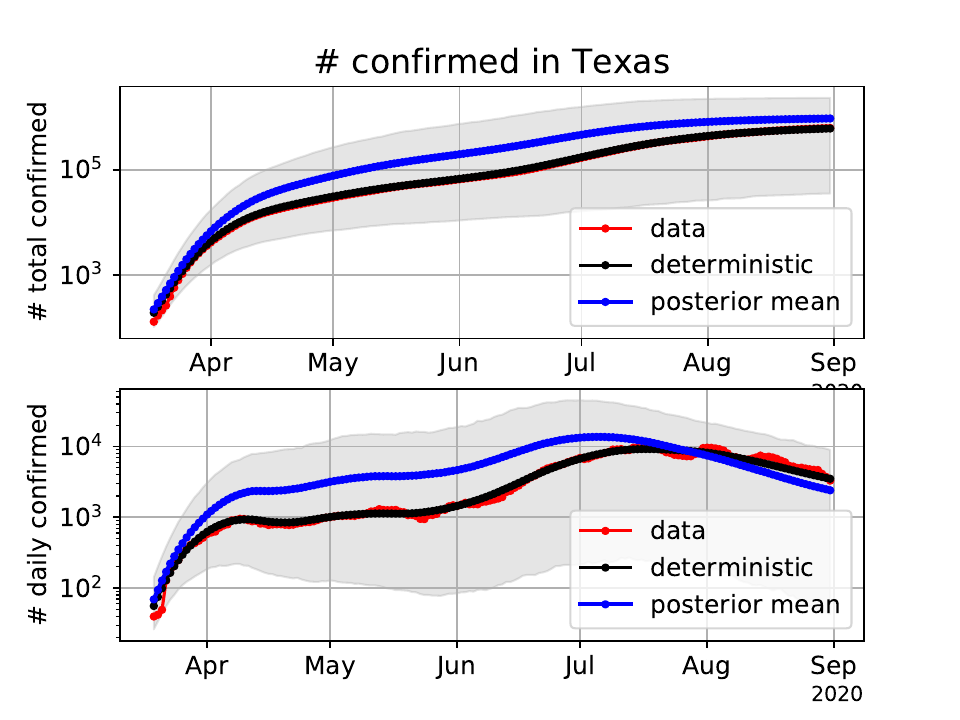}\hspace*{-0.5cm}
	\includegraphics[width=0.51\linewidth]{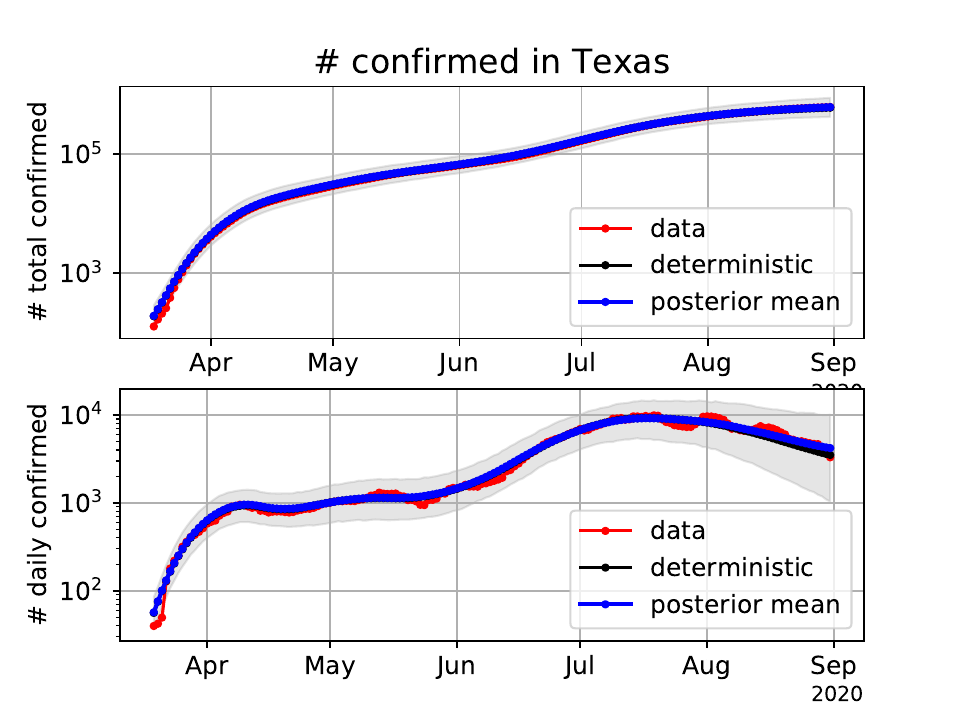}
	
	\vspace*{-0.3cm}
	
	\includegraphics[width=0.51\linewidth]{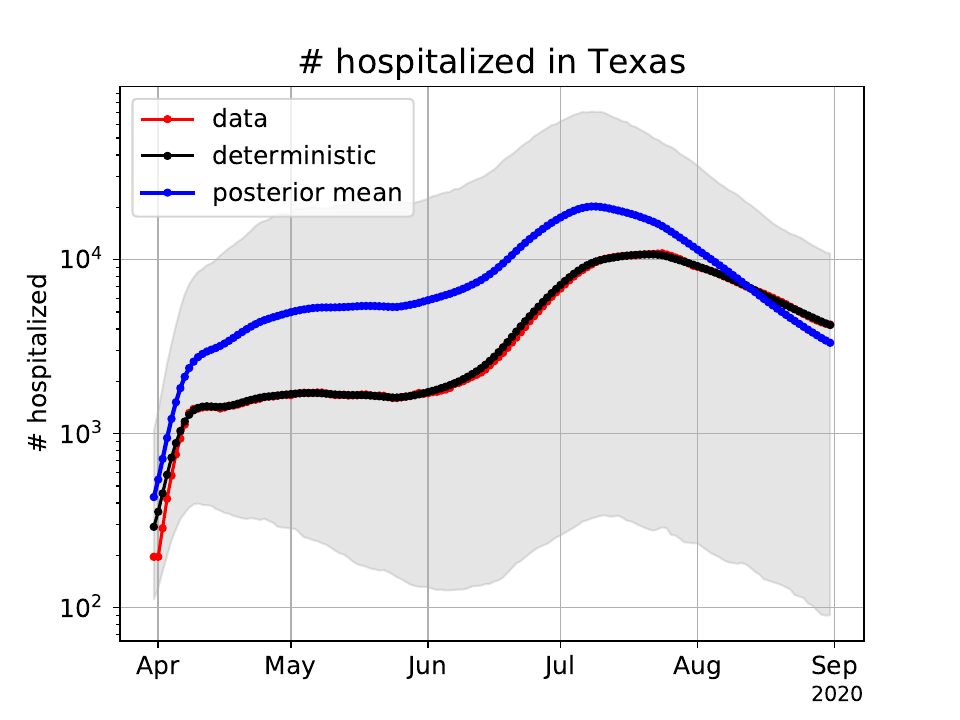}\hspace*{-0.5cm}
	\includegraphics[width=0.51\linewidth]{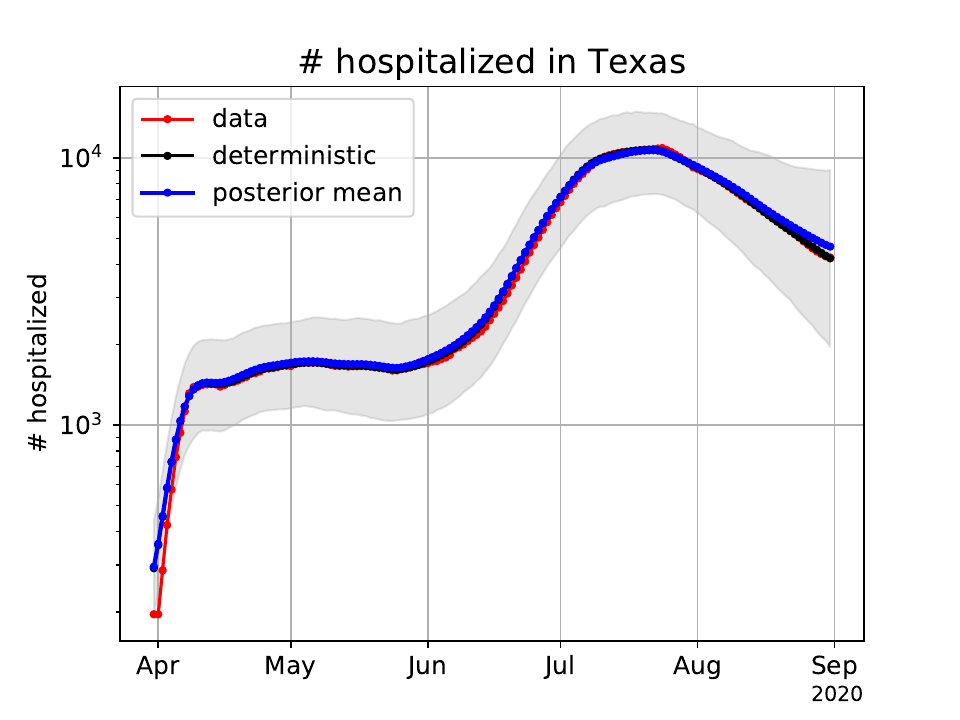}
	
	\caption{The same as in Figure \ref{fig:bayesian-NJ} for Texas here.}\label{fig:bayesian-TX}
\end{figure}

We set the the scaling parameter $s_g=1000$ and $s_I = 1$ in the prior covariance \eqref{eq:C-g} for $\bsg$, and $s_h = 1/10$ for the prior covariance \eqref{eq:C-h} for $\bsh$, and set the observation noise covariance matrix $\bC_\omega$ as the identity matrix. This leads to sufficient variation of the parameters that result in large variation (and no blowup) of the model output at different samples drawn from the prior distribution, as can be observed in the left part of Figure \ref{fig:bayesian-NJ} and \ref{fig:bayesian-TX} for New Jersey and Texas. We run the adaptive pSVGD Algorithm \ref{alg:apSVGD} with $N = 125$ samples in each of $K = 8$ cores, so 1000 samples in total. We stop the algorithm at $10\times 10$ iterations, where the projection basis functions $\Psi_r^{\ell_x}$ are updated for each 10 iterations. We remark that the move of the samples already becomes small towards the end of the iteration. We choose the dimension $r$ such that the eigenvalue $\lambda_{r+1} < 10^{-1}$. The decay of the eigenvalues at different pSVGD iterations are shown in Figure \ref{fig:bayesian-eigenvalues} for both states, from which we can observe a significant reduction of the eigenvalues by several orders of magnitude for the first few dimensions, which implies that the data informed parameter subspace is intrinsically low-dimensional, e.g., the first 10 dimension (out of the total 1520 for New Jersey and 1488 for Texas) catches over 99\% difference of the prior and posterior distribution measured by the KL divergence \eqref{eq:KL}.  

\begin{figure}[!htb]
	\centering
	%
	
	\includegraphics[width=0.51\linewidth]{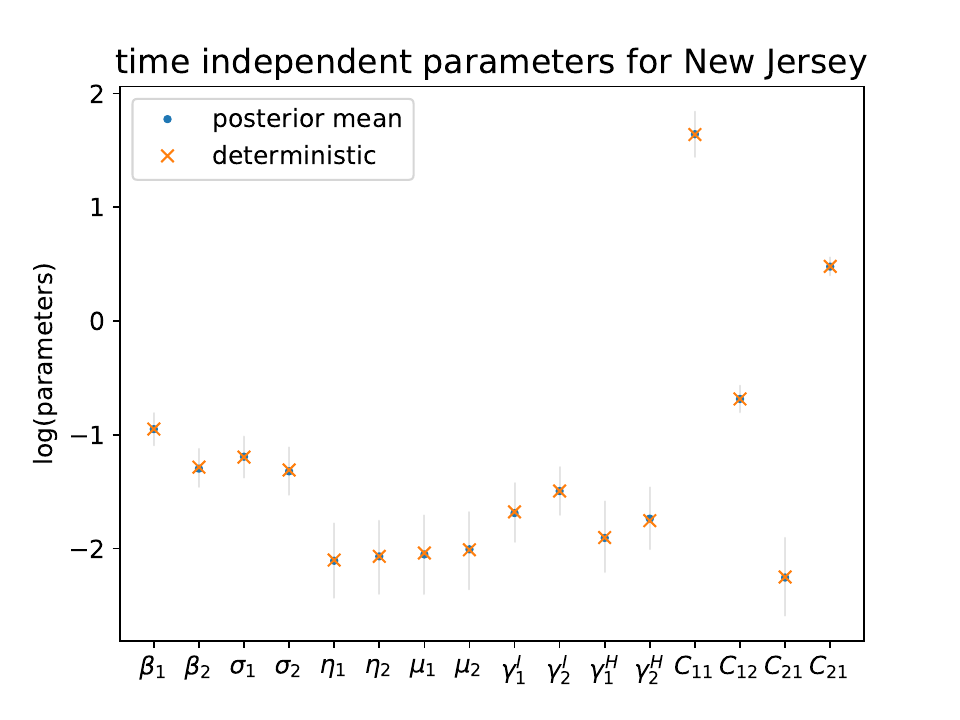}\hspace*{-0.5cm}
	\includegraphics[width=0.51\linewidth]{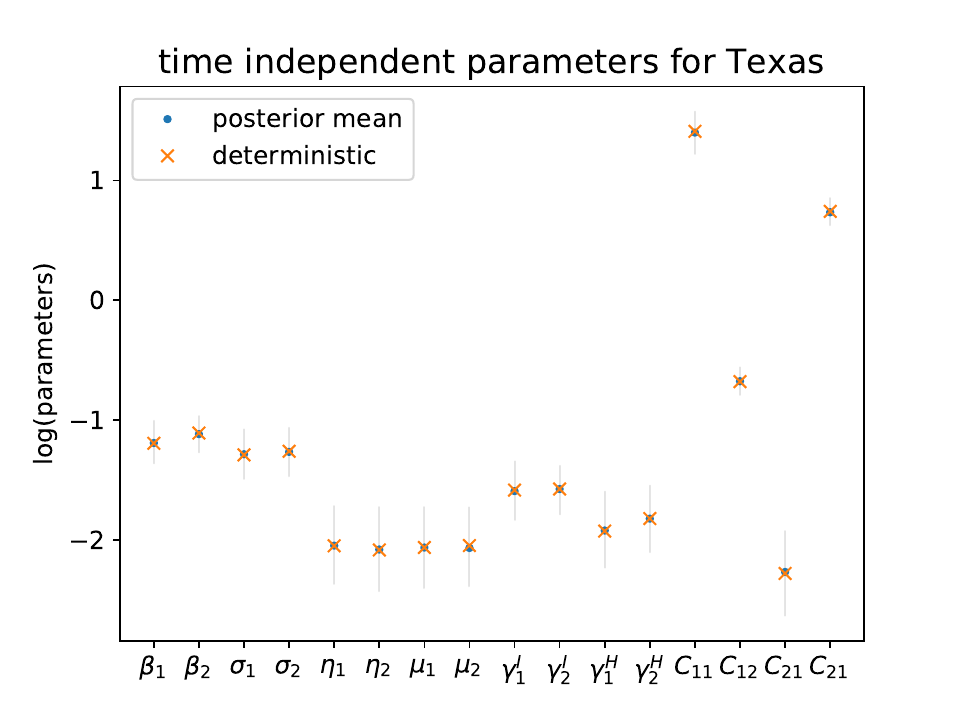}	
	
	\caption{Optimal values by deterministic inversion, and sample mean and 90\% credible interval (90\% samples fall in this interval) of the posterior samples by Bayesian inference for time independent parameters. Note that the values are shown for $\bsh$ in \eqref{eq:logh}, i.e., logarithm of the parameters, to have better visualization. Left: New Jersey. Right: Texas.}\label{fig:bayesian-parameters}
\end{figure}

\begin{figure}[!htb]
	\centering
	\includegraphics[width=0.51\linewidth]{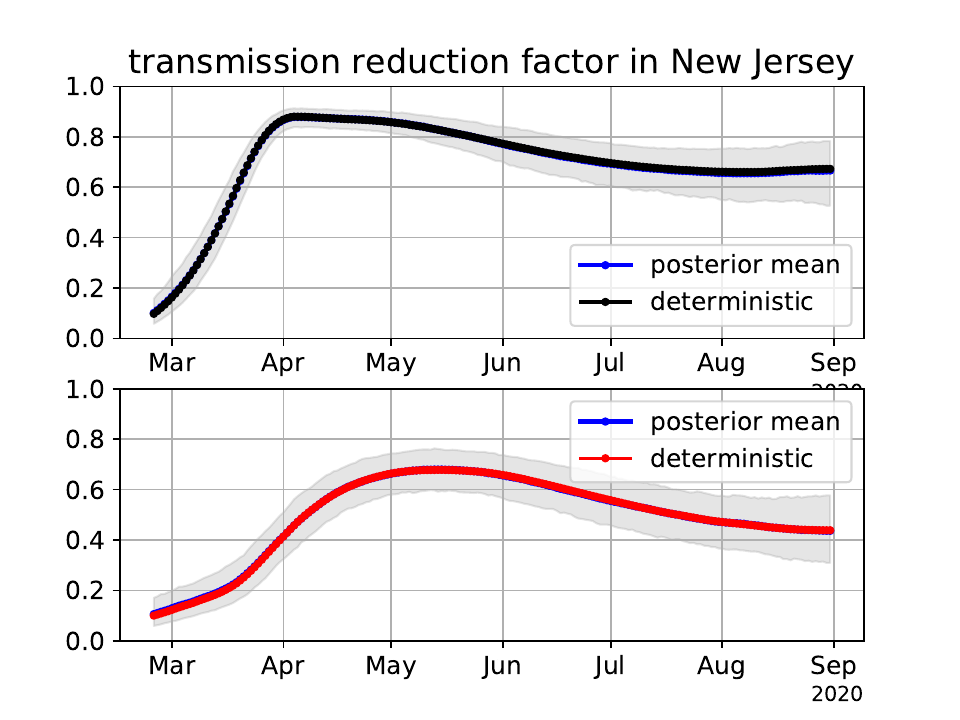}\hspace*{-0.5cm}	
	\includegraphics[width=0.51\linewidth]{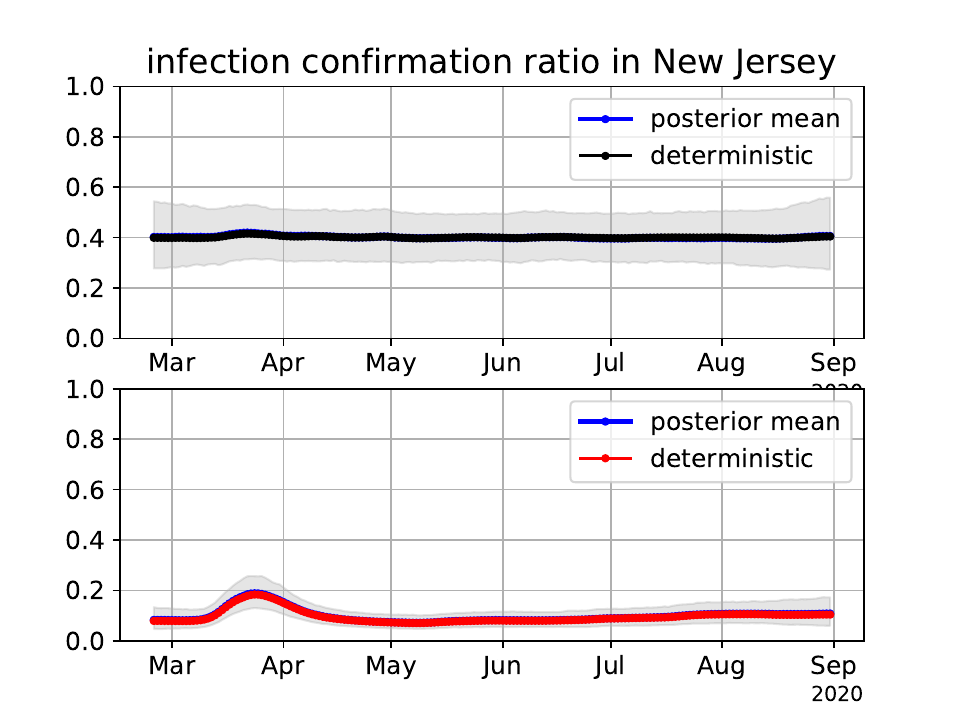}
	
	\includegraphics[width=0.51\linewidth]{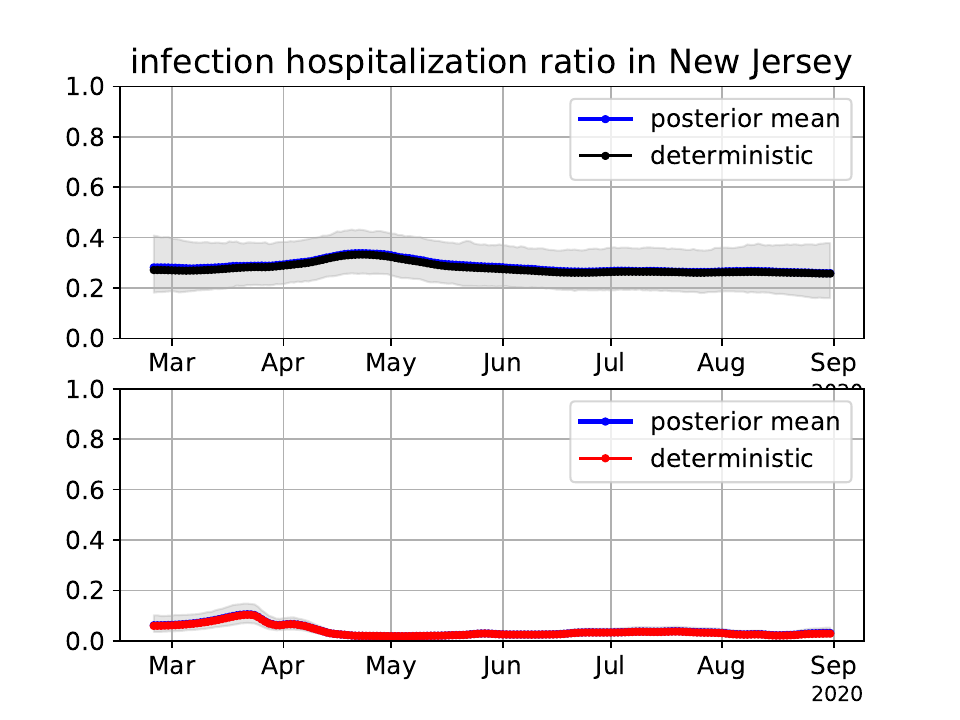}\hspace*{-0.5cm}
	\includegraphics[width=0.51\linewidth]{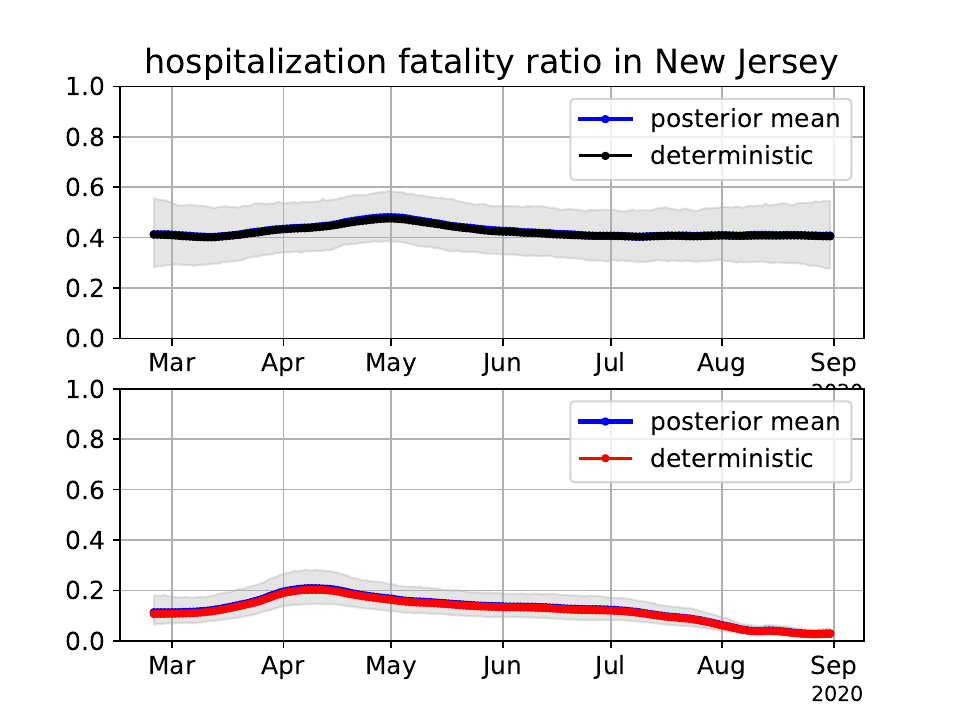}	
	
	\caption{Optimal sample by deterministic inversion, posterior sample mean, and 90\% credible interval (90\% samples fall in this interval) of the posterior samples by Bayesian inference for the time dependent parameters inside LTC (top) and outside LTC (bottom) in New Jersey.}\label{fig:bayesian-parameters-NJ}
\end{figure}

\begin{figure}[!htb]
	\centering
	\includegraphics[width=0.51\linewidth]{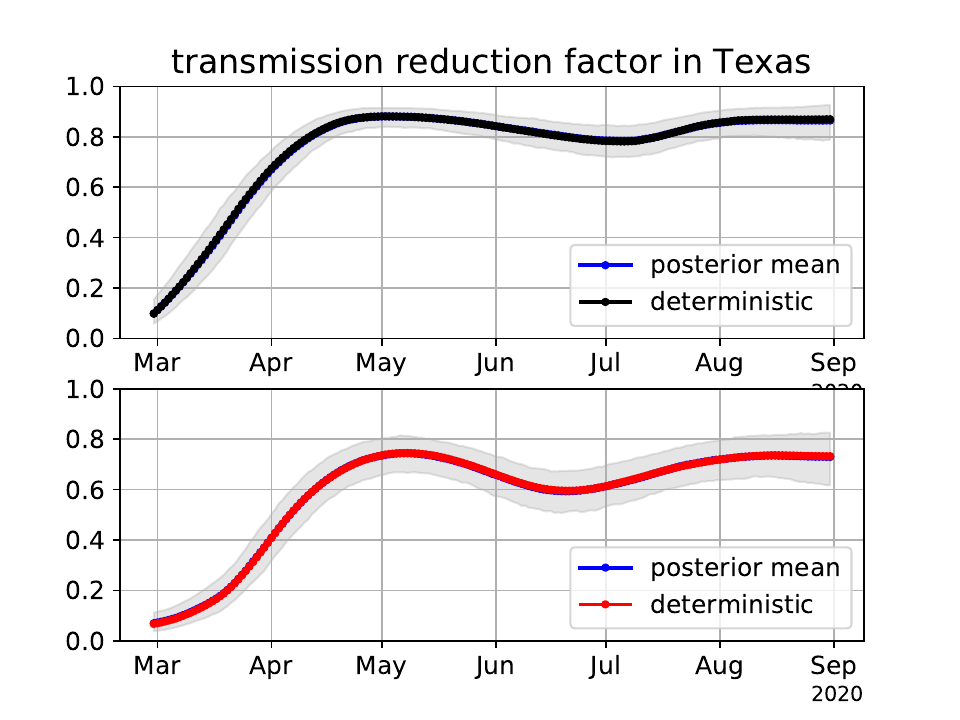}\hspace*{-0.5cm}	
	\includegraphics[width=0.51\linewidth]{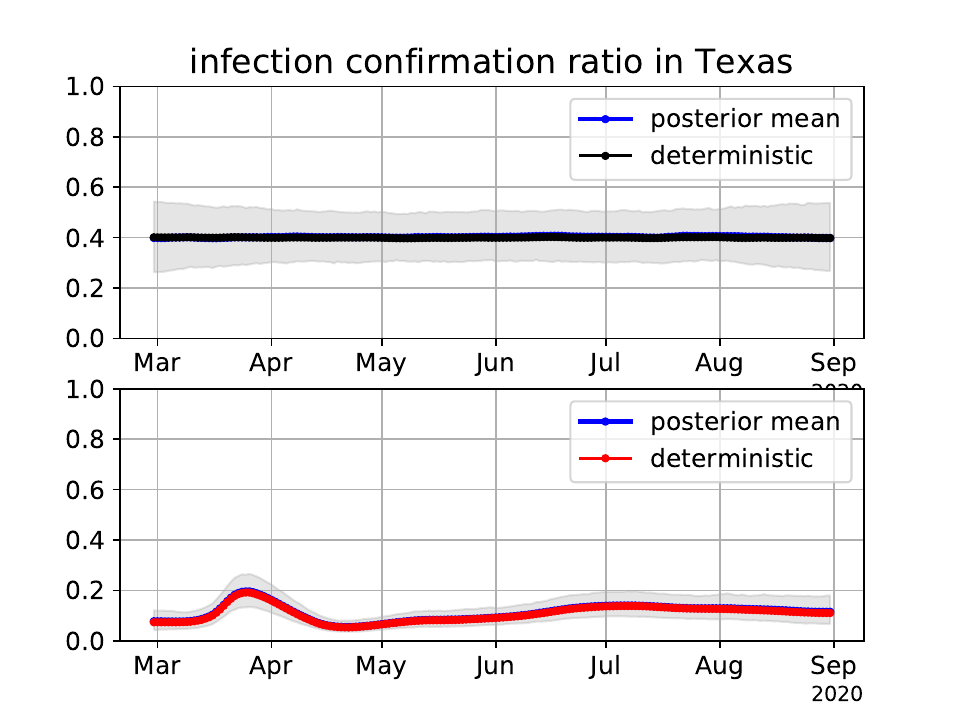}
	
	\includegraphics[width=0.51\linewidth]{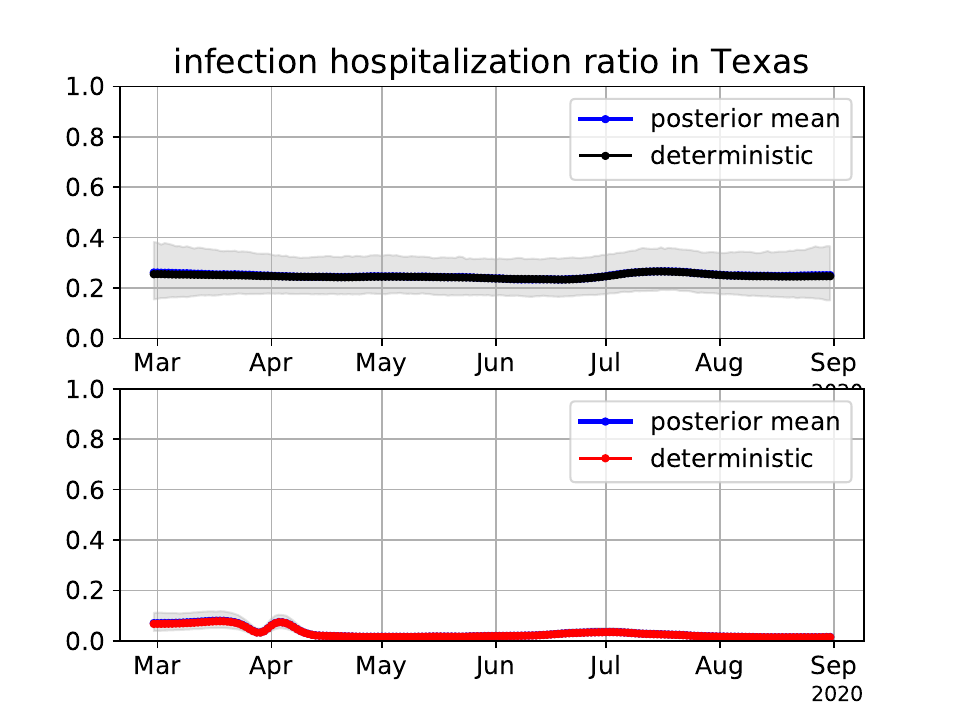}\hspace*{-0.5cm}
	\includegraphics[width=0.51\linewidth]{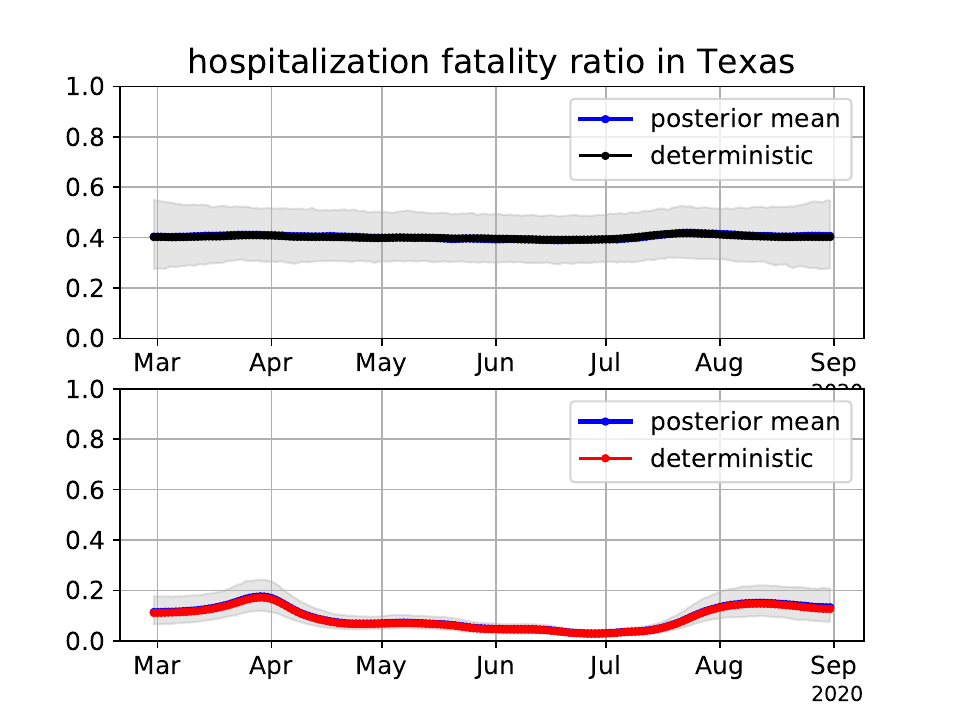}	
	
	\caption{The same as in Figure \ref{fig:bayesian-parameters-NJ} for Texas.}\label{fig:bayesian-parameters-TX}
\end{figure}

Figure \ref{fig:bayesian-parameters} displays the statistics of the posterior samples, including their mean and 90\% credible interval, in which 90\% samples fall, as well as the optimal values obtained from the deterministic inversion for the time independent parameters, while Figures \ref{fig:bayesian-parameters-NJ} and \ref{fig:bayesian-parameters-TX} display these quantities for the time dependent parameters. We observe that the posterior sample mean is close to the optimal values obtained by the deterministic inversion. Meanwhile, it is evident from the large 90\% credible interval of the posterior samples that there are uncertainties in the parameters that can not be ignored. These uncertainties become crucial for reliable forecast of new infected, hospitalized, and deceased cases into the future based on the model output, as well as for risk-averse optimal mitigation design to flatten the curve. 

\begin{figure}[!htb]
	\centering
	
	\includegraphics[width=0.51\linewidth]{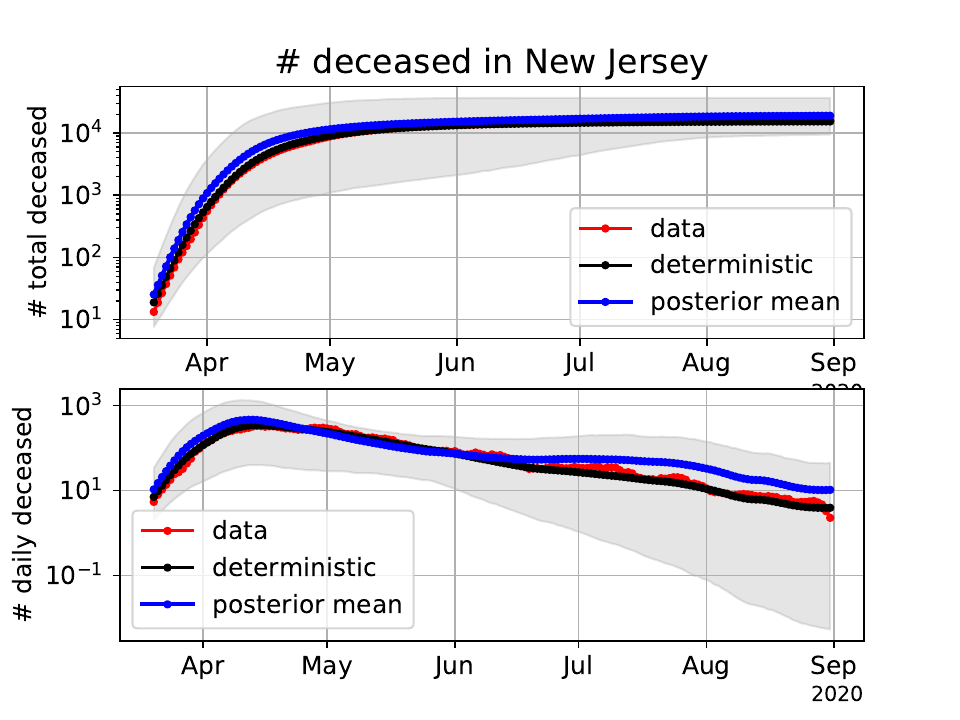}\hspace*{-0.5cm}
	\includegraphics[width=0.51\linewidth]{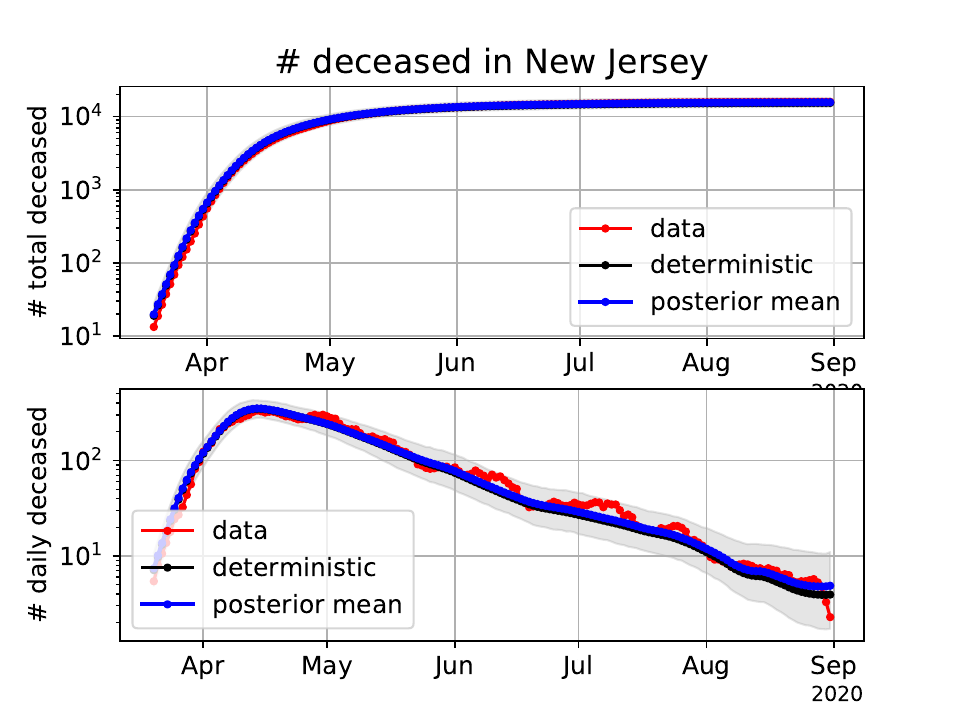}
	
	\vspace*{-0.3cm}
	
	\includegraphics[width=0.51\linewidth]{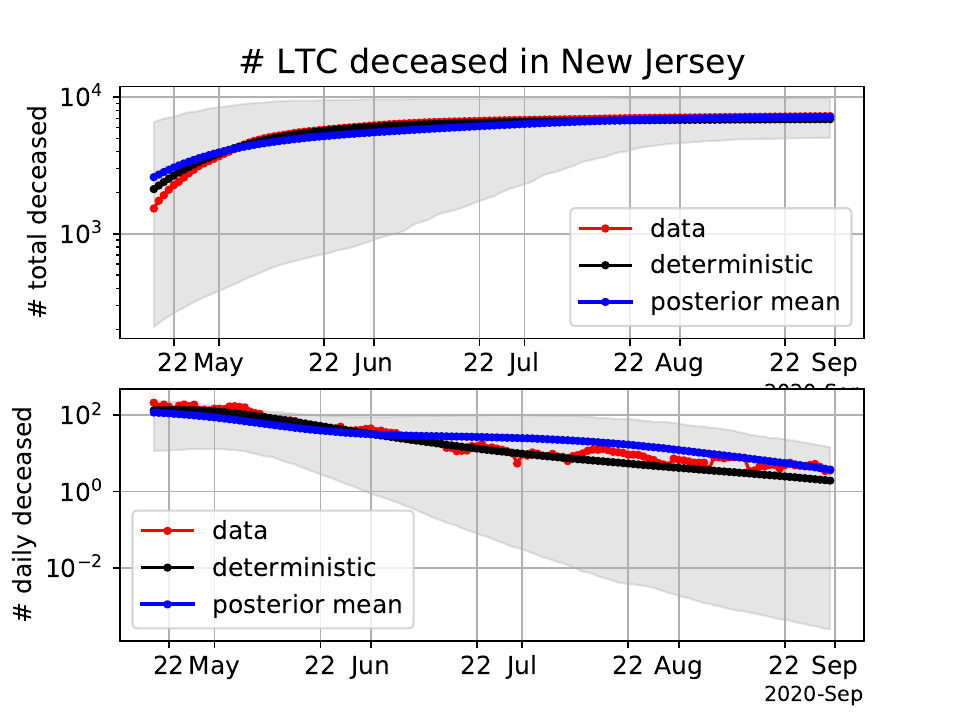}\hspace*{-0.5cm}
	\includegraphics[width=0.51\linewidth]{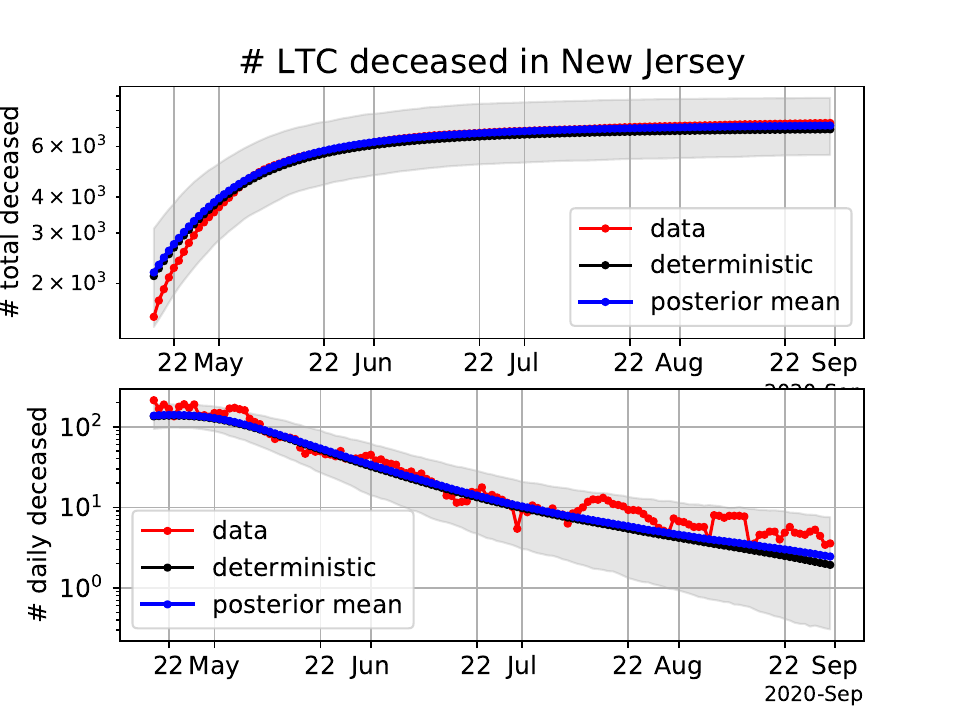}
	
	\caption{The same as in Figure \ref{fig:bayesian-NJ} for total deceased cases (top) and deceased cases inside LTC (bottom) in New Jersey.}\label{fig:bayesian-NJ-deceased}
\end{figure}

\begin{figure}[!htb]
	\centering	
	\includegraphics[width=0.51\linewidth]{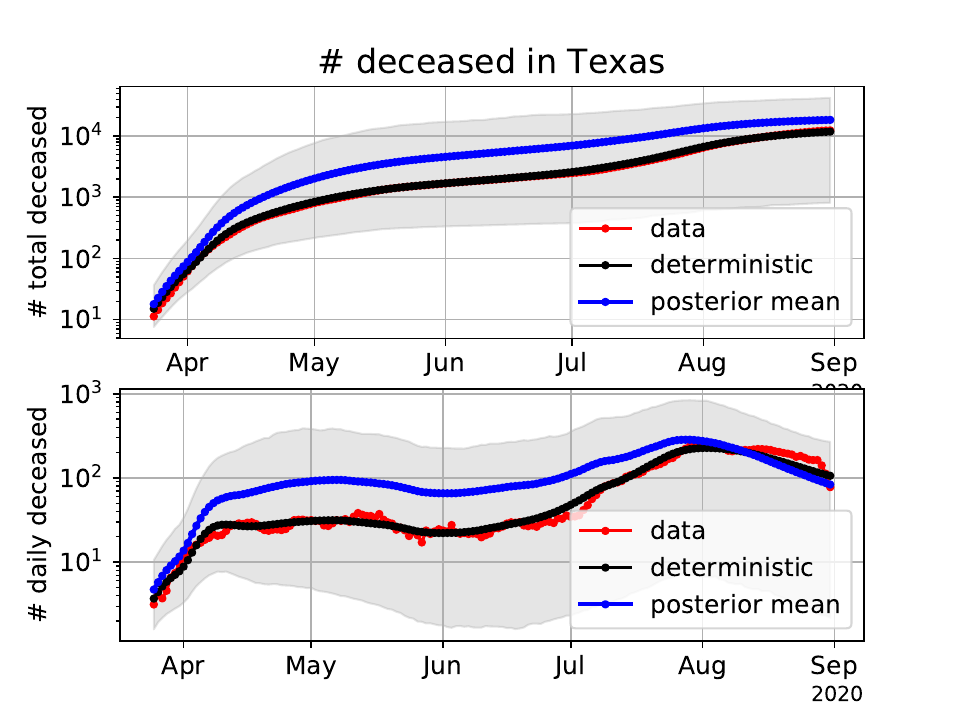}\hspace*{-0.5cm}
	\includegraphics[width=0.51\linewidth]{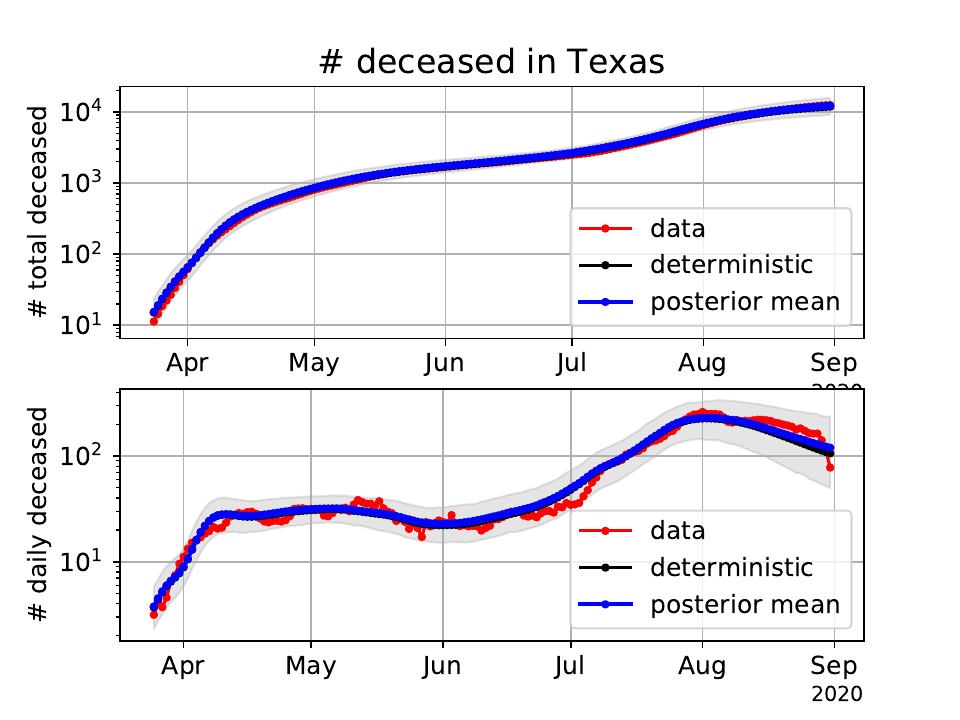}
	
	\vspace*{-0.3cm}
	
	\includegraphics[width=0.51\linewidth]{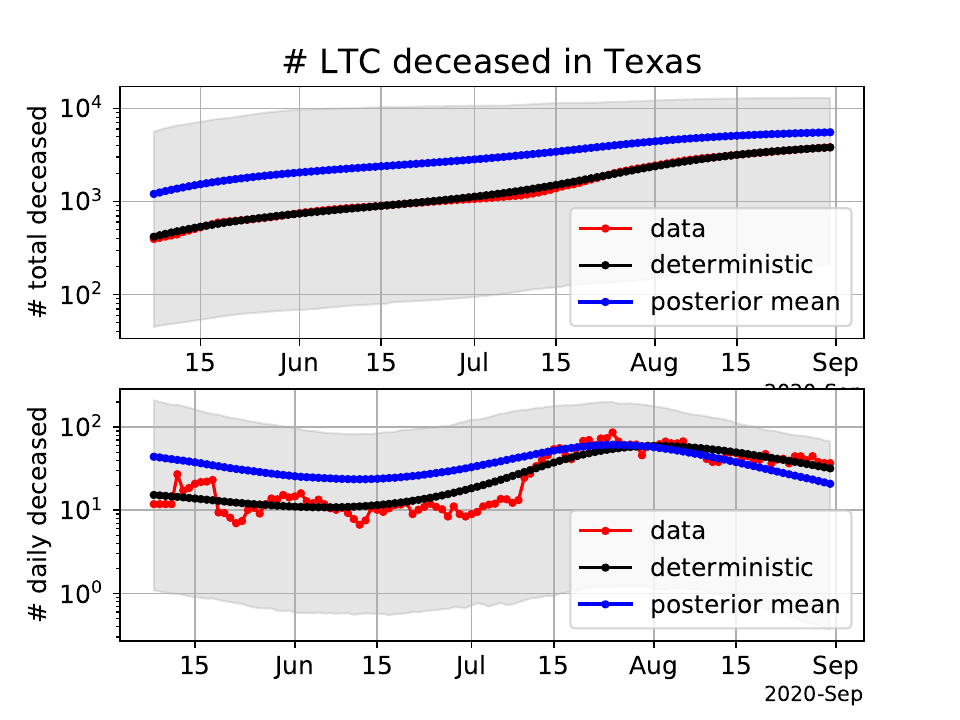}\hspace*{-0.5cm}
	\includegraphics[width=0.51\linewidth]{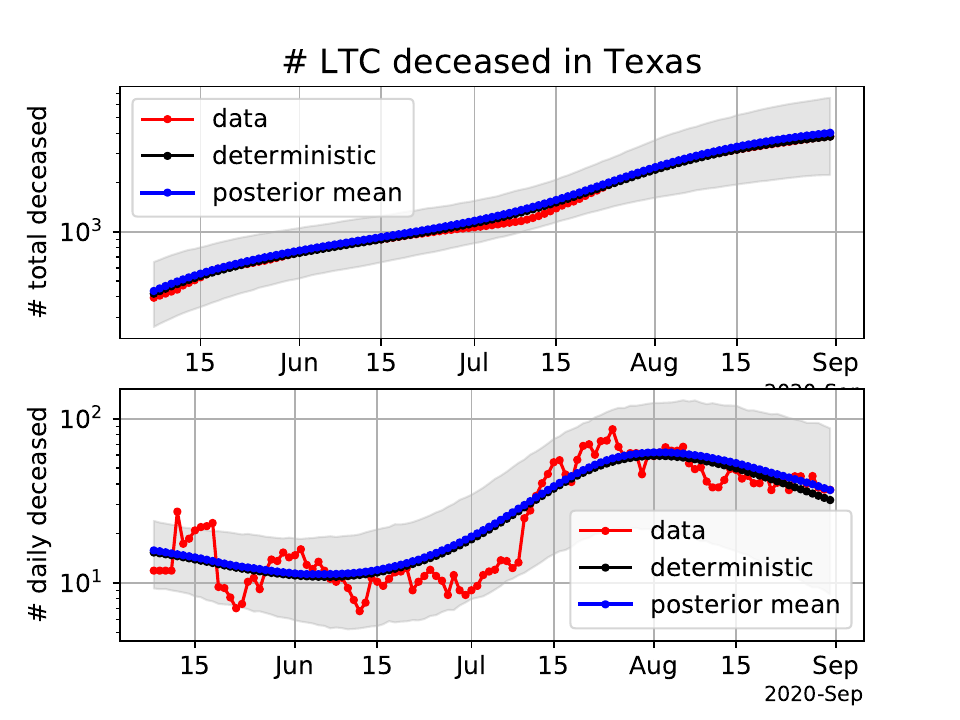}
	
	\caption{The same as in Figure \ref{fig:bayesian-NJ} for total deceased cases (top) and deceased cases inside LTC (bottom) in Texas.}\label{fig:bayesian-TX-deceased}
\end{figure}

Finally, we show the comparison of the statistics (mean and 90\% credible interval) of the model outputs at the prior samples and the posterior samples in Figures \ref{fig:bayesian-NJ} and \ref{fig:bayesian-TX} for confirmed and hospitalized cases, and in Figures \ref{fig:bayesian-NJ-deceased} and \ref{fig:bayesian-TX-deceased} for deceased cases. From these figures we can observe that the variation of the model outputs is considerably reduced from the prior to the posterior samples. Moreover, most of the data, except for a few data points that is largely uncertain or possibly incorrect/outlier, fall inside the 90\% credible interval of the model outputs, which demonstrate the reliability and advantage of the Bayesian inference that account for the parameter and data uncertainties. Despite this ability in quantifying the uncertainties in both the parameters and model outputs, we caution that the credible intervals of different parameters and model outputs are evidently different, e.g., the credible intervals for the number of total confirmed and deceased cases are much smaller than that of the hospitalized, and daily confirmed and deceased cases, while the credible intervals for the number of LTC deceased cases are relatively larger than that of the total number of deceased cases, etc., which may bring in over or under confidence on the inference results. This is probably due to the relatively simple choice of the prior distribution of the parameters and the distribution of the observation noise.

\subsection{Forecast under uncertainty}

\begin{figure}[!htb]
	\centering
	\includegraphics[width=0.51\linewidth]{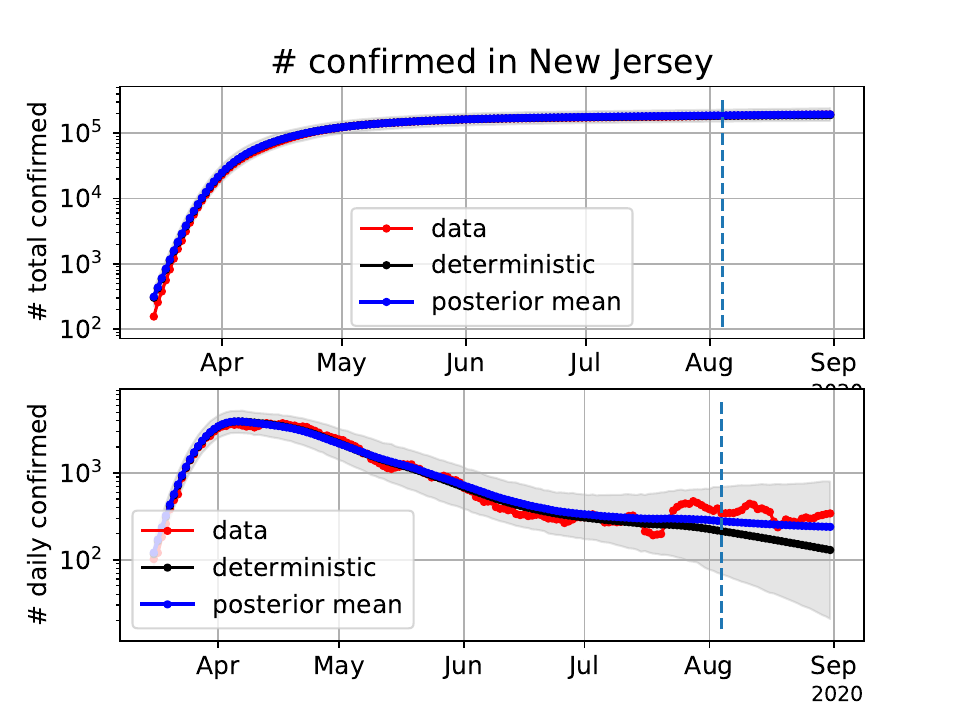}\hspace*{-0.5cm}
	\includegraphics[width=0.51\linewidth]{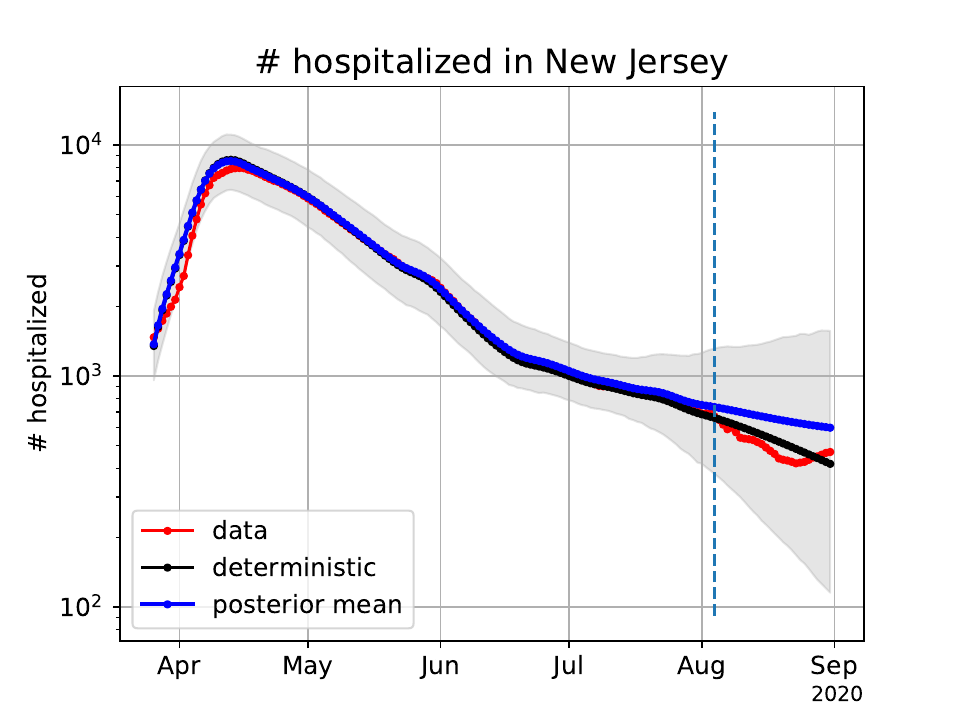}
	
	\vspace*{-0.3cm}
	
	\includegraphics[width=0.51\linewidth]{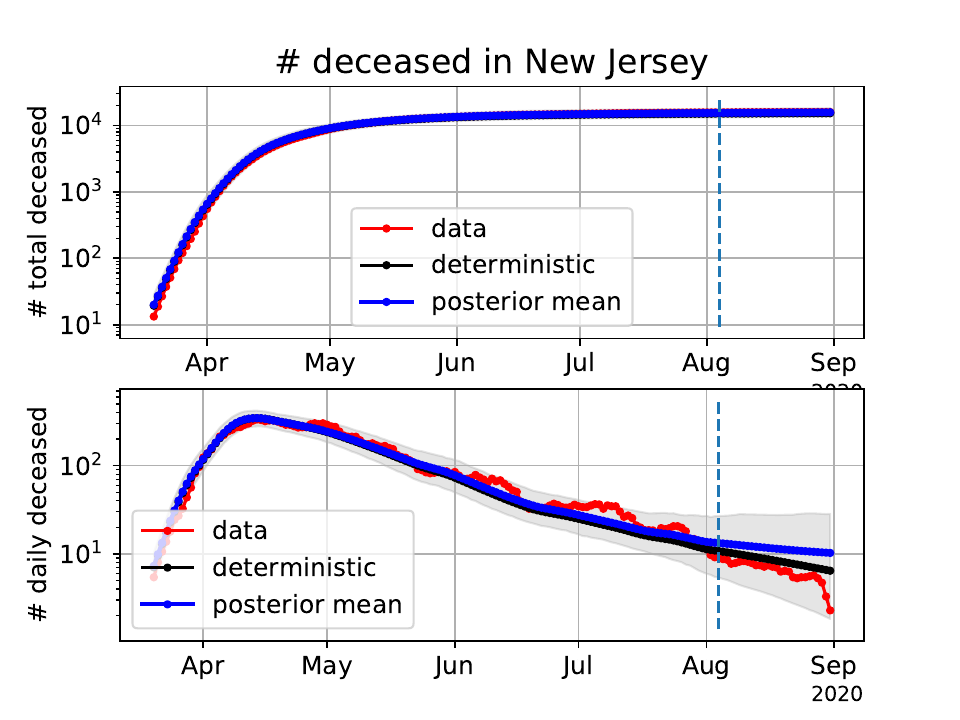}\hspace*{-0.5cm}
	\includegraphics[width=0.51\linewidth]{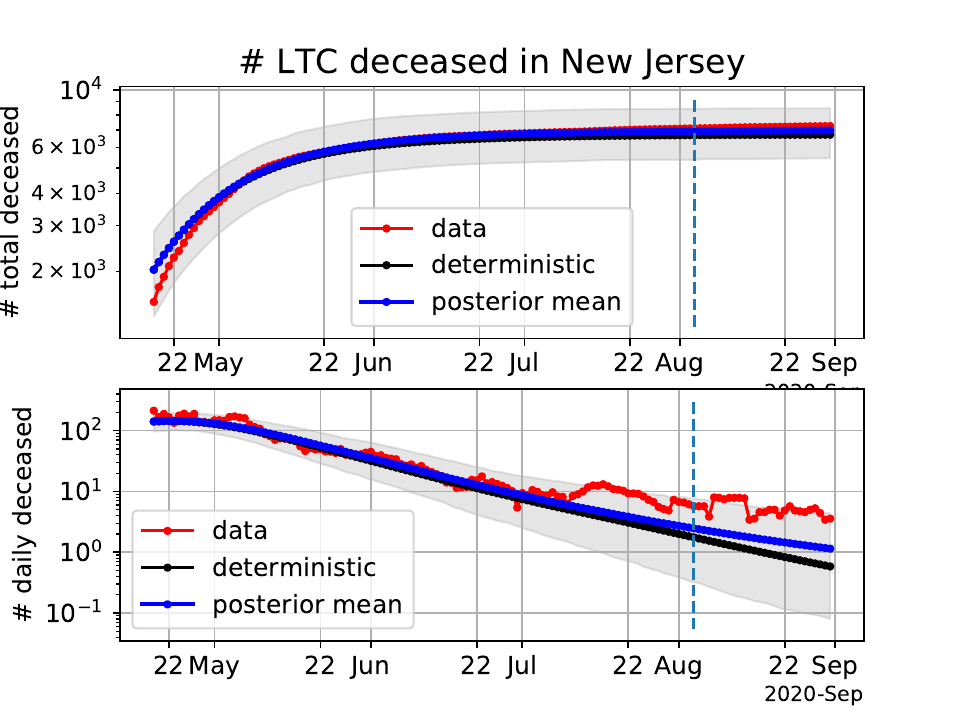}
	
	\caption{Comparison of the data and model outputs for confirmed cases (top-left), hospitalized cases (top-right), total deceased cases (bottom-left), and deceased cases inside LTC (bottom-right) by both the deterministic inversion and the Bayesian inference using data except for the last four weeks, as well as forecast results in the last four weeks, which are separated by the dashed line. The data here are the same as the seven day moving average data in Figure \ref{fig:data-NJ}. The deterministic model outputs are the results of model \eqref{eq:LTC-model} at the optimal parameter value obtained by the deterministic inversion \eqref{eq:inference-deterministic}. The sample mean and 90\% credible interval (90\% samples fall in this interval) of the model outputs are obtained as the results of model \eqref{eq:LTC-model} at the prior samples (left) and posterior samples (right) from the Bayesian inference.}\label{fig:projection-NJ}
\end{figure}

\begin{figure}[!htb]
	\centering
	\includegraphics[width=0.51\linewidth]{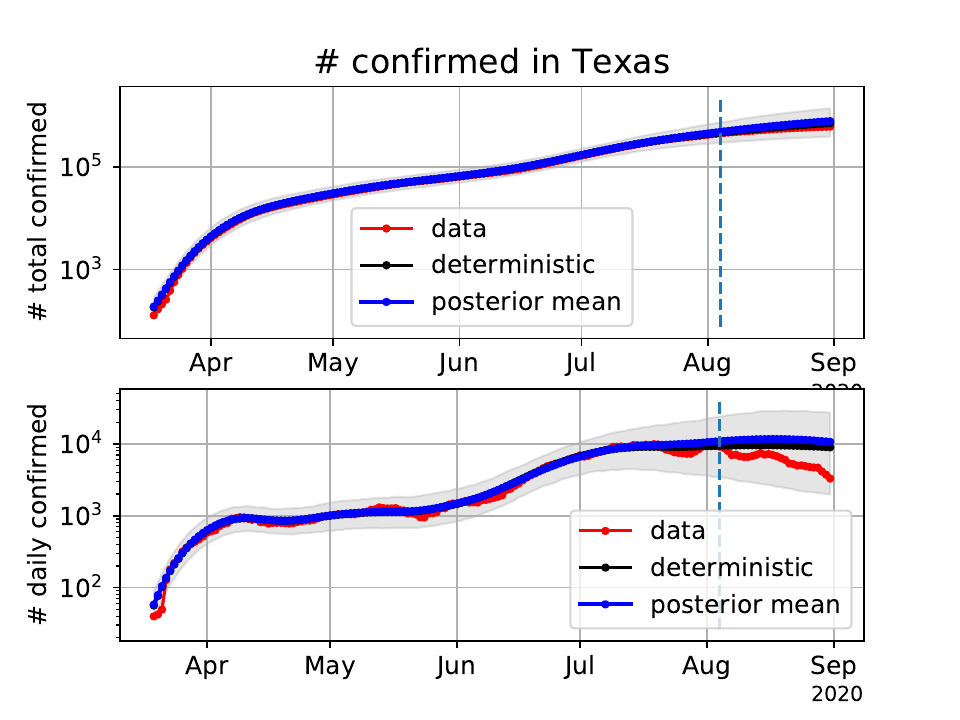}\hspace*{-0.5cm}
	\includegraphics[width=0.51\linewidth]{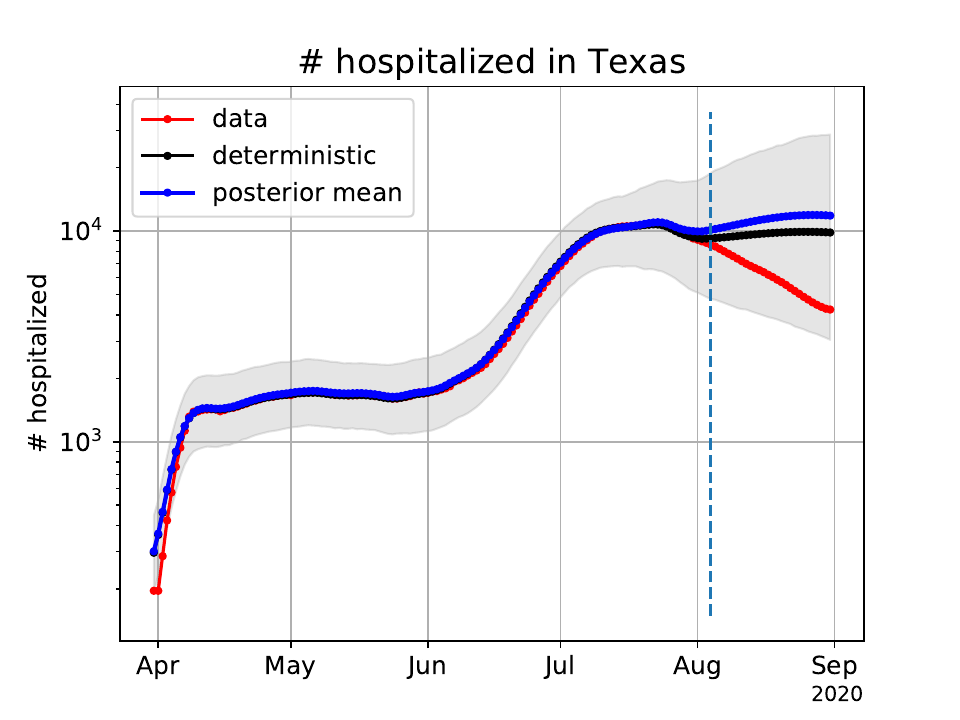}
	
	\vspace*{-0.3cm}
	
	\includegraphics[width=0.51\linewidth]{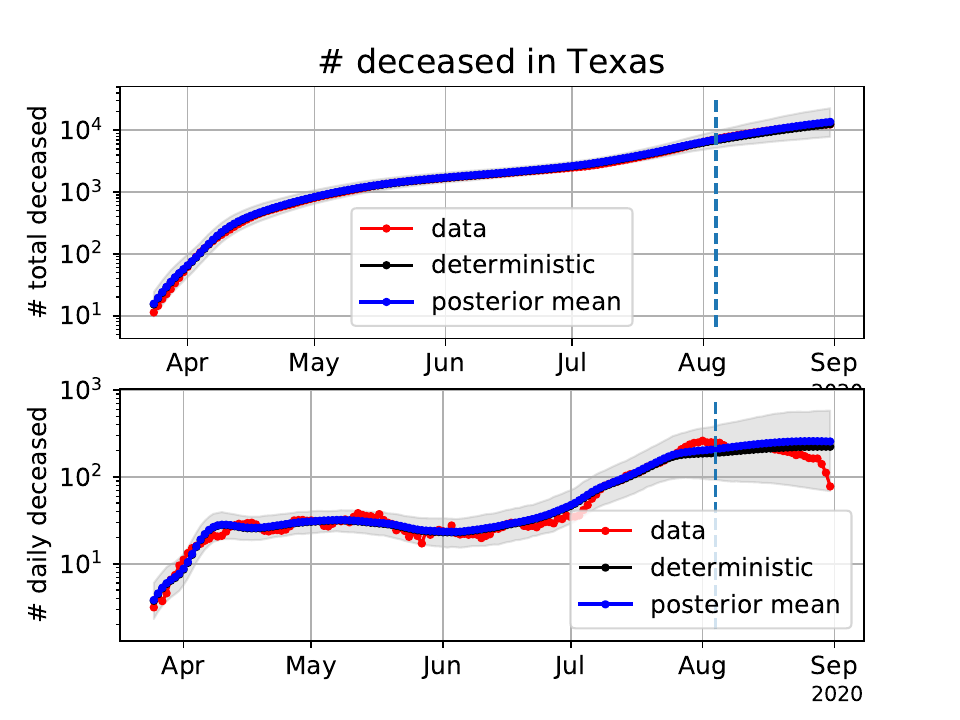}\hspace*{-0.5cm}
	\includegraphics[width=0.51\linewidth]{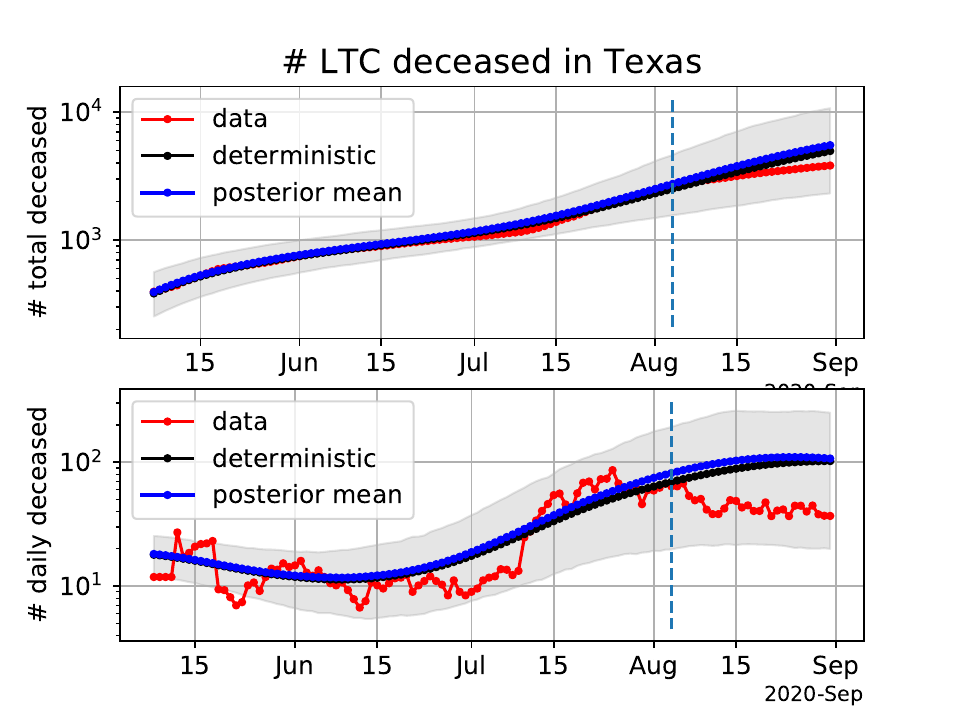}
	
	\caption{The same as in Figure \ref{fig:projection-NJ} for Texas here.}\label{fig:projection-TX}
\end{figure}

In this section, we compare the model forecast under uncertainty and compare the forecast results with the data. We divide the data into two parts, one part including all but the last four weeks, which are used for parameter inference, and the other part including only the last four weeks, which are used for comparison with the forecast. We run pSVGD for Bayesian inference of the parameters with 1000 samples for 100 iterations. We use the 1000 samples at the end of iteration to make forecast for the last four weeks, during which the time dependent parameters are assigned with values at the end of the inference period, i.e., one day before the last four weeks. The inference and forecast results for the confirmed, hospitalized, and deceased cases are shown in Figure \ref{fig:projection-NJ} for New Jersey and Figure \ref{fig:projection-TX} for Texas. We can observe a relatively good agreement of most of the data and the model forecast at both the optimal parameters by deterministic inversion and the sample posterior mean. However, some of the model forecast evidently depart from the data, e.g., the hospitalized cases in Texas with sharp change, due to the evident dynamic change of epidemics reflected by the time dependent parameters (e.g., IHR and HFR outside LTC) in the last four weeks, see Figure \ref{fig:bayesian-parameters-TX}. Nevertheless, the data still fall in the 90\% credible interval of the forecast. We can also observe that the longer the forecast is, the larger the uncertainty of the forecast results become.

\section{Conclusions}
\label{sec:conclusions}

We have proposed a Bayesian inference framework for learning the
dynamics of COVID-19, with application to the spread and outcome
inside and outside LTC facilities, which have experienced about 40\%
of total COVID-19 caused and/or related deaths. We designed an LTC
stratified SEIR compartmental model with extended compartments of
confirmed and unconfirmed positive infections, hospitalization, and
death to account for the available data on the number of positively
confirmed, hospitalized, and deceased cases, as well as the number of
deceased cases inside LTC facilities. To estimate the parameters in
the model, including the time-dependent transmission reduction
factors, various epidemic rates and proportions, as well as the
contact matrix---and more importantly quantify the uncertainties in
the parameters, model outputs, and data---we investigated both a
deterministic inversion and a Bayesian inference approach that builds
on the deterministic result. The deterministic inversion with our
model was demonstrated to effectively capture the main dynamics, while
the Bayesian inference method was demonstrated to achieve reliable
quantification of various uncertainties, in particular providing tight
90\% credible intervals that sufficiently cover the data. The high
dimensional Bayesian inversion was made tractable by our pSVGD
method's ability to exploit the two orders of magnitude effective
dimensionality reduction inherent in inferring the model parameters
from the available data. The posterior predictives produced by the
Bayesian inference 
play a critical role in
a solution of a risk-averse stochastic optimal control problem to
find optimal mitigation strategies for COVID-19, 
which
are presented in a follow-on article \cite{ChenGhattas20d}. 

Limitations of this work include: (1) The data we use inevitably
include incorrect and delayed numbers, 
as discussed for the data in Figure \ref{fig:data-NJ} and
\ref{fig:data-TX}. We accounted for these flaws
only by smoothing the data using seven day moving averages. If they
were available, additional 
data could improve the
results of the inferences, e.g., the number of daily confirmed cases
inside LTC, state dependent hospitalization ratios, time-dependent
positive/negative ratio of the viral test, etc. 
(2) The model we constructed is, as mentioned in Section
\ref{sec:model}, as simple as possible for our target of learning the
dynamics inside and outside LTC. Though it would have been
straightforward to incorporate in our model, we did not account for
the age- and risk-dependent heterogeneities in infection
hospitalization and fatality ratios; their effect can be observed in the widening
misfit in the number of deaths in recent months, probably due to the
shift of infections toward a younger and healthier population.

\section*{Acknowledgement}
We thank Nick Alger, Amal Alghamdi,  Joshua Chen, Ionut Farcas, Longfei Gao,
Dingchen Luo, Thomas O'Leary-Roseberry, and Bassel Saleh for helpful
discussions. 

\bibliographystyle{siamplain}
\bibliography{references}

\appendix

\section{Adjoint-based computation of the gradient}
\label{sec:adjoint}

Without loss of generality, we consider a system of parametric ordinary differential equations 
\begin{equation}\label{eq:ODEs}
\left\{
\begin{split}
y_t(t) & = f(t, y(t), \theta) \quad t \in (0, T],\\
y(0) & = y_0,
\end{split}
\right.
\end{equation}
where the state $y(t)\in \bR^n$, with initial condition $y_0$ that does not depend on the parameter $\theta \in \bR^d$. The term $y_t(t) = \frac{d}{dt} y(t)$ represents the time derivative of $y(t)$. The right hand side $f: \bR\times \bR^n \times \bR^d \to \bR^n$. Suppose we have an objective function $J(\theta)$ to minimize w.r.t.\ $\theta$, which can be written as 
\begin{equation}
J(\theta) = \int_{0}^T g(t, y(t), \theta) dt,
\end{equation}  
where $y(t)$ implicitly depends on the parameter $\theta$ through the ODE system \eqref{eq:ODEs}, $g: \bR\times \bR^n \times \bR^d \to \bR$ may explicitly depend on the parameter $\theta$, e.g., through a regularization term. The goal is to compute the gradient of the objective $\nabla_\theta J(\theta)$,
subject to the ODE system \eqref{eq:ODEs}. We introduce the adjoint variables $\lambda: [0, T] \to \bR^n$ 
and write 
\begin{equation}
J(\theta) = \int_{t=0}^T g(t, y(t), \theta) dt + \int_{t=0}^T \lambda^T(t) (y_t(t) -  f(t, y(t), \theta)) dt 
\end{equation} 
where the last two terms are zeros since \eqref{eq:ODEs} hold. Note that $y$ implicitly depends on $\theta$, we have
\begin{equation}\label{eq:Jgrad}
\begin{split}
\nabla_\theta J(\theta) & = \int_{t=0}^T \partial_y g(t, y(t), \theta) \partial_\theta y(t) + \partial_\theta g(t, y(t), \theta) dt\\
& + \int_{t=0}^T \lambda^T(t) (\partial_\theta y_t(t) - \partial_y f(t, y(t), \theta) \partial_\theta y(t) - \partial_\theta f(t, y(t), \theta)) dt 
\end{split}
\end{equation}
Note that $\partial_\theta y_t(t) =  \frac{d}{dt}\partial_\theta y(t)$ under the assumption that $y$ is continuous in the domain of $t$ and $\theta$. Using integration by parts for the first term in the second line, we have 
\begin{equation}
\int_{t=0}^T \lambda^T(t) \partial_\theta y_t(t) dt = \left(\lambda^T(t) \partial_\theta y(t)\right)\big|_{0}^T - \int_{t=0}^T \lambda_t^T(t) \partial_\theta y(t) dt. 
\end{equation}
Reorganizing \eqref{eq:Jgrad} by putting the terms with $\partial_\theta y(t)$ together we obtain 
\begin{equation}
\begin{split}
\nabla_\theta J(\theta) & = \int_{t=0}^T \left(-\lambda^T_t(t) -  \lambda^T(t)\partial_y f(t, y(t), \theta) + \partial_y g(t, y(t), \theta) \right) \partial_\theta y(t) dt\\
& + \int_{t=0}^T \left(\partial_\theta g(t, y(t), \theta) - \lambda^T(t) \partial_\theta f(t, y(t), \theta) \right) dt \\
& + \lambda^T(T) \partial_\theta y(T) 
\end{split}
\end{equation}
To eliminate the terms with $\partial_\theta y(t)$ and $\partial_\theta y(T)$,
we set the adjoint ODE system
\begin{equation}\label{eq:ODEsAdjoint}
\left\{
\begin{split}
\lambda^T_t(t)  & = -  \lambda^T(t)\partial_y f(t, y(t), \theta) + \partial_y g(t, y(t), \theta), \quad t \in [0, T),\\
\lambda(T) & = 0,
\end{split}
\right.
\end{equation}
which leads to the final result 
\begin{equation}\label{eq:JgradF}
\nabla_\theta J(\theta) = \int_{t=0}^T \left(\partial_\theta g(t, y(t), \theta) - \lambda^T(t) \partial_\theta f(t, y(t), \theta) \right) dt 
\end{equation}
for which we need to solve the ODE system \eqref{eq:ODEs} for $y(t)$, then solve the adjoint ODE system \eqref{eq:ODEsAdjoint} for $\lambda^T(t)$, and finally compute \eqref{eq:JgradF}. Note that the gradient $\partial_y f(t, y(t), \theta) \in \bR^{n\times n}$ , $\partial_y g(t, y(t), \theta) \in \bR^{n}$, $\partial_\theta g(t, y(t), \theta) \in \bR^{d}$, and $\partial_\theta f(t, y(t), \theta) \in \bR^{n\times d}$ 
can be obtained using symbolic differentiation when $f$ and $g$ 
have complicated dependence on $y$ and $\theta$.

\end{document}

%% file: LTC-BIP_shared.tex
\usepackage[square,numbers,sort]{natbib}

\usepackage{amsmath}
\usepackage{amssymb} 
\usepackage{algorithm}
\usepackage{algorithmic}
\usepackage{multirow}
\usepackage{pstool}
\usepackage{multicol}
\allowdisplaybreaks
\usepackage{hyperref}
\hypersetup{
	colorlinks=true,
	linkcolor=blue,
	filecolor=magenta,      
	urlcolor=cyan,
}

\extrafloats{100}

\newcommand{\bC}{{\mathbb{C}}}

\newcommand{\bE}{{\mathbb{E}}}

\newcommand{\bI}{{\mathbb{I}}}

\newcommand{\bR}{{\mathbb{R}}}

\newcommand{\cA}{\mathcal{A}}
\newcommand{\cB}{\mathcal{B}}

\newcommand{\cH}{\mathcal{H}}

\newcommand{\cN}{\mathcal{N}}

\newcommand{\cT}{\mathcal{T}}

\newcommand{\cQ}{\mathcal{Q}}

\newcommand{\argmin}{\operatornamewithlimits{argmin}}

\newcommand{\bsf}{\boldsymbol{f}}
\newcommand{\bsg}{\boldsymbol{g}}
\newcommand{\bsh}{\boldsymbol{h}}

\newcommand{\beq}{\begin{equation}}
\newcommand{\eeq}{\end{equation}}
\newcommand{\ba}{\begin{array}}
	\newcommand{\ea}{\end{array}}

\DeclareMathOperator\arctanh{arctanh}

\usepackage{framed,multirow}

\usepackage{latexsym}

\usepackage{url}
\usepackage{xcolor}
\definecolor{newcolor}{rgb}{.8,.349,.1}

\let\realurl\url
\renewcommand{\url}[1]{%
	\realurl{#1}%
	\wlog{URLX #1 }%
}

\headers{Bayesian inference of heterogeneous epidemic model}{P. Chen, K. Wu, and O. Ghattas}

\title{ Bayesian inference of heterogeneous epidemic models: Application to COVID-19 spread accounting for long-term care facilities
	\thanks{Submitted to the editors DATE.
\funding{This work was partially funded by the Department of Energy, Office of Science, Office of
	Advanced Scientific Computing Research, Mathematical Multifaceted Integrated Capability Centers
	(MMICCS) program under award DE-SC0019303; 
	and the National Science Foundation, 
	Division of Mathematical Sciences under award DMS-2012453.}}}

\author{Peng Chen\thanks{Oden Institute for Computational Engineering \& Sciences, The University of Texas at Austin, Austin, TX.
  (\email{peng@oden.utexas.edu}).}
\and Keyi Wu\thanks{Department of Mathematics, The University of Texas at Austin, Austin, TX.
  (\email{keyiwu@math.utexas.edu}).}
\and Omar Ghattas\thanks{Oden Institute for Computational Engineering \& Sciences,  Department of Geological Sciences, Jackson School of Geosciences, Walker Department of Mechanical Engineering, The University of Texas at Austin, Austin, TX.  (\email{omar@oden.utexas.edu}).}}

\usepackage{amsopn}
